\documentclass[
  b5paper,
  fontsize=10pt,
  twoside=semi,
  toc=listof,%
  toc=bibliography,%
  chapterprefix=true,%
  captions=tableheading,%
  numbers=endperiod%
]{scrbook}
\usepackage{titling}
\usepackage[noadjust]{cite}
\usepackage{bibentry}
\nobibliography*
\usepackage[T1]{fontenc}
\usepackage[english,spanish,es-lcroman]{babel}
\renewcaptionname{english}{\contentsname}{Table of Contents}
\usepackage{lmodern}
\usepackage{amsmath, amssymb, mathtools}
\usepackage{enumitem}
\usepackage{graphicx}
\usepackage{mathrsfs}
\usepackage{subcaption}
\usepackage{xfrac}
\usepackage{setspace}
\usepackage{calc}
\usepackage{tikz}
\usetikzlibrary{calc,decorations.pathreplacing}

\addtokomafont{sectioning}{\rmfamily}
\addtokomafont{chapter}{\linespread{1}}

\usepackage[headsepline]{scrlayer-scrpage}
\KOMAoptions{onpsinit=\linespread{1}}

\usepackage{etoolbox}
\makeatletter
\preto{\chapterheadendvskip}{%
  \if@mainmatter\else
    \vspace{-\baselineskip}\hfill%
    \tikz{\coordinate(h);%
      \draw[overlay,remember picture,line width=1.5pt]%
        ([yshift=-7pt]h)--++(-150mm,0);}%
    \par\fi%
}
\makeatother

\usepackage{amsthm}

\newtheorem{theorem}{Theorem}[section]
\newtheorem{corollary}[theorem]{Corollary}
\newtheorem{lemma}[theorem]{Lemma}
\newtheorem{proposition}[theorem]{Proposition}
\theoremstyle{definition}
\newtheorem{definition}[theorem]{Definition}
\newtheorem{assumption}[theorem]{Assumption}
\theoremstyle{remark}
\newtheorem{remark}[theorem]{Remark}

\newtheorem*{example}{Example}
\newcommand{\triang}{\hfill$\triangle$}
\AtEndEnvironment{example}{\triang}
\numberwithin{equation}{section}

\usepackage[refpage, english, noprefix, intoc]{nomencl}

\makenomenclature

\usepackage{xr}
\usepackage{subfiles}
\ifSubfilesClassLoaded{
  \externaldocument{\subfix{main}}
}{}

\usepackage[a-1b]{pdfx}
\hypersetup{linktocpage,unicode=true,bookmarks=true,bookmarksopen=false,breaklinks=false,pdfborder={0 0 0},colorlinks=false}
\usepackage[utf8]{inputenc}

\usepackage{xcolor}
\definecolor{cblue}{rgb}{0.16, 0.32, 0.75}
\definecolor{cred}{rgb}{0.7, 0.11, 0.11}
\hypersetup{%
	,linkcolor=cred
	,citecolor=cblue
}
\newcommand{\advisors}[1]{\def\theadvisors{#1}}
\renewcommand*{\mid}{:\,}
\renewcommand*{\d}{\mathrm{d}}

\DeclareMathOperator{\dom}{dom}

\newcommand{\argdot}{{\hspace{0.18em}\cdot\hspace{0.18em}}}
\renewcommand{\Re}{\mathrm{Re}}
\renewcommand{\Im}{\mathrm{Im}}

\begin{document}
\frontmatter
\author{Á. Aitor Balmaseda Martín}
\advisors{Prof.\ Alberto Ibort Latre \\[1mm] Prof.\ Juan Manuel Pérez Pardo}
\title{Quantum Control at the Boundary}
\date{November, 2021}
\pagestyle{plain}
\begin{titlepage}
	\begin{center}
		\begin{huge}
      \textbf{\thetitle\\}
    \end{huge}
    \begin{LARGE}
      \vspace{0.5cm}
      by\\
      \vspace{0.5cm}
    \end{LARGE}
    \begin{LARGE}
			\theauthor\\
    \end{LARGE}
    \vspace{3cm}	
    \begin{large}
      \textit{
      A dissertation submitted in partial fulfilment of the requirements for the degree of Doctor of Philosophy in\\[1em]
      Mathematical Engineering\\}
      \vspace{1.5cm}
      Universidad Carlos III de Madrid\\
      \vspace{1.5cm}
      Advisors:\\[1em]
      \theadvisors\\
      \vspace{2cm}
      \thedate\\
   \end{large}
		 
	\end{center}
\end{titlepage}
	
\thispagestyle{empty}
\vspace*{\fill}
\begin{center}
This thesis is distributed under license ``Creative Commons \textbf{Atributtion - Non Commercial - Non Derivatives}''.\\
\medskip
\includegraphics[width=4.2cm]{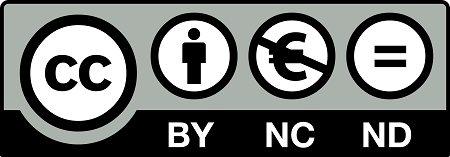} 
\end{center}

\cleardoublepage

\chapter{Agradecimientos}   
\selectlanguage{spanish}
  Tras avistar en el horizonte la tierra que anuncia el final de este primer viaje, es momento de echar la vista atrás y agradecer a quienes lo han hecho posible.

  Empezando por el principio, quiero agradecer a quienes han dirigido este trabajo: Alberto Ibort y Juan Manuel Pérez Pardo; sus voces se escuchan entre las líneas de esta tesis.
  Si Alberto no hubiese aparecido, ofreciéndome sus ideas en mi búsqueda de Trabajo de Fin de Máster, yo no me habría embarcado en este viaje.
  Si Alberto ha sido la brújula que ha señalado el rumbo a seguir estos años, Juanma ha sido el mapa con el que he aprendido a transitar los caminos que desembocan en esta tesis.
  Gracias a los dos, por vuestra guía.

  También hay aquí algo de Fernando Lledó: gracias por regar la semilla de la que brotó mi Trabajo de Fin de Máster antes de florecer esta tesis.

  Gracias también a David Krejčiřík por su hospitalidad en Praga (y en Marsella) y por recordarme que no todos los operadores son autoadjuntos.
  A Fabio Di Cosmo y a Davide Lonigro: el eco de nuestras discusiones también resuena en estas páginas.

  Más allá de las palabras aquí escritas, en los espacios que las separan se entreven apoyos más sutiles pero no menos importantes.
  Gracias a mis padres, Elena y Ángel, por enseñarme primero a andar por el camino que hoy transito y después acompañarme en sus giros; también a mi hermana y mi sobrino, por alegrar sus tramos, y a Carmen y Mila, que han sufrido sus altibajos empujando siempre en la buena dirección.

  A Elías, por sujetarme la cabeza sobre los hombros y los pies sobre el suelo. Por poner la luz que no ponía el sol del invierno praguense.
  Por seguir acompañándome.
  Por todo, gracias.

  Termino recordando, inevitablemente, ese viento variable que ha soplado tanto en la buena dirección como en todas la demás.
  Gracias a esos compañeros de aventuras, por descubrirme una inesperada verdad: quemarlo todo puede no ser la solución óptima.
  También a Dani, por ese enrale revitalizante y por todos estos años: gracias.

  \vspace*{-1.9mm}
  \begin{flushright}
    A.\ Balmaseda\\
    Madrid, octubre de 2021
  \end{flushright}
  \pagebreak
  \begin{small}
    \selectlanguage{english}
    This work was partially supported by the ``Ministerio de Economía, Industria y Competitividad'' research project MTM2017-84098-P, the QUITEMAD project P2018/TCS-4342 funded by ``Comunidad Autónoma de Madrid'' and by ``Universidad Carlos III de Madrid'' through its Ph.D. program grant PIPF UC3M 01-1819 and through its mobility grant in 2020.
  \end{small}

\chapter{Published and submitted content}
\thispagestyle{empty}
The main results of this thesis are based on the following list of publications.

\begin{itemize}
  \item \bibentry{Balmaseda2018}.\\
    \url{https://arxiv.org/abs/2108.00495}.

    I am the author of this work, and it is partially included in Chapters \ref{ch:H_form_domain}, \ref{ch:quantum_circuits} and \ref{ch:control_systems}.
    The material from this source included in this thesis is not singled out with typographic means and references.

  \item \bibentry{BalmasedaPerezPardo2019}.\\
    \href{https://link.springer.com/chapter/10.1007/978-3-030-24748-5_5}{DOI:10.1007/978-3-030-24748-5\_5}.

    I am an author of this work, and it is partially included in Chapters \ref{ch:H_form_domain}, \ref{ch:quantum_circuits} and \ref{ch:control_systems}.
    The material from this source included in this thesis is not singled out with typographic means and references.

  \item \bibentry{BalmasedaDiCosmoPerezPardo2019}.
    \href{https://www.mdpi.com/2073-8994/11/8/1047}{DOI:10.3390/sym11081047}.

    I am an author of this work, and it is partially included in Chapter \ref{ch:quantum_circuits}.
    The material from this source included in this thesis is not singled out with typographic means and references.

  \item \bibentry{BalmasedaLonigroPerezPardo2021}.\\
    \url{https://arxiv.org/abs/1910.04135}

    I am an author of this work, and it is partially included in Chapters \ref{ch:H_form_domain}-\ref{ch:control_systems}.
    The material from this source included in this thesis is not singled out with typographic means and references.
\end{itemize}

\chapter{Resumen}
\selectlanguage{spanish}
  El principal objetivo de esta tesis es presentar y probar la viabilidad de un método no estándar para controlar el estado de un sistema cuántico, cuya dinámica está gobernada por la ecuación de Schrödinger, modificando sus condiciones de frontera en lugar de interaccionar con el sistema a través de campos externos para dirigir su estado.

  El control del estado de sistemas cuánticos está volviéndose más y más significativo dado el asombroso avance experimental que está teniendo lugar motivado por la carrera para alcanzar tecnologías cuánticas efectivas.
  Un requisito natural para cualquier procesador de información cuántico es el control de los estados de un cúbit.
  Por ejemplo, los cúbits basados en espín pueden controlarse a través de puertas universales basadas en la rotación del espín un cierto ángulo sobre un eje dado \cite{PressLaddZhangEtAl2008}.
  Una demostración de la viabilidad del control y de la lectura de alta fidelidad de un cúbit basado en espín puede encontrarse en \cite{Morello2015,PlaTanDehollainEtAl2013}.

  El marco matemático necesario para trabajar con el control de sistemas cuánticos basados en el espín es la Teoría de Control Geométrica (véase \cite{DAlessandro2007} para un repaso de este enfoque).
  También puede mencionarse \cite{Moseley2004,BonnardGlaserSugny2012}, dónde se realiza un estudio del control de sistemas cuánticos de espín utilizando dicha teoría.
  Una revisión del estado de la cuestión en control (óptimo) de sistemas cuánticos puede verse en \cite{GlaserBoscainCalarcoEtAl2015}.
  Sin embargo, la Teoría de Control Geométrica encuentra problemas serios a la hora de tratar sistemas infinito dimensionales, provenientes de las dificultades intrínsecas a la geometría en dimensión infinita.
  Éste constituye un problema estándar que se aborda también, por ejemplo, en Teoría de Campos y por lo tanto puede ser sorteado.
  Sin embargo, lo cierto es que tan sólo un pequeño número de resultados han aparecido en las fuentes bibliográficas, sobre todo referentes a la controlabilidad de sistemas bilineales (cf.\ sec.\ \ref{sec:bilinear_quantum_control}).
  Véanse por ejemplo los trabajos de Beauchard \emph{et al.}\ \cite{Beauchard2005,BeauchardCoron2006} y Chambrion \emph{et al.}\ \cite{ChambrionMasonSigalottiEtAl2009}.

  El enfoque habitual del control de sistemas cuánticos se basa en el uso de campos externos para manipular el estado del sistema.
  Desde un punto de vista tecnológico, aparecen dificultades al intentar controlar un sistema cuántico de este modo, que tienen que ver con las complicaciones para manipular un sistema hecho de unas pocas partículas mientras se intenta mantener las correlaciones cuánticas.
  Como consecuencia, el sistema cuántico debe mantenerse a muy baja temperatura y las interacciones debe producirse muy rápidamente \cite{Bacciagaluppi2016}, lo cual es inconveniente para las aplicaciones.

  El paradigma de Control Cuántico en la Frontera aborda el control de sistemas cuánticos de una forma diametralmente opuesta al enfoque estándar.
  En lugar de buscar el control del sistema cuántico interactuando directamente con el mismo a través de un campo externo, el control se consigue manipulando las condiciones de contorno del sistema.
  El espectro de un sistema cuántico, por ejemplo un electrón moviéndose en un pozo de potencial, depende de las condiciones de contorno (típicamente Dirichlet o Neumann).
  Por lo tanto, modificando dichas condiciones de contorno puede modificarse el estado del sistema, lo cual permitiría en última instancia controlarlo \cite{Ibort2010}.
  Este tipo de interacción, que es más débil en cierto sentido, hace esperar que sea más sencillo mantener las correlaciones cuánticas.
  Llamamos sistema (cuántico) de control en la frontera a los sistemas de control definidos dentro de este paradigma.

  El marco establecido por el Control Cuántico en la Frontera ha sido usado para mostrar cómo generar estados entrelazados en sistemas compuestos modificando las condiciones de frontera del sistema \cite{IbortMarmoPerezPardo2014}.
  La relación entre dicho paradigma y la topología del sistema ha sido explorada en \cite{PerezPardoBarberoLinanIbort2015} y recientemente usada para describir las propiedades físicas de sistemas con paredes móviles \cite{FacchiGarneroMarmoEtAl2016,FacchiGarneroMarmoEtAl2018,FacchiGarneroLigabo2018,FacchiGarneroLigabo2018a,Garnero2018}.
  Sin embargo, a pesar del interés intrínseco que tienen, algunos problemas básicos como la controlabilidad de sistemas sencillos dentro de este paradigma no han sido aún estudiados.

  La mayoría de los sistemas cuánticos que aparecen en las aplicaciones para desarrollar ordenadores cuánticos, tales como las trampas de iones o los circuitos superconductores, son infinito dimensionales.
  Mientras tanto, los modelos usados para describirlos son aproximaciones finito dimensionales \cite{Wendin2017,LawEberly1996}, lo que introduce errores en dicha descripción.
  Es por tanto natural buscar mejores modelos matemáticos para estos sistemas.
  Buenos candidatos para estos modelos pueden construirse a partir de lo que llamamos Circuitos Cuánticos, una generalización de los grafos cuánticos (esto es, grafos métricos equipados con un operador diferencial en sus aristas, junto con condiciones de contorno apropiadas que definan una extensión autoadjunta de dicho operador \cite{KostrykinSchrader2003,Kuchment2004}).
  Por un lado, los Circuitos Cuánticos comparten con los dispositivos físicos la estructura tipo grafo y la dinámica en ellos está gobernada por operadores diferenciales (en concreto, consideraremos el operador de Laplace-Beltrami y Laplacianos magnéticos junto con algunos potenciales que representan campos magnéticos y eléctricos en los Circuitos Cuánticos).
  Por otro lado, los Circuitos Cuánticos son suficientemente sencillos como para encontrar soluciones de la ecuación de evolución numéricamente (o incluso analíticamente).
  Todo esto hace interesante considerar la dinámica y la controlabilidad de sistemas definidos en Circuitos Cuánticos.
  Algunos ejemplos concretos de Circuitos Cuánticos pueden encontrarse al final del Capítulo \ref{ch:quantum_circuits}, véase la Figura \ref{fig:examples}.

  Más aún, los grafos cuánticos pueden verse como un caso concreto de Circuito Cuántico, por lo que los resultados presentados en esta tesis pueden aplicarse directamente a ellos.
  Y, además de el interés matemático de estudiar la dinámica de sistemas definidos en grafos cuánticos, creemos que éstos pueden ser utilizados para modelizar circuitos superconductores.
  Más concretamente, es sabido que las junturas de Josephson (que son un dispositivo fundamental en los circuitos superconductores, cf.\ \cite{Wendin2017,WendinShumeiko2007,DevoretSchoelkopf2013}) pueden modelizarse usando condiciones de frontera \cite{KulinskiiPanchenko2015,TruemanWan2000}.
  Nótese que los circuitos superconductores son una de las prometedoras nuevas tecnologías en el desarrollo de ordenadores cuánticos.

  Para mostrar la viabilidad del método de Control Cuántico en la Frontera, definimos una familia de sistemas de control cuántico en la frontera sobre Circuitos Cuánticos.
  Para ello, utilizamos la caracterización de las extensiones autoadjuntas del operador de Laplace-Beltrami desarrolladas en \cite{IbortLledoPerezPardo2015}, que parametriza las extensiones autoadjuntas en función de operadores unitarios actuando en el espacio de datos en la frontera.
  Estos operadores unitarios son adecuados para representar los grupos de simetría de la variedad subyacente \cite{IbortLledoPerezPardo2015a}, hecho que fue usado en \cite{BalmasedaDiCosmoPerezPardo2019} para caracterizar las condiciones de frontera compatibles con las estructuras de grafo que consideramos aquí.

  Antes de poder abordar el problema de controlabilidad, es necesario afrontar el problema de existencia de soluciones de la ecuación de Schrödinger con condiciones de contorno dependientes del tiempo, problema que tiene su propia importancia.
  Situaciones similares a la que presentaremos han sido consideradas anteriormente para algunos sistemas cuánticos concretos \cite{DellAntonioFigariTeta2000,DiMartinoAnzaFacchiEtAl2013,PerezPardoBarberoLinanIbort2015}.
  La dependencia temporal del dominio del Hamiltoniano del sistema añade dificultades extra a las habituales complicaciones que surgen de la naturaleza discontinua de los operadores no acotados, lo que compromete la existencia de soluciones para la ecuación de Schrödinger.
  En este sentido, los resultados más generales de que disponemos son debidos a J.\ Kisyński \cite{Kisynski1964}.
  Basándose en una idea aproximativa de K.\ Yosida \cite{Yosida1948}, Kisyński propone condiciones suficientes para la existencia de operadores de green para problemas de Cauchy abstractos en espacios de Banach, incluyendo el caso en el que los operadores que definen la evolución tienen dominios dependientes del tiempo.
  Estos resultados pueden ser aplicados al caso concreto de la ecuación de Schrödinger, tal y como hace el propio Kisyński, cf.\ \cite[Sec.\ 8]{Kisynski1964}.
  Más concretamente, proporciona condiciones suficientes para la existencia de propagadores unitarios para sistemas cuánticos cuyos Hamiltonianos tienen dominio dependiente del tiempo pero no así las formas sesquilineales asociadas a éstos.
  Este caso también fue estudiado por B.\ Simon \cite{Simon1971}, cuyo enfoque se basa en el mismo método aproximativo de Yosida, pero abordado el problema de manera ligeramente distinta.

  Para simplificar la exposición, en lugar de aplicar los resultados más generales de Kisyński directamente a nuestros sistemas cuánticos de control en la frontera, introduciremos una familia de sistemas de control estándar relacionada, que llamamos sistemas de inducción cuántica, cuyos Hamiltonianos tendrán dominio de forma constante.
  La relación entre estas dos familias de sistemas es usada frecuentemente en la discusión, lo que nos permite probar los resultados primero para los sistemas de inducción cuántica y transferirlos a los sistemas de control en la frontera.
  Siguiendo esta estrategia, usaremos los resultados de Simon y de Kisyński para Hamiltonianos con dominio de forma constante, obteniendo condiciones suficientes para que ambos sistemas tengan una dinámica bien definida.

  En el estudio de las propiedades dinámicas de los Hamiltonianos dependientes del tiempo con dominio de forma constante, la escala de espacios de Hilbert asociada a el Hamiltoniano, $\mathcal{H}^+ \subset \mathcal{H} \subset \mathcal{H}^-$, aparece como una herramienta básica tanto en el desarrollo de Simon como en el de Kisyński.
  A lo largo de esta tesis, usamos nociones tales como la existencia de dinámica o la controlabilidad tanto en $\mathcal{H}$ como en $\mathcal{H}^-$; por ello, usaremos los adjetivos \emph{fuerte} y \emph{débil} para referirnos a las nociones en $\mathcal{H}$ y en $\mathcal{H}^-$ respectivamente.

  Una vez resuelto el problema de la existencia de soluciones de la ecuación de Schrödinger, podremos abordar el problema de controlabilidad.
  Usaremos como una herramienta útil para este fin el Teorema \ref{thm:stability_bound}, un resultado de estabilidad que generaliza los resultados de A.D.\ Sloan \cite{Sloan1981} constituyendo uno de los resultados originales principales expuestos en esta tesis.
  Que sepamos, esta generalización es el resultado de estabilidad más general en el contexto de la ecuación de Schrödinger dependiente del tiempo.
  La demostración de este teorema se basa en las técnicas usadas por Simon y, como parte de dicha demostración, obtenemos una ligera generalización de algunos resultados en \cite{Simon1971} (cf.\ Lemma~\ref{lemma:propagators_bound_simon}).
  La importancia de este resultado de estabilidad no reside tan solo en su papel a la hora de demostrar la controlabilidad de los sistemas de control bajo estudio, permitiendo demostrar la viabilidad del Control Cuántico en la Frontera; tiene consecuencias de mayor alcance para la Teoría de Control Cuántico ya que permite obtener cotas \emph{a priori} del error cometido al controlar el estado del sistema.

  Demostramos la controlabilidad aproximada de los sistemas de inducción cuántica basándonos en un resultado de T.\ Chambrion \emph{et al.}\ en \cite{ChambrionMasonSigalottiEtAl2009}, donde estudian la controlabilidad de una clase de sistemas bilineales de control cuántico con controles constantes a trozos.
  La estructura que tienen los sistemas de inducción cuántica es similar, pero no exactamente igual, a la de los problemas estudiados por Chambrion \emph{et al.}
  Por ello, definimos un sistema auxiliar para el cual el teorema de Chambrion \emph{et al.} asegura controlabilidad aproximada y después, usando el resultado de estabilidad, mostramos que esta controlabilidad aproximada se extiende al sistema de inducción cuántica original.
  Si embargo, la clase de controles considerada por Chambrion \emph{et al.} no puede ser llevada automáticamente a sistemas con Hamiltonianos cuyos dominios dependen del tiempo, ya que las soluciones de la ecuación de Schrödinger (en el sentido fuerte) podrían no existir.
  Por lo tanto, la controlabilidad aproximada obtenida para los sistemas de inducción cuántica usando este procedimiento es en el sentido débil (es decir, en $\mathcal{H}^-$).
  Usando de nuevo el resultado de estabilidad podemos demostrar controlabilidad aproximada en sentido fuerte para los sistemas de inducción cuántica con controles suaves, y ésta puede ser transferida a los sistemas de control cuántico en la frontera usando la relación entre ambos.

  Estos resultados de controlabilidad, demostrados en el Capítulo~\ref{ch:control_systems}, junto con los resultados de estabilidad en el Capítulo~\ref{ch:H_form_domain} constituyen las contribuciones originales más relevantes de esta tesis.

  Por último, remarcamos el hecho de que la caracterización de las extensiones autoadjuntas que usamos aquí puede ser aplicada a otros operadores diferenciales como el operador de Dirac \cite{IbortPerezPardo2015,PerezPardo2017,IbortLlavonaLledoEtAl2021}, por lo que los resultados presentados en esta tesis podrían extenderse a tales casos.
  Más aún, dicha caractrerización está adaptada al cálculo numérico \cite{IbortPerezPardo2013,LopezYelaPerezPardo2017}, por lo que podrían desarrollarse algoritmos con aplicaciones al control óptimo de estos sistemas a partir de los resultados aquí mostrados.
\selectlanguage{english}

\cleardoublepage
\tableofcontents            
\cleardoublepage

\mainmatter
\pagestyle{scrheadings}


\chapter{Introduction}
  The main goal of this dissertation is to present and prove the viability of a non-standard method for controlling the state of a quantum system, whose dynamic is governed by the Schrödinger equation, by modifying its boundary conditions instead of relying on the action of external fields to drive the state of the system.

  Control of the states of a quantum system is becoming more and more significative because of the outstanding experimental achievements that are taking place as the race to reach effective quantum information technologies picks up.
  A basic demand for quantum information systems is the control of the state of a single qubit.
  For instance, spin based qubits can be controlled using a universal gate based on rotations of the spin by any angle around a given axis \cite{PressLaddZhangEtAl2008}.
  Proof of concept for the manipulation and high-fidelity readout of individual spin qubits was shown for instance in \cite{Morello2015,PlaTanDehollainEtAl2013}.

  The mathematical background to deal with quantum spin control has been Geometric Control Theory (see \cite{DAlessandro2007} for a review on the subject).
  We may mention \cite{Moseley2004,BonnardGlaserSugny2012} where a study of geometric control of quantum spin systems is shown.
  A review on the state-of-the-art on quantum (optimal) control can be found in \cite{GlaserBoscainCalarcoEtAl2015}.
  However, Geometric Control Theory has serious problems when dealing with infinite-dimensional systems because of the analytical difficulties encountered when dealing with geometry in infinite dimensions.
  This is a standard problem faced in Field Theory too and, even if it can be addressed, the truth is that only a few results are known, mainly on controllability of bilinear systems (see Sec.\ \ref{sec:bilinear_quantum_control}).
  See for instance Beauchard et al.\ \cite{Beauchard2005,BeauchardCoron2006} and Chambrion et al.\ \cite{ChambrionMasonSigalottiEtAl2009}.

  The standard approach to quantum control bases on the use of an external field to manipulate the system.
  From a technological point of view, some difficulties appear when controlling a quantum system in this way, related with the complications of manipulating a system made of few particles while maintaining the quantum correlations.
  As a consequence the quantum systems need to be kept under very low temperatures and the interactions with them have to be performed very fast \cite{Bacciagaluppi2016}, which is inconvenient for applications.

  The Quantum Control at the Boundary scheme takes an approach which is radically different to the standard one.
  Instead of seeking the control of the quantum system by directly interacting with it through an external field, the control is achieved by manipulating the boundary conditions of the system.
  The spectrum of a quantum system, for instance an electron moving in a box, depends on the boundary conditions imposed on it (usually Dirichlet or Neumann boundary conditions).
  Hence, a modification of such boundary conditions modifies the state of the system allowing for its manipulation and, eventually, its control \cite{Ibort2010}.
  This kind of interaction is weaker, which makes one to expect that it may help maintaining the quantum correlations.
  We call (quantum) boundary control systems to the control systems defined within this control scheme.

  The Quantum Control at the Boundary paradigm has been used to show how to generate entangled states in composite systems by modifications of the boundary conditions \cite{IbortMarmoPerezPardo2014}.
  The relation of the paradigm and topology change has been explored in \cite{PerezPardoBarberoLinanIbort2015} and recently used to describe the physical properties of systems with moving walls \cite{FacchiGarneroMarmoEtAl2016,FacchiGarneroMarmoEtAl2018,FacchiGarneroLigabo2018,FacchiGarneroLigabo2018a,Garnero2018}.
  But, in spite of its intrinsic interest, some basic issues such as the controllability of simple systems within this scheme have never been addressed.

  Most of the quantum systems that appear in the applications to develop quantum computers, like ion traps or superconducting circuits, are infinite-dimensional; however, the models used to describe them are finite-dimensional approximations \cite{Wendin2017,LawEberly1996}.
  This introduces errors in their description, and it is natural to look for better mathematical models for them.

  Good candidates to provide such models are what we call Quantum Circuits, a generalisation of Quantum Graphs (i.e., metric graphs equipped with a differential operator defined on its edges together with appropriate boundary conditions to determine a self-adjoint extension \cite{KostrykinSchrader2003,Kuchment2004}).
  On one side, Quantum Circuits share with the physical aforementioned devices the graph-like structure and their dynamics is governed by differential operators.
  In particular, we will consider the Laplace-Beltrami operator and magnetic Laplacians, together with some potentials that represent magnetic and electric fields in the Quantum Circuit.
  On the other side, Quantum Circuits are simple enough such that solutions can be found numerically or even analytically: this makes it worth to study dynamics and controllability on Quantum Circuits.
  Some explicit examples of the construction of Quantum Circuits can be found at the end of Chapter~\ref{ch:quantum_circuits}, see Fig.\ \ref{fig:examples}.

  Moreover, Quantum Graphs can be seen as a particular case of Quantum Circuits and therefore the results presented on this dissertation can be applied to them.
  In addition to the mathematical interest of studying dynamics on Quantum Graphs, we believe that these systems can be used to model superconducting circuits.
  In particular, it is known that Josephson Junctions, a device of major importance in superconducting circuits \cite{Wendin2017,WendinShumeiko2007,DevoretSchoelkopf2013}, can be modelled using boundary conditions \cite{KulinskiiPanchenko2015,TruemanWan2000}; superconducting circuits are the basis of one of the promising new technologies in the development of quantum computers.

  In order to show the viability of the Quantum Control at the Boundary method, we define a family of boundary control systems on Quantum Circuits.
  For this purpose, we exploit the characterisation of self-adjoint extensions of the Laplace-Beltrami operator developed in \cite{IbortLledoPerezPardo2015}.
  This characterisation parametrises self-adjoint extensions in terms of unitary operators acting on the space of boundary data; these unitaries are suitable to carry representations of the symmetry groups of the underlying manifold \cite{IbortLledoPerezPardo2015a}.
  This fact was used in  \cite{BalmasedaDiCosmoPerezPardo2019} to characterise the boundary conditions compatible with the graph structure that we consider here.

  Before being able to address the problem of controllability, we need to address the problem of existence of solutions for the Schrödinger equation with time-dependent boundary conditions, a problem that has its own significance.
  Situations similar to the ones presented here have been considered before in some particular quantum systems \cite{DellAntonioFigariTeta2000,DiMartinoAnzaFacchiEtAl2013,PerezPardoBarberoLinanIbort2015}.
  The time-dependence of the domain of the Hamiltonian adds extra difficulties to the usual ones coming from the non-continuous nature of unbounded operators, compromising the existence of solutions of the Schrödinger equation.
  In this sense, the most general result available is due to J.\ Kisyński \cite{Kisynski1964}.
  Basing on an approximative idea from K.\ Yosida \cite{Yosida1948}, Kisyński provides sufficient conditions for the existence of green operators for \emph{abstract Cauchy problems on Banach spaces}.
  The case of time-dependent domains of the operators defining the evolution is covered and sufficient conditions for the existence of dynamics are established. These results can be applied to the particular case of the Schrödinger equation.
  In particular, \cite[Sec.\ 8]{Kisynski1964} provides sufficient conditions for the existence of unitary propagators for a quantum system whose Hamiltonian has time-dependent domain such that the closed Hermitian sesquilinear form associated with the Hamiltonian has time-independent domain.
  This case is also studied by B.\ Simon \cite{Simon1971}, relying on the same approximating method by Yosida but with a slightly different approach.

  In order to simplify things, instead of applying Kisyński's most general result to our quantum boundary control system, we introduce a related family of standard control systems, which we call induction control systems, whose Hamiltonians have time-dependent domain but constant form domain.
  This relation is exploited through all the dissertation, allowing us to prove results first for the induction control systems and then transfer them to the bondary control systems.
  Following this strategy, we use Simon's and Kisyński's results for Hamiltonians with constant form domain to provide sufficient conditions for induction control systems to have well defined dynamics and transfer these results to quantum boundary control systems.
  Hamiltonians with time-independent form domain allow for the introduction of a powerful analytical tool to address these problems: the scale of Hilbert spaces associated to the Hamiltonian, $\mathcal{H}^+ \subset \mathcal{H} \subset \mathcal{H}^-$.
  Through this dissertation, notions such as the existence of dynamics or the controllability are going to be used on both the Hilbert space $\mathcal{H}$ and $\mathcal{H}^-$.
  This leads us to use the adjective \emph{strong} for referring to notions applied on $\mathcal{H}$, such as \emph{strong dynamics} or \emph{strong controllability}, and \emph{weak} for the notions applied on $\mathcal{H}^-$.

  Once the existence of solutions for the Schrödinger equation is established, the controllability problem can be addressed.
  A useful tool for this purpose is Theorem \ref{thm:stability_bound}, a stability result which generalises the results by A.D.\ Sloan \cite{Sloan1981} and constitutes one of the main original contributions on this dissertation.
  This non-trivial generalisation constitutes, to our knowledge, the most general stability result in the context of time-dependent Schrödinger equations with form-constant Hamiltonians.
  The proof of this theorem bases on the techniques used by Simon and, in proving the result, we obtain a slight generalisation of some results in \cite{Simon1971} (cf.\ Lemma~\ref{lemma:propagators_bound_simon}).
  The importance of this stability result is not only because it allows us to prove the controllability of the control systems under study, hence showing the viability of the Quantum Control at the Boundary scheme.
  It has far reaching consequences for Quantum Control Theory as it allows to obtain \emph{a priori} estimates on the error committed when driving the state of the system.

  We show the controllability of induction control systems relying on a controllability result by T.\ Chambrion et al.\ in \cite{ChambrionMasonSigalottiEtAl2009}, where controllability for a class of bilinear quantum control systems with piecewise constant control functions is proven.
  The structure of induction control systems is similar the structure of the quantum control systems where Chambrion et al.'s result holds.
  However, these results cannot be applied directly and we need to define an auxiliary system for which Cambrion et al.'s theorem ensures approximate controllability.
  Then, using the stability result, this approximate controllability can be transferred to the induction control system.
  The kind of controls provided by Chambrion et al.'s theorem cannot be directly applied to the situation with time-dependent domains, as solutions of Schrödinger's equation may not exist.
  As a consequence, the approximate controllability obtained for induction control systems using the procedure described above holds only in the weak sense (cf.\ Definitions \ref{def:strong_controllability} and \ref{def:weak_controllability}).
  Applying again the stability result we are able to prove approximate controllability for the induction control system with smooth controls in the strong sense.
  This approximate controllability is transferred to quantum boundary control systems through their relation with induction systems.

  These results, proven in Chapter \ref{ch:control_systems}, together with the stability results in Chapter \ref{ch:H_form_domain} constitute the main original contributions in this dissertation.
  Although pointed out both by Simon in his book \cite{Simon1971} and by Sloan in \cite{Sloan1981}, the details on the relation between B.\ Simon's and J.\ Kisyński's existence results are shown for the first time in Section \ref{subsec:simon_vs_kisynski}.

  Finally, let us remark that the characterisation of self-adjoint extensions that we use in this work can be used for other differential operators like Dirac operators \cite{IbortPerezPardo2015,PerezPardo2017,IbortLlavonaLledoEtAl2021}.
  Therefore, it is likely that similar results can be proven also in these other situations.
  Moreover, this characterisation is suitable for numerics \cite{IbortPerezPardo2013,LopezYelaPerezPardo2017}, so that further developments of the control problem with applications to optimal control may also be enabled by the results presented here.

  This dissertation is organised as follows.
  Chapter \ref{ch:hilbert_spaces} introduces the notation and needed ideas of the theory of Hilber spaces, operators and sesquilinear forms.
  The basics of Operator Theory on Hilbert spaces are presented in Section \ref{sec:hilbert_operators}, while Section \ref{sec:forms} presents the notions on sesquilinear forms needed through the text.
  The construction of scales of Hilbert spaces (also known as rigged Hilbert spaces or Gelfand triples) is reviewed in Section \ref{sec:hilbert_scales}.
  Finally, Section \ref{sec:op_form_functions} establishes some regularity results for operator-valued and (sesquilinear) form-valued functions that are going to be used in the rest of this work.

  Chapter \ref{ch:quantum_control} summarises some key concepts from Control Theory.
  In Section \ref{sec:quantum_control} the notions of approximate controllability, \emph{weak} and \emph{strong}, for quantum control systems are introduced.
  T.\ Chambrion et al.'s main result in \cite{ChambrionMasonSigalottiEtAl2009} is stated in Section \ref{sec:bilinear_quantum_control}, while in Section \ref{sec:quantum_control_boundary} the scheme of Quantum Control at the Boundary is introduced.

  Chapter \ref{ch:H_form_domain} deals with the existence of solutions for the Schrödinger equation associated to Hamiltonians with time-dependent domains such that the domain of the associated family of Hermitian sesquilinear forms is constant.
  In Section \ref{sec:quantum_dynamics} Hamiltonians with constant form domain are introduced, as well as the \emph{weak} and \emph{strong} Schrödinger equations.
  Section \ref{sec:existence_dynamics} presents both B.\ Simon's and J.\ Kisyński's approaches for proving the existence of solutions for the Schrödinger equation, establishing the relation between them and stating the existence theorems used in the rest of this dissertation.
  In Section \ref{sec:stability} the stability result is proven, showing that converging Hamiltonians with constant form domain induce similar dynamics, in a sense to be specify later.
  Form-linear Hamiltonians, a subfamily of the Hamiltonians with constant form domain, is defined in Section \ref{sec:form_linear}.

  Chapter \ref{ch:laplacian} studies the self-adjoint extensions of the Laplace-Beltrami operator on a Riemannian manifold.
  Sobolev spaces on a Riemannian manifold and the Laplace-Beltrami operator are introduced in Section \ref{sec:sobolev_laplace}.
  The method developed in \cite{IbortLledoPerezPardo2015}, which exploits the relation between closed semibounded Hermitian forms with semibounded self-adjoint operators for characterising self-adjoint extensions of the Laplace-Beltrami operator, is reviewed in Section \ref{sec:sa_extensions}.

  Chapter \ref{ch:quantum_circuits} describes Quantum Circuits; in Section \ref{sec:quantum_circ_def} the boundary conditions admissible on a Quantum Circuit are discussed, fixing the self-adjoint extensions of the Laplace-Beltrami operator (and also those of the magnetic Laplacian) compatible with the topology of the circuit.
  The boundary conditions defining self-adjoint extensions which will be used to define the family of quantum boundary control system under study are also defined.
  In Section \ref{sec:quasi_delta_BC}, the relation between the self-adjoint extensions of the Laplace-Beltrami operator and those of the magnetic Laplacian is analysed.

  Finally, Chapter \ref{ch:control_systems} studies the controllability problem for both the induction control system and the boundary control system.
  Both systems are defined in Section \ref{sec:control_systems_def}, where the relation between them is exploited to prove the existence of solutions for Schrödinger equation using the results in Chapter \ref{ch:H_form_domain}.
  Then, Section \ref{sec:controllability} addresses the controllability problem for both systems, yielding sufficient conditions for approximate controllability.
\ifSubfilesClassLoaded{
  \bibliographystyle{plain}
  \bibliography{../tesis}
  }{}
\end{document}

\documentclass[../main.tex]{subfiles}
\ifSubfilesClassLoaded{
    \setcounter{chapter}{1}
  }{}
\begin{document}
\chapter{Introduction to Hilbert spaces, operators and forms} \label{ch:hilbert_spaces}
  As a preparation for the upcoming exposition, this chapter is devoted to introduce the basic ideas and notations on Hilbert spaces that will be needed in the following chapters.
  In Section \ref{sec:hilbert_operators}, standard results on Operator Theory on Hilbert spaces are presented with the aim of fixing the notation, while Section \ref{sec:forms} introduces sesquilinear forms.
  Section \ref{sec:hilbert_scales} reviews the construction of scales of Hilbert spaces and states some results which are going to be extensively used throughout this work.
  Finally, Section \ref{sec:op_form_functions} establishes some results on the regularity of form-valued functions and how it relates with the regularity of the associated operator-valued functions.

\section{Operators on Hilbert spaces} \label{sec:hilbert_operators}
  \nomenclature[H]{$\mathcal{H}$}{Complex, separable Hilbert space}%
  \nomenclature[I]{$\langle \cdot, \cdot \rangle$}{Inner product of the Hilbert space $\mathcal{H}$}%
  \nomenclature[N]{$\Vert\cdot\Vert$}{Norm of the Hilbert space $\mathcal{H}$}%
  Let $\mathcal{H}$ be a complex, separable Hilbert space with inner product $\langle \cdot, \cdot \rangle$ and norm $\|\cdot\|$.
  An operator $T$ on $\mathcal{H}$ is a linear map $T: \dom T \subset \mathcal{H} \to \mathcal{H}$ densely defined, that is, $\dom T$ is a dense subspace of $\mathcal{H}$.
  Details and proofs of the following results can be found, for instance, on \cite{ReedSimon1980}.

  \begin{definition}
    Let $\mathcal{H}_1$, $\mathcal{H}_2$ be complex, separable Hilbert spaces with associated norms $\|\cdot\|_{\mathcal{H}_1}$ and $\|\cdot\|_{\mathcal{H}_2}$.
    An operator $T: \dom T \subset \mathcal{H}_1 \to \mathcal{H}_2$ is said to be \emph{bounded} if there exists a positive constant $M$ such that
    \begin{equation*}
      \|T\Phi\|_{\mathcal{H}_2} \leq M \|\Phi\|_{\mathcal{H}_1} \quad \Phi \in \dom T.
    \end{equation*}
    We denote by $\mathcal{B}(\mathcal{H}_1, \mathcal{H}_2)$ the linear space of bounded operators $T: \mathcal{H}_1 \to \mathcal{H}_2$, endowed with the norm
    \begin{equation*}
      \|T\|_{\mathcal{B}(\mathcal{H}_1,\mathcal{H}_2)} = \sup_{\Phi \in \dom T \smallsetminus \{0\}} \frac{\|T\Phi\|_{\mathcal{H}_2}}{\|\Phi\|_{\mathcal{H}_1}}.
    \end{equation*}
    \nomenclature[B]{$\mathcal{B}(\mathcal{H}_1, \mathcal{H}_2)$}{Space of bounded, linear operators from $\mathcal{H}_1$ to $\mathcal{H}_2$}%
    \nomenclature[N]{$\Vert \cdot \Vert_{\mathcal{B}(\mathcal{H}_1,\mathcal{H}_2)}$}{Norm in the space of bounded operators from $\mathcal{H}_1$ to $\mathcal{H}_2$}%
    When it leads to no confusion, we will simplify the notation by omitting the subscript on the operator norm.
  \end{definition}

  Bounded operators are continuous and therefore can be extended by continuity from its dense domain to the whole Hilbert space, which makes unnecessary to specify a domain for them.
  For this reason, we will always consider that the domain of a bounded operator is $\mathcal{H}$.
  This is not the case for unbounded operators, for which the domain cannot be extended in the same way.

  The following result, known as Uniform Boundedness Principle or Banach-Steinhaus Theorem, holds in the more general case of Banach spaces (see \cite[Thm.\ III.9]{ReedSimon1980}).
  However, for the purposes of this work we state it for operators defined on a Hilbert space.

  \begin{theorem}[Uniform Boundedness Principle]
    Let $\mathcal{H}$ be a Hilbert space and $Y$ be a normed, linear space.
    Let $\mathcal{F}$ be a family of bounded operators from $\mathcal{H}$ to $Y$.
    If, for each $\Phi \in \mathcal{H}$ fixed, the set $\{\|T\Phi\|_Y \mid T \in \mathcal{F}\}$ is bounded, then $\{\|T\| \mid T \in \mathcal{F}\}$ is bounded.
  \end{theorem}

  \begin{definition}
    Let $T$ and $T_1$ be operators from $\mathcal{H}_1$ to $\mathcal{H}_2$.
    We say that $T$ is an \emph{extension} of $T_1$ if $\dom T_1 \subset \dom T$ and $T\Phi = T_1 \Phi$ for every $\Phi \in \dom T_1$, and we denote it by $T_1 \subset T$.
  \end{definition}

  Despite the fact that unbounded operators are not continuous, some of them can be seen as continuous operators on their domains.
  \begin{definition}
    Let $T: \mathcal{H}_1 \to \mathcal{H}_2$ be an operator with dense domain.
    $T$ is said to be \emph{closed} if for every Cauchy sequence $\{\Phi_n\}_n \subset \mathcal{H}_1$ such that $\{T\Phi_n\}_n$ is a Cauchy sequence in $\mathcal{H}_2$ it holds
    \begin{equation*}
      \Phi = \lim_{n \to \infty} \Phi_n \in \dom T \quad\text{and}\quad \lim_{n \to \infty} T\Phi_n = T \Phi.
    \end{equation*}
    An operator is said to be \emph{closable} if it has a closed extension.
    For a closed operator $T$, its closure, denoted by $\overline{T}$, is defined as its smallest closed extension.
  \end{definition}
  \begin{remark}
    For an operator $T$, being closed is equivalent to $\dom T$ being closed with respect to the graph norm
    \begin{equation*}
      \|\cdot\|_T = \sqrt{\|\cdot\|_{\mathcal{H}_1}^2 + \|T\cdot\|_{\mathcal{H}_2}^2}.
    \end{equation*}
    \nomenclature[N]{$\Vert\cdot\Vert_T$}{Graph norm of the operator $T$}%
    Note that, even if it is always possible to complete $\dom T$ with respect to the graph norm, this closure needs not be the domain of a closed extension of the operator $T$.
  \end{remark}
  The following theorem, relating closedness and boundedness of an operator, holds also for operators in Banach spaces (see \cite[Thm.\ III.12]{ReedSimon1980}).
  \begin{theorem}[Closed Graph Theorem]
    Let $T: \mathcal{H}_1 \to \mathcal{H}_2$ be an operator with $\dom T = \mathcal{H}_1$.
    Then $T$ is bounded if and only if $T$ is closed.
  \end{theorem}

  A core concept of this work is the notion of self-adjoint operators, but before going into the definition of self-adjointness we need to define the adjoint of an (unbounded) operator.
  \begin{definition}
    Let $T$ be an operator on $\mathcal{H}$ with dense domain, and let $\Phi \in \mathcal{H}$.
    $\Psi$ is in the domain of the \emph{adjoint operator}, $T^\dagger$, if there exists $\chi \in \mathcal{H}$ such that
    \begin{equation*}
      \langle \Psi, T\Phi \rangle = \langle \chi, \Phi \rangle
    \end{equation*}
    for every $\Phi \in \dom T$.
    For $\Psi \in \dom T^\dagger$, we define $T^\dagger\Psi \coloneqq \chi$.
  \end{definition}
  Note that $\chi$ is uniquely determined since the domain $\dom T$ is dense on $\mathcal{H}$.
  It can be shown \cite[Thm. VIII.1]{ReedSimon1980} that a densely defined operator $T$ is closable if and only if $\dom T^\dagger$ is dense, in which case $\overline{T} = (T^\dagger)^\dagger$.

  \begin{definition}
    Let $T$ be a closed operator on $\mathcal{H}$.
    $\lambda \in \mathbb{C}$ is said to be in the \emph{resolvent set}, $\rho(T)$, if $\lambda \mathbb{I} - T$ is a bijection of $\dom T$ onto $\mathcal{H}$ with bounded inverse.
    $R_\lambda(T) = (\lambda \mathbb{I} - T)^{-1}$ is called the \emph{resolvent} of $T$ at $\lambda$.
    The \emph{spectrum} of $T$, $\sigma(T)$, is the complement of the resolvent set: $\sigma(T) = \mathbb{C} \smallsetminus \rho(T)$.
  \end{definition}

  \begin{definition}
    An operator $T$ is said to be \emph{symmetric} if $T \subset T^\dagger$; that is,
    \begin{equation*}
      \langle \Psi, T\Phi \rangle = \langle T\Psi, \Phi \rangle,\qquad \text{for all }\Psi,\Phi \in \dom T.
    \end{equation*}
  \end{definition}
  Therefore, for a symmetric operator $T$, $\dom T \subset \dom T^\dagger$ and $T^\dagger$ is densely defined.
  Hence, $T$ is closable and its closure is $(T^\dagger)^\dagger$.
  \begin{definition}
    An operator $T$ is called \emph{self-adjoint} if $T=T^\dagger$.
    That is, if and only if $T$ is symmetric and $\dom T = \dom T^\dagger$.
  \end{definition}

  The difference between self-adjoint and symmetric operators can only be found on unbounded operators: the adjoint of a bounded operator is bounded, and therefore bounded symmetric operators are self-adjoint.
  In general, it is easier to check if an operator is symmetric since it only involves calculations with the operator itself.
  On the other hand, the definition of self-adjointness involves calculating the domain of the adjoint operator which is a subtle task.
  However, it is self-adjointness the notion having a relevant role for generalizing the nice properties from Hermitian matrices such as the spectral theorem and the real spectrum.
  Also, it plays a central role in Quantum Mechanics, since self-adjoint operators generate unitary groups (Stone's Theorem \cite[Thm.\ VIII.12]{ReedSimon1980}).

  \begin{definition}
    Let $T$ be a self-adjoint operator.
    We say that $T$ is \emph{semibounded from below} (or lower semibounded) if there is $m \in \mathbb{R}$ such that
    \begin{equation} \label{eq:def_semibounded}
      \langle \Phi, T\Phi \rangle \geq m \|\Phi\|^2, \quad \text{for all } \Phi \in \dom T,
    \end{equation}
    and we write $T \geq m$.
    The biggest $m$ such that Eq.\ \eqref{eq:def_semibounded} holds is the \emph{lower semibound} of $T$.
    $T$ said to be \emph{positive} if $m \geq 0$ and \emph{strictly positive} if $m > 0$.
  \end{definition}
  \begin{remark}
    In a similar way, one can define \emph{upper semiboundedness} and \emph{(strictly) negative} operators.
    However, these concepts are not going to be used in the rest of this dissertation.
    For this reason, we will use \emph{semibounded} and \emph{semibound} as synonyms of \emph{lower semibounded} and \emph{lower semibound}.
    We will also use \emph{lower bound} as a synonym of \emph{lower semibound}.
  \end{remark}

  \begin{definition}
    Let $T_1, T_2$ be self-adjoint operators.
    We write $T_1 \geq T_2$ if $T_1 - T_2$ is a positive operator, and $T_1 > T_2$ if $T_1 - T_2$ is strictly positive.
  \end{definition}

\section{Sesquilinear forms} \label{sec:forms}
  Self-adjoint operators semibounded from below are closely related to closed sesquilinear forms, see for instance \cite[Ch.\ VI]{Kato1995}, \cite[Sec.\ VIII.6]{ReedSimon1980}.
  Throughout this work, we observe the convention establishing that a sesquilinear form is anti-linear on its first argument and linear on the second one.
  We will only consider sesquilinear forms $h: \dom h \times \dom h \to \mathbb{C}$ defined on dense subspaces of the Hilbert space $\mathcal{H}$: i.e., $\dom h$ is a dense subspace of $\mathcal{H}$.
  Extensions of sesquilinear forms are defined analogously to the extensions of operators.
  \begin{definition}
    A sesquilinear form $h$ is said to be \emph{bounded} if
    \begin{equation*}
      |h(\Psi, \Phi)| \leq K \|\Psi\| \|\Phi\|, \quad \text{for all } \Psi,\Phi \in \dom h,
    \end{equation*}
    and the smallest $K$ satisfying the preceding inequality is called the \emph{bound} of $h$.
  \end{definition}
  \begin{remark}
    Note that, like in the case of bounded operators, bounded sesquilinear forms can be extended by continuity and therefore without loss of generality one can consider $\dom h = \mathcal{H}$.
  \end{remark}

  For bounded sesquilinear forms, an analogous of the Uniform Boundedness Principle can be proven.
  \begin{theorem} \label{thm:unif_boundedness_forms}
    Let $\mathcal{H}$ be a Hilbert space and let $\mathcal{F}$ be a family of bounded sesquilinear forms.
    If for every $\Psi, \Phi \in \mathcal{H}$ the set $\{|h(\Psi, \Phi)| \mid h \in \mathcal{F}\}$ is bounded, then
    \begin{equation*}
      \{|h(\Psi, \Phi)| \mid \|\Psi\| \leq 1, \|\Phi\| \leq 1, \Psi,\Phi \in \mathcal{H},\, h \in \mathcal{F}\}
    \end{equation*}
    is bounded.
  \end{theorem}
  \begin{proof}
    For fixed $\Psi \in \mathcal{H}$, consider the family of bounded operators from $\mathcal{H}$ to $\mathbb{C}$
    \begin{equation*}
      \mathcal{T}_\Psi = \{\Phi \mapsto h(\Psi,\Phi) \mid h \in \mathcal{F}\}.
    \end{equation*}
    By assumption, for every $\Psi, \Phi \in \mathcal{H}$, there is a constant $K_{\Psi,\Phi} > 0$ such that $|h(\Psi, \Phi)| \leq K_{\Psi,\Phi}$.
    Therefore, by the Uniform Boundedness Principle the set
    \begin{equation*}
      \{\|T\| \mid T \in \mathcal{T}_\Psi\} = \{|h(\Psi, \Phi)| \mid \|\Phi\| \leq 1, \Phi \in \mathcal{H},\, h \in \mathcal{F}\}
    \end{equation*}
    is bounded.
    Applying again the Uniform Boundedness Principle to the family of bounded operators $\{\Psi \mapsto \overline{h(\Psi,\Phi)} \mid \|\Phi\| \leq 1, \Phi \in \mathcal{H},\, h \in \mathcal{F}\}$, the result follows.
  \end{proof}

  \begin{definition}
    A sesquilinear form $h$ is called \emph{Hermitian} if
    \begin{equation*}
      h(\Psi, \Phi) = \overline{h(\Phi, \Psi)} \qquad \text{for all } \Psi,\Phi \in \dom h.
    \end{equation*}
    An Hermitian sesquilinear form $h$ is \emph{semibounded from below} if there is $m \in \mathbb{R}$ such that
    \begin{equation*}
      h(\Phi,\Phi) \geq m \|\Phi\|^2.
    \end{equation*}
    The biggest value $m$ satisfying the preceding inequality is called the \emph{lower semibound} (or simply lower bound) of $h$.
    We say that $h$ is \emph{positive} if $m \geq 0$, and \emph{strictly positive} if $m > 0$.
  \end{definition}
  As for operators, we will only consider lower semibounded sesquilinear forms, and will use \emph{semibounded} and \emph{semibound} as synonyms of \emph{lower semibounded} and \emph{lower semibound}.
  We will also use \emph{lower bound} as a synonym of \emph{lower semibound}.

  An important concept to relate sesquilinear forms with operators is the notion of closed quadratic form.

  \begin{definition}\label{def:graph-norm}
    Let $h$ be a semibounded Hermitian sesquilinear form with dense domain $\dom h$, and let $m \geq 0$ be such that $-m$ is the semibound of $h$.
    We define the \emph{graph norm} of the sesquilinear form $h$ by
    \begin{equation*}
      \|\Phi\|_h:=\sqrt{(1+m)\|\Phi\|^2+h(\Phi,\Phi)},\quad \Phi\in\dom h.
    \end{equation*}
    We say $h$ is \emph{closed} if $\dom h$ is closed with respect to the graph norm $\|\cdot\|_h$.
    The sesquilinear form $h$ is said to be \emph{closable} if there exists a closed sesquilinear form that extends it.
  \end{definition}

  We recall next an important result, cf.\ \cite[Sec.\ VI.2]{Kato1995}.
  \begin{theorem}[Representation Theorem]\label{thm:repKato}
    Let $h$ be an Hermitian, closed, semibounded sesquilinear form defined on the dense domain $\dom h \subset \mathcal{H}$.
    Then there exists a unique, self-adjoint, semibounded operator $T$ with domain $\mathcal{D}$ and the same lower bound such that:
    \begin{enumerate}[label=\textit{(\roman*)},nosep, leftmargin=*]
      \item $\Phi \in \mathcal{D}$ if and only if $\Phi \in \dom h$ and there exists
        $\chi \in \mathcal{H}$ such that
      \begin{equation*}
          h(\Psi, \Phi) = \langle \Psi, \chi \rangle, \qquad \forall \Psi \in \dom h.
      \end{equation*}
      \item $h(\Psi, \Phi) = \langle \Psi, T\Phi \rangle$ for any $\Psi \in \dom h$,
        $\Phi \in \mathcal{D}$.
      \item $\mathcal{D}$ is a core for $h$, that is,
        $\overline{\mathcal{D}}^{\|\cdot\|_h} = \dom h$.
    \end{enumerate}
  \end{theorem}
  Note that this theorem establishes a one-to-one correspondence between closed, semibounded Hermitian sesquilinear forms and semibounded self-adjoint operators.
  This motivates the following definition.
  \begin{definition} \label{def:representing_op}
    Let $h$ be a closed, semibounded Hermitian sesquilinear form.
    The operator $T$ defined in Theorem \ref{thm:repKato} is said to be the operator \emph{representing} $h$.
    Conversely, $h$ is called the sesquilinear form \emph{represented by} $T$.
  \end{definition}

\section{Scales of Hilbert spaces} \label{sec:hilbert_scales}
  Through the text, the notion of scales of Hilbert spaces (also known as rigged Hilbert spaces or Gelfand triples) is used repeatedly.
  The basic ideas of the construction are recalled briefly on this subsection.
  Details and proofs of the following statements can be found for instance in \cite[Ch. I]{Berezanskii1968}.

  Let $\mathcal{H}^+ \subset \mathcal{H}$ be a dense subspace of the Hilbert space $\mathcal{H}$, and let $\langle \cdot, \cdot \rangle_+$ be an inner product endowing $\mathcal{H}^+$ with the structure of a Hilbert space and such that the associated norm, $\|\cdot\|_+$, satisfies
  \begin{equation*}
    \|\Phi\| \leq \|\Phi\|_+, \qquad \Phi \in \mathcal{H}^+.
  \end{equation*}
  By the Riesz Representation Theorem, the restriction of the inner product of $\mathcal{H}$ can be represented using the inner product in $\mathcal{H}^+$. That is, there exists an operator $\hat{J}: \mathcal{H} \to \mathcal{H}^+$ such that
  \begin{equation*}
    \langle \Psi, \Phi \rangle = \langle \hat{J} \Psi, \Phi \rangle_+, \qquad
    \Psi \in \mathcal{H}, \Phi \in \mathcal{H}^+.
  \end{equation*}
  This operator is injective, and allows to define another inner product on $\mathcal{H}$,
  \begin{equation*}
    \langle \cdot, \cdot \rangle_- \coloneqq \langle \hat{J}\cdot, \hat{J}\cdot \rangle_+.
  \end{equation*}
  Let $\mathcal{H}^-$ be the completion of $\mathcal{H}$ with respect to the norm $\|\cdot\|_-$ associated to $\langle \cdot, \cdot \rangle_-$.
  The operator $\hat{J}$ can be extended by continuity to an isometric bijection $J: \mathcal{H}^- \to \mathcal{H}^+$.
  \nomenclature[H]{$\mathcal{H}^\pm$}{Spaces in the construction of a scale of Hilbert spaces}
  The spaces $\mathcal{H}, \mathcal{H}^\pm$ form the scale of Hilbert spaces $\mathcal{H}^+ \subset \mathcal{H} \subset \mathcal{H}^-$.

  Finally, since
  \begin{equation*}
    |\langle \Psi, \Phi \rangle| = |\langle J\Psi, \Phi \rangle_+| \leq \|J\Psi\|_+ \|\Phi\|_+
    = \|\Psi\|_- \|\Phi\|_+, \quad
    \Psi \in \mathcal{H}, \Phi \in \mathcal{H}^+,
  \end{equation*}
  the inner product on $\mathcal{H}$ can be continuously extended to a pairing
  \nomenclature[P]{$(\cdot, \cdot)$}{Canonical pairing between $\mathcal{H}^+$ and $\mathcal{H}^-$}
  \begin{equation*}
    (\cdot, \cdot): \mathcal{H}^- \times \mathcal{H}^+ \cup \mathcal{H}^+ \times \mathcal{H}^- \to \mathbb{C}.
  \end{equation*}
  Note also that, by definition,
  \nomenclature[I]{$\langle \cdot, \cdot \rangle_{\pm}$}{Inner product on $\mathcal{H}^\pm$}
  \begin{equation*}
    \langle \Psi, \Phi \rangle_{\pm} = (\Psi, J^{\mp} \Phi), \qquad \Psi,\Phi \in \mathcal{H}^\pm.
  \end{equation*}

  With these ideas at hand, let us introduce now the scale of Hilbert spaces associated to a sesquilinear form.
  Let $h: \mathcal{H}^+ \times \mathcal{H}^+ \to \mathbb{C}$ be an Hermitian, strictly positive sesquilinear form such that $\mathcal{H}^+$ is complete with respect to the norm induced by the inner product $\langle \cdot, \cdot \rangle_+ \coloneqq h(\cdot, \cdot)$ and satisfying
  \begin{equation*}
    \|\Phi\|^2 \leq h(\Phi, \Phi), \qquad \Phi \in \mathcal{H}^+.
  \end{equation*}
  It is clear then that following the previous construction we can define the scale of Hilbert spaces $\mathcal{H}^+ \subset \mathcal{H} \subset \mathcal{H}^-$.

  Let $H$ be the positive, self-adjoint operator representing $h$; that is,
  \begin{equation*}
    \langle \Psi, \Phi \rangle_+ = h(\Psi, \Phi) = \langle \Psi, H\Phi \rangle, \quad
  \end{equation*}
  for all $\Psi \in \mathcal{H}^+$ and $\Phi \in \dom H$.
  Note that, if $H$ is strictly positive, $H^{-1} \in \mathcal{B}(\mathcal{H})$ is well defined and $\hat{J} = H^{-1}$.
  Therefore, the operators $H, H^{-1}$ can be extended to $\tilde{H} = J^{-1} \in \mathcal{B}(\mathcal{H}^+, \mathcal{H}^-)$ and $\tilde{H}^{-1} = J \in \mathcal{B}(\mathcal{H}^-, \mathcal{H}^+)$.
  To simplify the notation, we will denote these extensions and the original operators by the same symbols, $H$ and $H^{-1}$.

  Note also that $H$ being self-adjoint and strictly positive make $H^{\pm\sfrac{1}{2}}$ well-defined.
  Hence, it follows
  \begin{equation*}
    \langle \Psi, \Phi \rangle_{\pm} = \langle H^{\pm\sfrac{1}{2}}\Psi, H^{\pm\sfrac{1}{2}}\Phi \rangle
  \end{equation*}
  and
  \begin{equation*}
    H^{\pm\sfrac{1}{2}} \mathcal{H}^\pm = \mathcal{H}, \qquad H^{\pm\sfrac{1}{2}}\mathcal{H} = \mathcal{H}^{\mp}.
  \end{equation*}

  This discussion leads to the following definition.

  \begin{definition} \label{def:hilbert-scales-h}
    Let $\mathcal{H}$ be a Hilbert space with associated inner product $\langle \cdot, \cdot \rangle$, $\mathcal{H}^+ \subset \mathcal{H}$ be a dense subspace and let $h: \mathcal{H}^+ \times \mathcal{H}^+ \to \mathbb{C}$ be an Hermitian, strictly positive, closed sesquilinear form.
    Denote by $H$ the strictly positive self-adjoint operator representing $h$ (cf.\ Def.\ \ref{def:representing_op}), and define the inner products
    \begin{equation*}
      \langle \Psi, \Phi \rangle_\pm = (\Psi, H^{\pm 1}\Phi) \qquad \Psi, \Phi \in \mathcal{H}^{\pm},
    \end{equation*}
    where $\mathcal{H}^-$ is the closure of $\mathcal{H}$ with respect to the norm $\|\cdot\|_-$ induced by $\langle \cdot, \cdot \rangle_-$.
    The \emph{scale of Hilbert spaces associated to $h$} is the scale
    \begin{equation*}
      (\mathcal{H}^+, \langle \cdot, \cdot \rangle_+)
      \subset (\mathcal{H}, \langle \cdot, \cdot \rangle)
      \subset (\mathcal{H}^-, \langle \cdot, \cdot \rangle_-).
    \end{equation*}
    The \emph{scale of Hilbert spaces associated to a semibounded sesquilinear form} $\tilde{h}$ with semibound $-m$ is the scale associated to the strictly positive sesquilinear form $h(\Psi, \Phi) = \tilde{h}(\Psi, \Phi) + m + 1$.
  \end{definition}

  It is also possible to define the scale of Hilbert spaces associated to a semibounded self-adjoint operator.
  \begin{definition}
    Let $H$ be a lower semibounded self-adjoint operator, densely defined on $\mathcal{H}$.
    The scale of Hilbert spaces associated to $H$ is the scale associated to the closed Hermitian sesquilinear form represented by $H$ (cf.\ Def.\ \ref{def:representing_op}).
  \end{definition}

  Let us finish this section with a useful result on scales of Hilbert spaces sharing a common $\mathcal{H}^+$.
  \begin{theorem} \label{thm:eqiv+_equiv-}
    Let $\mathcal{H}^+$ be a dense subset of a Hilbert space $\mathcal{H}$ and let $\langle \cdot, \cdot \rangle_{+,1}$ and $\langle \cdot, \cdot \rangle_{+,2}$ be inner products given rise to the scale of Hilbert spaces
    \begin{equation*}
      (\mathcal{H}^+, \langle \cdot, \cdot \rangle_{+,i}) \subset \mathcal{H} \subset (\mathcal{H}^-, \langle \cdot, \cdot \rangle_{-,i}),\quad i=1,2.
    \end{equation*}
    Denote by $\|\cdot\|_{\pm,i}$ the norm on $\mathcal{H}^\pm$, and let $c > 0$.
    The following are equivalent:
    \begin{enumerate}[label=\textit{(\roman*)},nosep,leftmargin=*]
      \item \label{item:equiv+}
        For all $\Phi \in \mathcal{H}^+$, $c^{-1} \|\cdot\|_{+,1} \leq \|\cdot\|_{+,2} \leq c \|\cdot\|_{+,1}$.
      \item \label{item:equiv-}
        For all $\Phi \in \mathcal{H}^-$, $c^{-1} \|\cdot\|_{-,1} \leq \|\cdot\|_{-,2} \leq c \|\cdot\|_{-,1}$.
    \end{enumerate}
  \end{theorem}
  \begin{proof}
    Let $A_i$ be the strictly positive self-adjoint operator such that
    \begin{equation*}
      \langle \Psi, \Phi \rangle_{+,i} = \langle A_i \Psi, A_i \Phi \rangle,\quad
      i = 1, 2.
    \end{equation*}
    By the Closed Graph Theorem, the operator defined by $T \coloneqq A_1 A_2^{-1}$ is a bounded operator on $\mathcal{H}$ and, given $\Phi\in\mathcal{H}^+$,
    \begin{equation*}
      \|\Phi\|_{+,1} = \|A_1A_2^{-1}A_2\Phi\|
      \leq \|T\| \|A_2\Phi\| = \|T\| \|\Phi\|_{+,2}.
    \end{equation*}
    Analogously one can get  $\|\Psi\|_{+,2} \leq \|T^{-1}\| \|\Psi\|_{+,1}$ for $\Psi \in \mathcal{H}^+$.

    Using the adjoint of $T$, $T^\dagger = A_2^{-1} A_1$, one can prove similar inequalities for the norms $\|\cdot\|_{-,i}$, $i=1,2$:
    \begin{equation} \label{eq:Tdagger_minus_norm}
      \|\Phi\|_{-,1} \leq \|T^{-\dagger}\| \|\Phi\|_{-,2}, \quad
      \|\Phi\|_{-,2} \leq \|T^\dagger\| \|\Phi\|_{-,1}.
    \end{equation}

    If \ref{item:equiv+} holds, one has
    \begin{equation*}
      \|T\Phi\| = \|A_2^{-1} \Phi\|_{+,1} \leq c \|A_2^{-1}\Phi\|_{+,2} = c \|\Phi\|.
    \end{equation*}
    Similarly, it follows $\|T^{-1}\Phi\| \leq c$, and therefore $\|T^{\pm \dagger}\| = \|T^{\pm 1}\| \leq c$.
    By Eq.~\eqref{eq:Tdagger_minus_norm}, \ref{item:equiv-} follows.

    The other implication, i.e.\ \ref{item:equiv-} implies \ref{item:equiv+}, is shown analogously.
  \end{proof}

\section{Regularity of operator-valued and form-valued functions} \label{sec:op_form_functions}
\subsection{Form-valued functions}
  Let $I \subset \mathbb{R}$ be an interval and let $\mathcal{V} = \{v_t\}_{t \in I}$ be a family of bounded, sesquilinear forms defined on the Hilbert space $\mathcal{H}^+$.
  The aim of this section is to investigate the relationship between the regularity of the functions $t \in I \mapsto v_t(\Psi,\Phi) \in \mathbb{C}$ for $\Psi, \Phi \in \mathcal{H}^+$ fixed and the regularity of the form-valued functions $t \in I \mapsto v_t \in \mathcal{V}$.

  For some arguments on this section the notion of \emph{equicontinuity} is going to be important.
  Let us remember the basic definition, while for further details we refer for instance to \cite[Sec.\ I.6]{ReedSimon1980}.

  \begin{definition}
    Let $\mathcal{F}$ be a family of functions from a given normed space $(X, \|\cdot\|_X)$ to another normed space $(Y, \|\cdot\|_Y)$.
    We say that $\mathcal{F}$ is an equicontinuous family if and only if
    \begin{equation*}
      (\forall\varepsilon)(\forall x \in X)(\exists\delta)(\forall f \in \mathcal{F}) \|x - x'\| < \delta \quad
      \text{implies} \quad
      \|f(x) - f(x')\| < \varepsilon.
    \end{equation*}
  \end{definition}

  Let us start by introducing some lemmas which are going to be useful for proving these results.
  \begin{lemma} \label{lemma:sup-cont}
    Let $\mathcal{F}$ be an equicontinuous family of functions from $\mathcal{H}$ to $\mathbb{C}$.
    Then the function $f: \Phi \in \mathcal{H} \mapsto \sup_{F \in \mathcal{F}} |F(\Phi)| \in \mathbb{R}$ is continuous.
  \end{lemma}
  \begin{proof}
    Let $\Phi, \Phi_0 \in \mathcal{H}$.
    It holds
    \begin{alignat*}{2}
      \sup_{F \in \mathcal{F}} |F(\Phi)| &= \sup_{F \in \mathcal{F}} [|F(\Phi)| - |F(\Phi_0)| + |F(\Phi_0)|]\\
      &\leq \sup_{F \in \mathcal{F}} [|F(\Phi)| - |F(\Phi_0)|] + \sup_{F \in \mathcal{F}} |F(\Phi_0)|,
    \end{alignat*}
    Combining this inequality with the one obtained interchanging the roles of $\Phi$ and $\Phi_0$, it follows
    \begin{equation*}
      \left|\sup_{F \in \mathcal{F}} |F(\Phi)| - \sup_{F \in \mathcal{F}} |F(\Phi_0)|\right|
      \leq \sup_{F \in \mathcal{F}} \left||F(\Phi)| - |\vphantom{\lim_2}F(\Phi_0)|\right|
    \end{equation*}

    Therefore, one has
    \begin{equation*}
      |f(\Phi) - f(\Phi_0)| = \left|\sup_{F \in \mathcal{F}} |F(\Phi)| - \sup_{F \in \mathcal{F}} |F(\Phi_0)|\right|
      \leq \sup_{F \in \mathcal{F}} \left|\vphantom{\lim_2}|F(\Phi)| - |F(\Phi_0)|\right|.
    \end{equation*}
    From this inequality and the equicontinuity of $\mathcal{F}$ the result follows.
  \end{proof}

  \begin{lemma} \label{lemma:locally_unif_bound}
    For $t \in I \subset \mathbb{R}$, let $v_t: \mathcal{H}^+ \times \mathcal{H}^+$ be a bounded sesquilinear form.
    Let $t_0 \in I$ and suppose that $\lim_{t \to t_0} v_t(\Psi, \Phi)$ exists for every $\Psi, \Phi \in \mathcal{H}^+$.
    Then there is a neighbourhood $B_{t_0}$ of $t_0$ and a constant $K$ such that
    \begin{equation*}
      |v_t(\Psi,\Phi)| \leq K \|\Psi\|_+ \|\Phi\|_+, \qquad
      \forall \Psi,\Phi \in \mathcal{H}^+, \quad \forall t \in B_{t_0}.
    \end{equation*}
    Moreover, $L(\Psi, \Phi) = \lim_{t \to t_0} v_t(\Psi, \Phi)$ defines a bounded sesquilinear form on $\mathcal{H}^+$ with
    \begin{equation*}
      |L(\Psi,\Phi)| \leq K \|\Psi\|_+ \|\Phi\|_+.
    \end{equation*}
  \end{lemma}
  \begin{proof}
    For every $\Psi, \Phi \in \mathcal{H}^+$, the existence of $\lim_{t \to t_0} v_t(\Psi, \Phi)$ implies that there is a neighbourhood $B_{t_0}$ of $t_0$ and a constant $K_{\Psi,\Phi} > 0$ such that for $t \in B_{t_0}$ it holds $|v_t(\Psi, \Phi)| \leq K_{\Psi,\Phi}$.
    Therefore, by Theorem \ref{thm:unif_boundedness_forms}, there is $K > 0$ such that for every $t \in B_{t_0}$ and every $\Psi, \Phi \in \mathcal{H}^+$ it holds
    \begin{equation*}
      |v_t(\Psi,\Phi)| \leq K \|\Psi\|_+ \|\Phi\|_+.
    \end{equation*}
    It is straightforward to check that $L(\Psi, \Phi)$ is a sesquilinear form, and since the previous bound holds independently of $t \in B_{t_0}$, it follows $|L(\Psi,\Phi)| \leq K \|\Psi\|_+ \|\Phi\|_+$.
  \end{proof}

  \begin{lemma} \label{lemma:form_uniform_limit}
    For $t \in I \subset \mathbb{R}$, let $v_t: \mathcal{H}^+ \times \mathcal{H}^+$ be a bounded sesquilinear form such that $\lim_{t \to t_0} v_t(\Psi, \Phi)$ exists for every $\Psi, \Phi \in \mathcal{H}^+$ and denote by $L$ the bounded sesquilinear form defined by $L(\Psi, \Phi) = \lim_{t \to t_0} v_t(\Psi, \Phi)$.
    Then it holds
    \begin{equation*}
      \adjustlimits\lim_{t \to t_0} \sup_{\Psi,\Phi \in \mathcal{H}^+ \smallsetminus \{0\}}  \frac{|v_t(\Psi, \Phi) - L(\Psi,\Phi)|}{\|\Phi\|_+ \|\Psi\|_+} = 0.
    \end{equation*}
  \end{lemma}
  \begin{proof}
    Fix $\Psi_0, \Phi_0 \in \mathcal{H}^+$ such that $\|\Psi_0\|_+ \leq 1$ and $\|\Phi_0\|_+ \leq 1$, and denote by $B_{t_0}$ the neighbourhood of Lemma \ref{lemma:locally_unif_bound}.
    By the definition of the limit, we have that for any $\varepsilon > 0$, there exists $\delta_0$ such that $B_{t_0}(\delta_0) \coloneqq \{t \in \mathbb{R} \mid |t - t_0| < \delta_0\} \subset B_{t_0}$ and for $t \in B_{t_0}(\delta_0)$ it holds
    \begin{equation*}
      |v_t(\Psi_0, \Phi_0) - L(\Psi_0, \Phi_0)| < \frac{\varepsilon}{2}.
    \end{equation*}

    Therefore, for $t \in B_{t_0}(\delta_0)$, it holds
    \begin{equation} \label{eq:form_unif_lim_1}
      |v_t(\Psi_0, \Phi) - L(\Psi_0, \Phi)| \leq
      \sup_{t \in B_{t_0}(\delta_0)} |v_t(\Psi_0,\Phi-\Phi_0) - L(\Psi_0,\Phi-\Phi_0)|
      + \frac{\varepsilon}{2}.
    \end{equation}

    Consider the function $f_{\Phi_0}: \mathcal{H}^+ \to [0, \infty)$ defined by
    \begin{equation*}
      f_{\Phi_0}(\Phi) \coloneqq \sup_{t \in B_{t_0}(\delta_0)} |v_t(\Psi_0,\Phi-\Phi_0) - L(\Psi_0,\Phi-\Phi_0)|.
    \end{equation*}
    By Lemma \ref{lemma:locally_unif_bound}, the family $\{\Phi \mapsto v_t(\Psi_0,\Phi-\Phi_0) - L(\Psi_0,\Phi-\Phi_0) \mid t \in B_{t_0}\}$ is equicontinuous, and therefore Lemma \ref{lemma:sup-cont} implies $f_{\Phi_0}$ is a continuous map.
    This implies $f_{\Phi_0}$ is also weakly-continuous and therefore $U_{\Phi_0} \coloneqq f_{\Phi_0}^{-1}(\{x\in \mathbb{R}: |x| < {\varepsilon}/{2}\})$ is an open neighbourhood of $\Phi_0$ in the weak topology of $\mathcal{H}^+$.
    By Equation \eqref{eq:form_unif_lim_1} and the definition of $U_{\Phi_0}$, for every $\Phi \in U_{\Phi_0}$ and every $t \in B_{t_0}(\delta_0)$, it holds
    \begin{equation*}
      |v_t(\Psi_0, \Phi) - L(\Psi_0, \Phi)| \leq \varepsilon.
    \end{equation*}

    That is, for every $\Phi_0 \in \mathcal{H}^+$, $\|\Phi_0\|_+ \leq 1$ and $\varepsilon > 0$ there is $\delta_{\Phi_0}$ and a neighbourhood of $\Phi_0$ in the weak topology, $U_{\Phi_0}$, such that for every $\Phi \in U_{\Phi_0}$ and $|t - t_0| < \delta_{\Phi_0}$ it holds
    \begin{equation*}
      |v_t(\Psi_0, \Phi) - L(\Psi_0, \Phi)| < \varepsilon.
    \end{equation*}
    The family $\{U_{\Phi}: \|\Phi\|_+ \leq 1,\,\Phi \in \mathcal{H}^+\}$ is an open covering of the closed unit ball on $\mathcal{H}^+$ and, by the weak compacity of the unit ball, there is a finite subcovering $\{U_n\}_{n = 1}^N$.
    It follows that, for every $U_n$, $1 \leq n \leq N$, there is $\delta_n$ such that for $\Phi \in U_n$ and $|t - t_0| < \delta_n$
    \begin{equation*}
      |v_t(\Psi_0, \Phi) - L(\Psi_0, \Phi)| \leq \varepsilon.
    \end{equation*}
    Therefore, for $|t - t_0| < \max_n \delta_n$ and every $\Phi \in \mathcal{H}^+$ with $\|\Phi\|_+ \leq 1$ one has
    \begin{equation*}
      |v_t(\Psi_0, \Phi) - L(\Psi_0, \Phi)| \leq \varepsilon.
    \end{equation*}
    Repeating the argument for $\Psi$ yields the limit is uniform on $\Psi, \Phi$ in the closed unit ball, from which the result follows.
  \end{proof}

  We have now all the pieces needed to show the main result of this subsection.
  \begin{proposition} \label{prop:form_equicont_equidiff}
    For $t \in I \subset \mathbb{R}$, let $v_t: \mathcal{H}^+ \times \mathcal{H}^+ \to \mathbb{C}$ be a family of bounded sesquilinear forms.
    Then:
    \begin{enumerate}[label=\textit{(\roman*)},nosep,leftmargin=*,ref=\ref{prop:form_equicont_equidiff}\textit{(\roman*)}]
      \item \label{prop:equicontinuity_forms}
        If for every $\Psi, \Phi \in \mathcal{H}^+$ the map $t \mapsto v_t(\Psi, \Phi)$ is continuous, it holds
        \begin{equation*}
          \adjustlimits\lim_{t \to t_0} \sup_{\Psi,\Phi \in \mathcal{H}^+ \smallsetminus \{0\}}  \frac{|v_t(\Psi, \Phi) - v_{t_0}(\Psi,\Phi)|}{\|\Phi\|_+ \|\Psi\|_+} = 0.
        \end{equation*}
        If in addition $I$ is compact, there is $M>0$ such that $|v_t(\Psi,\Phi)| \leq M \|\Psi\|_+ \|\Phi\|_+$ for every $t \in I$ and every $\Psi,\Phi \in \mathcal{H}^+$.
      \item \label{prop:equidifferentiability_forms}
        If for every $\Psi, \Phi \in \mathcal{H}^+$ the map $t \mapsto v_t(\Psi, \Phi)$ is differentiable, then
        \begin{equation*}
          \adjustlimits\lim_{t \to t_0} \sup_{\Psi,\Phi \in \mathcal{H}^+ \smallsetminus \{0\}}  \frac{1}{\|\Phi\|_+ \|\Psi\|_+}  \left|\frac{v_t(\Psi, \Phi) - v_{t_0}(\Psi,\Phi)}{t - t_0} - \dot{v}_{t_0}(\Psi,\Phi)\right| = 0,
        \end{equation*}
        where $\dot{v}_t(\Psi,\Phi)$ denotes the derivative of $t \mapsto v_t(\Psi,\Phi)$.
        If, in addition, $I$ is compact and $t \mapsto \dot{v}_t(\Psi,\Phi)$ is continuous for every $\Psi,\Phi$, there is $M > 0$ such that $|\dot{v}_t(\Psi,\Phi)| < M \|\Psi\|_+ \|\Phi\|_+$ for every $t \in I$.
    \end{enumerate}
  \end{proposition}
  \begin{proof}
    By Lemma \ref{lemma:form_uniform_limit} the continuity of $t \mapsto v_t(\Psi,\Phi)$ for each $\Psi, \Phi \in \mathcal{H}^+$ implies
    \begin{equation*}
          \adjustlimits\lim_{t \to t_0} \sup_{\Psi,\Phi \in \mathcal{H}^+ \smallsetminus \{0\}}  \frac{|v_t(\Psi, \Phi) - v_{t_0}(\Psi,\Phi)|}{\|\Phi\|_+ \|\Psi\|_+} = 0,
    \end{equation*}
    and its differentiability implies
    \begin{equation*}
      \adjustlimits\lim_{t \to t_0} \sup_{\Psi,\Phi \in \mathcal{H}^+ \smallsetminus \{0\}}  \frac{1}{\|\Phi\|_+ \|\Psi\|_+}  \left|\frac{v_t(\Psi, \Phi) - v_{t_0}(\Psi,\Phi)}{t - t_0} - \dot{v}_{t_0}(\Psi,\Phi)\right| = 0.
    \end{equation*}
    The uniform bounds follow from Lemma \ref{lemma:locally_unif_bound} and the compacity of $I$.
  \end{proof}

\subsection{Operator-valued functions}
  Given a scale of Hilbert spaces $\mathcal{H}^+ \subset \mathcal{H} \subset \mathcal{H}^-$, by the Riesz Representation Theorem any bounded, sesquilinear form $v: \mathcal{H}^+ \times \mathcal{H}^+ \to \mathbb{C}$ can be interpreted as the unique operator $V \in \mathcal{B}(\mathcal{H}^+, \mathcal{H}^-)$ defined by $v(\Psi,\Phi) = (\Psi, V\Phi)$.
  In this brief subsection we transfer the previous results for form-valued functions to the operator-valued functions $t \in I \mapsto V_t \in \mathcal{B}(\mathcal{H}^+, \mathcal{H}^-)$.

  \begin{proposition} \label{prop:op_norm_differentiability}
    Let $\mathcal{H}^+ \subset \mathcal{H} \subset \mathcal{H}^-$ be a scale of Hilbert spaces, and let $I \subset \mathbb{R}$ be a compact interval.
    For $t \in I$, let $v_t: \mathcal{H}^+ \times \mathcal{H}^+ \to \mathbb{C}$ be a family of Hermitian, bounded, sesquilinear forms, and $V(t): \mathcal{H}^+ \to \mathcal{H}^-$ be defined as the unique operators such that $(\Psi, V(t)\Phi) = v_t(\Psi, \Phi)$, $t \in I$.
    Then, it holds:
    \begin{enumerate}[label=\textit{(\roman*)},nosep,leftmargin=*,ref=\ref{prop:op_norm_differentiability}\textit{(\roman*)}]
      \item \label{prop:form-norm}
        For every $t \in I$, the operator norm of $V(t) \in \mathcal{B}(\mathcal{H}^+,\mathcal{H}^-)$ is        \begin{equation*}
          \|V(t)\|_{+,-} = \sup_{\Psi, \Phi \in \mathcal{H}^+ \smallsetminus \{0\}} \frac{|v_t(\Psi, \Phi)|}{\|\Psi\|_+ \|\Phi\|_+}.
        \end{equation*}
    \end{enumerate}
    If, in addition the map $t \mapsto v_t(\Psi, \Phi)$ is continuously differentiable for every $\Psi, \Phi \in \mathcal{H}^+$, then:
    \begin{enumerate}[resume,label=\textit{(\roman*)},nosep,leftmargin=*,ref=\ref{prop:op_norm_differentiability}\textit{(\roman*)}]
      \item \label{prop:form-norm-continuity}
        The map $t \mapsto V(t)$ is continuous with respect to the $\|\cdot\|_{+,-}$ norm and the family $\{V(t)\}_{t \in I}$ is uniformly bounded.
      \item \label{prop:form-differentiability}
        The derivative $\frac{\d}{\d t} V(t)$ exists in the $\|\cdot\|_{+,-}$-norm sense and it is uniformly bounded, that is there is a constant $K$ such that for every $t \in I$,
        \begin{equation*}
          \left\|\frac{\d}{\d t} V(t)\right\|_{+,-} \leq K.
        \end{equation*}
    \end{enumerate}
  \end{proposition}
  \begin{proof}
    By definition of the operator norm, one has that
    \begin{equation*}
      \|V(t)\|_{+,-} = \sup_{\Phi \in \mathcal{H}^+ \smallsetminus \{0\}} \frac{\|V(t) \Phi\|_{-}}{\|\Phi\|_{+}}
      = \sup_{\substack{\Phi \in \mathcal{H}^+ \smallsetminus \{0\} \\ \Psi \in \mathcal{H}^- \smallsetminus \{0\}}} \frac{|\langle \Psi, V(t) \Phi \rangle_-|}{\|\Psi\|_-\|\Phi\|_{+}},
    \end{equation*}
    where we have used that, for any $\xi \in \mathcal{H}^-$ it holds $\|\xi\|_- = \sup_{\Psi \neq 0} \frac{|\langle \Psi, \xi \rangle_-|}{\|\Psi\|_-}$.

  Using the isomorphism $J: \mathcal{H}^- \to \mathcal{H}^+$ and the pairing $(\cdot, \cdot)$ associated to the scale of Hilbert spaces $\mathcal{H}^+ \subset \mathcal{H} \subset \mathcal{H}^-$ (cf.\ Section \ref{sec:hilbert_scales}), the previous equation can be written as
    \begin{equation*}
      \|V(t)\|_{+,-} = \sup_{\substack{\Phi \in \mathcal{H}^+ \smallsetminus \{0\} \\ \Psi \in \mathcal{H}^- \smallsetminus \{0\}}} \frac{|(J\Psi, V(t)\Phi)|}{\|\Psi\|_-\|\Phi\|_+}
      = \sup_{\Psi, \Phi \in \mathcal{H}^+ \smallsetminus \{0\}} \frac{|(\Psi, V(t)\Phi)|}{\|\Psi\|_+ \|\Phi\|_+},
    \end{equation*}
    where we have used the fact that $J$ is an isometric bijection from $\mathcal{H}^-$ to $\mathcal{H}^+$.
    By the definition of $V(t)$, it follows
    \begin{equation*}
      \|V(t)\|_{+,-} = \sup_{\Psi, \Phi \in \mathcal{H}^+ \smallsetminus \{0\}} \frac{|v_t(\Psi, \Phi)|}{\|\Psi\|_+ \|\Phi\|_+}.
    \end{equation*}

    Property \emph{(ii)} follows immediately from Proposition \ref{prop:form_equicont_equidiff} and \emph{(i)}.

    For proving \emph{(iii)}, note that, by Lemma \ref{lemma:locally_unif_bound}, $\dot{v}_t(\Psi, \Phi) \coloneqq \frac{\d}{\d t} h_t(\Psi, \Phi)$ defines a bounded sesquilinear form in $\mathcal{H}^+$.
    Let $\dot{V}(t)$ be the unique operator in $\mathcal{B}(\mathcal{H}^+, \mathcal{H}^-)$ such that $(\Psi, \dot{V}(t)\Phi) = \dot{v}_t(\Psi,\Phi)$ for every $\Psi,\Phi \in \mathcal{H}^+$.
    We claim that $\dot{V}(t) = \frac{\d}{\d t} V(t)$ in the sense of $\mathcal{B}(\mathcal{H}^+,\mathcal{H}^-)$, that is
    \begin{equation*}
      \lim_{t \to t_0} \left\| \frac{V(t) - V(t_0)}{t - t_0} - \dot{V}(t_0) \right\|_{+,-} = 0.
    \end{equation*}
    By \emph{(i)}, this is equivalent to
    \begin{equation*}
      \lim_{t \to t_0} \sup_{\substack{\Psi, \Phi \in \mathcal{H}^+ \\ \|\Psi\|_+ = 1 = \|\Phi\|_+}}
      \left|\frac{v_t(\Psi, \Phi) - v_{t_0}(\Psi, \Phi)}{t - t_0} - \dot{v}_{t_0}(\Psi, \Phi)\right| = 0,
    \end{equation*}
    which holds by Proposition \ref{prop:equidifferentiability_forms}, as does the uniform bound.
  \end{proof}
\end{document}

\documentclass[../main.tex]{subfiles}
\ifSubfilesClassLoaded{
    \setcounter{chapter}{2}
  }{}
\begin{document}
\chapter{Introduction to Quantum Control Theory} \label{ch:quantum_control}
  In this brief chapter, we review some basic notions of Control Theory and introduce the Quantum Control at the Boundary scheme.
  In particular, we recall the basics of control problems for quantum systems, which are nothing but dynamical systems on Hilbert spaces whose evolution is given by a linear equation.
  Section \ref{sec:quantum_control} introduces the basic notions and ideas of quantum control, discussing the difficulties that arise as a consequence of the infinite-dimensional nature of the Hilbert spaces involved.
  In Section \ref{sec:bilinear_quantum_control}, the main result in \cite{ChambrionMasonSigalottiEtAl2009} on the controllability of a family of (infinite-dimensional) quantum systems is presented.
  Finally, the Quantum Control at the Boundary scheme is presented in Section \ref{sec:quantum_control_boundary}.

\section{Control of quantum systems} \label{sec:quantum_control}
  The standard theory of Quantum Control has had a great success when applied to finite-dimensional quantum systems and finite-dimensional approximations of infinite-dimensional systems, among other things, due to the development of the geometric theory of control \cite{DAlessandro2007,AgrachevSachkov2004,Jurdjevic1996}.
  However, applying the same ideas to infinite-dimensional quantum control systems brings some complications, related with the analytical difficulties arising from the appearance of unbounded operators and the complications of infinite-dimensional geometry.
  Hence, the transition from finite-dimensional to infinite-dimensional control is not an easy task and, in fact, many of the results that apply for finite-dimensional quantum systems do not carry over to the infinite-dimensional case.

  A remarkable example of this is the notion of exact controllability, which is not suitable for infinite-dimensional systems \cite{BallMarsdenSlemrod1982,Turinici2000}.

  Note also that a direct extension of the results for finite-dimensional approximations of infinite-dimensional quantum systems is not possible in general.
  For instance, consider the case of the harmonic oscillator.
  While it can be shown that the quantum harmonic oscillator is not exactly controllable (cf.\ \cite{MirrahimiRouchon2004}), its finite-dimensional projection onto the subspace generated by its first $n$ eigenstates is exactly controllable for every $n$ \cite{RamakrishnaSalapakaDahlehEtAl1995}.

  The standard setting for quantum control consists of considering a quantum system whose (pure) states are vectors on a complex separable Hilbert space $\mathcal{H}$.
  Typically $\mathcal{H} = L^2(\Omega)$ in the infinite-dimensional setting, where $\Omega$ is a Riemannian manifold, and $\mathcal{H} = \mathbb{C}^N$ with the standard inner product when we consider a finite-dimensional situation.
  The dynamics of the quantum system is given by the Schrödinger equation,
  \begin{equation} \label{eq:schrodinger}
    \frac{\d}{\d t} \Psi(t) = -i H(t) \Psi(t),\qquad \Psi(t) \in \mathcal{H},
  \end{equation}
  where $H(t)$ is a family of self-adjoint operators known as the Hamiltonian of the system.
  In the case of a time-independent Hamiltonian, $H(t) = H_0$ with $H_0$ a fixed self-adjoint operator, the initial value problem given by Eq.\ \eqref{eq:schrodinger} and $\Psi(0) = \Psi_0 \in \mathcal{H}$ has a unique solution:
  \begin{equation*}
    \Psi(t) = e^{-it H_0}\Psi_0,
  \end{equation*}
  where $U_t = e^{-it H_0}$ is the unique strongly continuous one-parameter group of unitary operators generated by $H_0$.
  For Hamiltonians which are piecewise constant but have time-independent domain, a one-parameter group of unitary operators providing the solution to the initial value problem can be constructed applying the idea above iteratively.
  However, for genuine time-dependent Hamiltonians the existence of solutions is more convoluted (see Chapter~\ref{ch:H_form_domain} for a detailed discussion).
  \begin{definition}\label{def:control_system}
    Let $\mathcal{C}$ be a suitable set called \emph{space of controls}.
    Let $\{\mathcal{D}_c\}_{c\in\mathcal{C}}$ be a family of dense subsets of $\mathcal{H}$ and let $\{H(c)\}_{c\in\mathcal{C}}$ be a family of self-adjoint operators such that for each $c\in\mathcal{C}$ the operator $H(c)$ is defined on $\mathcal{D}_c$.
    A \emph{quantum control system} is the family of dynamical systems defined by the Schrödinger equation \eqref{eq:schrodinger} with time-dependent Hamiltonians $H(u(t))$, where the measurable \emph{control functions} $u:\mathbb{R}\to\mathcal{C}$ are in some appropriate class of functions.
  \end{definition}
  \begin{remark}
    In classical control theory, the space of controls $\mathcal{C}$ is defined to be a subset of $\mathbb{R}^n$ for some $n \in \mathbb{N}$.
    However, this is not the case for boundary control systems (see Definition \ref{def:quantumBCS}).
  \end{remark}

  Let us now introduce a particular quantum control system which is appropriate for modelling a quantum system coupled with a device whose dynamics allows to control the system.
  \begin{example}[Affine-control system]
    Consider a quantum system with constant Hamiltonian $H_0$.
    It can be \emph{controlled} by interacting with an external aparatus that will be described by operators $H_k$, assumed to be self-adjoint too, leading to the affine-control system:
    \begin{equation} \label{eq:schrodinger_affine}
      \frac{\d}{\d t} \Psi(t) = -i\left(H_0 + \sum_{k=1}^r u_k(t) H_k\right) \Psi(t),
    \end{equation}
    where the measurable functions $u_k(t)$ represent the intensity of the devices interacting with the given system.
    In this case the space of controls $\mathcal{C}$ is a subset of $\mathbb{R}^n$, and the control functions are usually considered piecewise constant (see Sec.\ \ref{sec:bilinear_quantum_control}).
  \end{example}

  The controllability problem for a quantum control system consists of deciding whether there exists a control function such that the associated evolution drives the system from any initial state to any target state.

  The problems appearing for infinite-dimensional systems, together with the fact that in most applications having absolute precision is not important, motivates the introduction of the concept of approximate controllability.
  \begin{definition}\label{def:strong_controllability}
    We say that a quantum control system is \emph{approximately controllable} if for every $\Psi_0, \Psi_T \in \mathcal{H}$ and any $\epsilon >0$ there exists $T>0$ and a measurable control function $u:[0,T]\to\mathcal{C}$ such that the Schrödinger equation has a solution $\Psi(t)$ that satisfies
    \begin{equation*}
      \Psi(0) = \Psi_0 \quad
      \text{and} \quad
      \|\Psi_T - \Psi(T)\| < \epsilon.
    \end{equation*}
  \end{definition}

  The controllability problem for quantum systems defined on a Hilbert space $\mathcal{H}$ which is part of a scale of Hilbert spaces $\mathcal{H}^+ \subset \mathcal{H} \subset \mathcal{H}^-$ is studied in Chapter \ref{ch:control_systems}.
  For this purpose, it is useful to consider approximate controllability not only with respect to the norm on $\mathcal{H}$, but also with respect to the norm $\|\cdot\|_-$ on $\mathcal{H}^-$.
  For convenience, we refer to this notion of controllability as weak controllability.
  \begin{definition}\label{def:weak_controllability}
    We say that a quantum control system is \emph{weakly approximately controllable} if for every $\Psi_0, \Psi_T \in \mathcal{H}$ and any $\epsilon >0$ there exists $T>0$ and a measurable control function $u:[0,T]\to\mathcal{C}$ such that the weak Schrödinger equation (i.e.\ the Schrödinger equation in $\mathcal{H}^-$, cf.\ Def.\ \ref{def:weak_Schrodinger}) has a solution $\Psi(t)$ that satisfies
    \begin{equation*}
      \Psi(0) = \Psi_0 \quad
      \text{and}\quad
      \|\Psi_T - \Psi(T)\|_- < \epsilon.
    \end{equation*}
  \end{definition}

  \begin{remark}
    The controllability problem for quantum control systems can also be formulated as a control problem with dynamics on the group $\mathcal{U}(\mathcal{H})$ of unitary operators on the Hilbert space:
    \begin{equation} \label{eq:schrodinger_unitaries}
      \frac{\d}{\d t}U_t = -i H(t)U_t.
    \end{equation}
    In this case, looking for exact controllability is equivalent to look for a one-parameter family of unitary operators $U_t$ satisfying Eq.\ \eqref{eq:schrodinger_unitaries} such that $U_0 = \mathbb{I}$ is the identity on $\mathcal{H}$, and $U_T$ in a subset $\mathcal{T} \subset \mathcal{U}(\mathcal{H})$ satisfying that for any $\Psi_0, \Psi_T \in \mathcal{H}$ there is $U \in \mathcal{T}$ such that $U\Psi_0 = \Psi_T$.
    Similar formulations can be made for weak and strong approximate controllability.
  \end{remark}

\section{Bilinear control of quantum systems: controllability results} \label{sec:bilinear_quantum_control}
  During the last decades, some results on the control of infinite-dimensional quantum systems have been obtained and successfully applied, for instance, to the control of molecules \cite{ChambrionMasonSigalottiEtAl2009, BoscainCaponigroChambrionEtAl2012}.
  While these results apply under fairly general conditions, the family of controls considered are piecewise constant functions.
  Doing this avoids the problem of existence of solutions for the Schrödinger equation.
  However, this kind of controls cannot be applied to some other relevant quantum systems \cite{ErvedozaPuel2009} and, in particular, cannot be applied to the situation with time-dependent domains as it forbids the existence of solutions of the Schrödinger equation in the strong sense (see Definition \ref{def:strong_Schrodinger} and Definition \ref{def:weak_Schrodinger}).

  Let us briefly review the main result in \cite{ChambrionMasonSigalottiEtAl2009}, which is one of the main tools on the controllability proofs in Chapter \ref{ch:control_systems}.
  Chambrion et al.\ study the approximate controllability of bilinear control systems defined as follows.
  \begin{definition} \label{def:normal_bilinear_CS}
    Let $r>0$ and assume that:
    \begin{enumerate}[%
      label=\textit{(H\arabic*)},
      nosep,
      labelindent=0.5\parindent,
      leftmargin=*
      ]
      \item\label{H1} $H_0, H_1$ are self-adjoint operators.
      \item\label{H2} There exists an orthonormal basis $\{\Phi_n\}_{n \in \mathbb{N}}$ of $\mathcal{H}$ made of eigenvectors of $H_0$.
      \item\label{H3} $\Phi_n \in \dom H_1$ for every $n \in \mathbb{N}$.
    \end{enumerate}
    A \emph{normal bilinear control system} is a quantum control system with controls given by $\mathcal{C} = \{c\in\mathbb{R} | c<r\}$ and family of Hamiltonians $\{H(c)\}_{c\in \mathcal{C}}$ determined by
    \begin{equation*}
      H(c) = H_0 + c H_1.
    \end{equation*}
  \end{definition}

  For normal bilinear control systems Chambrion et al.\ prove the following result.
  \begin{theorem}[Chambrion et al.\ {\cite[Thm. 2.4]{ChambrionMasonSigalottiEtAl2009}}] \label{thm:chambrion_controllability}
    Consider a normal bilinear control system, with $r > 0$ as described above.
    Let $\{\lambda_n\}_{n \in \mathbb{N}}$ denote the eigenvalues of $H_0$, each of them associated with the eigenfunction $\Phi_n$.
    Then, if the elements of the sequence $\{\lambda_{n+1} - \lambda_n\}_{n \in \mathbb{N}}$ are $\mathbb{Q}$-linearly independent and if $\langle \Phi_{n+1}, H_1 \Phi_n \rangle \neq 0$ for every $n \in \mathbb{N}$, the system is approximately controllable.
  \end{theorem}

  \begin{remark}
    The conditions on the eigenvalues and eigenfunctions are not restrictive: as it is shown in \cite{BoscainCaponigroChambrionEtAl2012, MasonSigalotti2010, PrivatSigalotti2010,PrivatSigalotti2010b} they are satisfied generically and they can also be bypassed, cf.\ \cite[Section 6.1]{ChambrionMasonSigalottiEtAl2009}.
    We will assume hereafter that these conditions are met.
  \end{remark}

\section{Quantum Control at the Boundary} \label{sec:quantum_control_boundary}
  In the previous discussion on the problem of control of quantum systems, it is considered that the system interacts with some external device whose dynamics allows to modify the evolution of our original system.
  However, the description of interacting quantum systems is a fundamental principle in Quantum Mechanics described using von Neumann's composition axiom: if $\mathcal{H}_A, \mathcal{H}_B$ represent the Hilbert spaces associated to two systems $A,B$ respectively, the composite system has associated Hilbert space $\mathcal{H}_{AB} = \mathcal{H}_A \otimes \mathcal{H}_B$ and the Hamiltonian of the system is given by
  \begin{equation} \label{eq:hamiltonian_composite}
    H_{AB} = H_A \otimes \mathbb{I}_B + \mathbb{I}_A \otimes H_B + H_\text{int},
  \end{equation}
  where $H_\text{int}$ is a self-adjoint operator describing the interaction among the two given systems.
  The way to pass from the Schrödinger equation corresponding to the operator $H_{AB}$ in Eq.\ \eqref{eq:hamiltonian_composite} to the Hamiltonian of the affine-control system given by \eqref{eq:schrodinger_affine}, is assuming that the external system has predetermined evolution described by a curve $\Psi_\text{ext}(t) \in \mathcal{H}_B$ and that for each $t \in [s,T]$ the Hilbert space $\mathcal{H}_{AB}$ is projected to the subspace
  \begin{equation*}
    \mathcal{H}_t = \mathcal{H}_A \otimes \mathbb{C} \Psi_\text{ext}(t) \subset \mathcal{H}_{AB}.
  \end{equation*}
  Denoting by $P_t$ such projector, one gets that the operator $H(t) = P_t H_{AB} P_t$ will have the form of the Hamiltonian in Eq.\ \eqref{eq:schrodinger_affine} (apart from the factor $\Psi_\text{ext}(t)$).

  This way of arguing, even if well established, is based upon the not completely natural assumption that the evolution of the system we are interacting with is completely determined (for instance, it is a classical system).
  Such hypothesis is not always appropriate and one should look for an alternative way of formulating the control problem.
  Here the Quantum Boundary Control paradigm emerges.

  \emph{Boundaries} are often introduced as a phenomenological way of modelling the interaction of a given system with an \emph{external world}, and this is even more so in quantum mechanical problems.
  For instance, if we are studying the motion of an electron in a box with impenetrable walls, we will model such system as a quantum system with Hilbert space $\mathcal{H} = L^2(\Omega)$ where $\Omega\subset \mathbb{R}^3$ is the region determined by the box.
  The boundary $\partial\Omega$ of $\Omega$ represents the \emph{walls} of the box, i.e., another extremely complicated system consisting of all the tightly bounded atoms surrounding $\Omega$ that prevent our electron moving past them.
  Therefore, considering Dirichlet boundary conditions on this problem is a phenomenological way of modelling a complex interaction between our system, the electron, and another system, the walls of the box.

  Thus, a natural way of reformulating the problem of quantum control would be by studying how these boundary conditions must change so that the state of the system would change in the desired way.

  As discussed later on (see Chapter \ref{ch:laplacian}), there is a natural identification between a class of suitable boundary conditions for the Laplace-Beltrami operator $\Delta$ on a Riemannian manifold and self-adjoint extensions of the symmetric minimal closure $\Delta_0$ of such operator.
  Hence, it is just natural to address the problem of quantum control as a problem of control whose space of controls is (a subset of) the space of all self-adjoint extensions of a given symmetric operator.
  This is the pivotal idea underlying this work.
  \begin{definition} \label{def:quantumBCS}
    Let $\mathcal{H}$ be a Hilbert space, and let $H$ be a symmetric operator admitting self-adjoint extensions.
    A \emph{quantum boundary control system} is a quantum control system whose space of controls $\mathcal{C}$ is the set of self-adjoint extensions of $H$.
  \end{definition}

  Some concrete quantum boundary control systems are defined in Chapter~\ref{ch:control_systems}, for which approximate controllability is proven.
\end{document}


\ifSubfilesClassLoaded{
  \setcounter{chapter}{3}
  }{}
\chapter{Hamiltonians with constant form domain} \label{ch:H_form_domain}
This chapter deals with the problem of the existence of dynamics for a quantum dynamical system, fixing the related notation and reviewing known results.
Both J.~Kisyński \cite{Kisynski1964} and B. Simon \cite{Simon1971} studied the existence of solutions of the Schrödinger equation for Hamiltonians with form-constant domains (cf.\ Def.\ \ref{def:hamiltonian_const_form}), both proving existence results based on an approximative procedure due to K.\ Yosida \cite{Yosida1948}.

This chapter is structured as follows.
First, in Section \ref{sec:quantum_dynamics}, we review the basics of quantum dynamical systems and introduce Hamiltonians with constant form domain, fixing the notation that will be used for the rest of this work.
In Section \ref{sec:existence_dynamics}, we present Simon's and Kisyński's approaches for proving the existence of soltions of the Schrödinger equation for Hamiltonians with constant form domain, exploring the similarities in both approaches.
After that, Section \ref{sec:stability} presents a stability result showing that similar Hamiltonians with constant form domain induce similar dynamics, in a precise sense to be defined in the text (cf. Thm.\ \ref{thm:stability_bound}).
Finally, Section \ref{sec:form_linear} introduces a particular family of Hamiltonians with constant form domain which we call \emph{form-linear Hamiltonians}.

\section{Quantum dynamics} \label{sec:quantum_dynamics}
  A quantum dynamical system is a linear dynamical system on a separable, complex Hilbert space $\mathcal{H}$ whose evolution is given by a family of self-adjoint, densely defined operators $\{H(t)\}_{t \in I}$ through the Schrödinger equation.
  This family of operators is called the Hamiltonian of the system and will be denoted simply by $H(t)$.

  \begin{definition}\label{def:strong_Schrodinger}
    Let $I\subset \mathbb{R}$ be a compact interval and let $H(t)$, $t\in I$, be a time-dependent Hamiltonian.
    The \emph{(strong) Schrödinger equation} is the linear evolution equation
    \begin{equation*}
      \frac{\d}{\d t}\Psi(t) = -iH(t)\Psi(t).
    \end{equation*}
  \end{definition}

  For the case on which the closed sesquilinear forms associated to $H(t)$ via Thm.\ \ref{thm:repKato} have constant domain (see Definition \ref{def:hamiltonian_const_form}), it will be useful to define the weak version of Schrödinger equation.
  \begin{definition}\label{def:weak_Schrodinger}
    Let $I\subset \mathbb{R}$ be a compact interval and let $H(t)$, $t\in I$, be a time-dependent Hamiltonian with constant form domain $\mathcal{H}^+$.
    For $t\in I$, let $h_t$ be the sesquilinear form uniquely associated with the self-adjoint operator $H(t)$.
    The \emph{weak Schrödinger equation} is the linear evolution equation
    \begin{equation*}
      \frac{\d}{\ dt} \langle \Phi, \Psi(t)\rangle = -ih_t(\Phi,\Psi(t)), \qquad
      \forall \Phi \in \mathcal{H}^+.
    \end{equation*}
  \end{definition}

  The strong Schrödinger equation is the usual Schrödinger equation; we will use the adjective \emph{strong} when we want to emphasise the difference with the weak Schrödinger equation.
  It is immediate to prove that solutions of the strong Schrödinger equation are also solutions of the weak Schrödinger equation when the latter is defined.

  The solutions of the initial value problems for the weak and strong Schrö\-din\-ger equations are given in terms of what are known as unitary propagators.

  \begin{definition}\label{def:unitarypropagator}
    A \emph{unitary propagator} is a two-parameter family of unitary operators $U(t,s)$, $s,t\in \mathbb{R}$, satisfying:
    \begin{enumerate}[label=\textit{(\roman*)},nosep,leftmargin=*]
      \item $U(t,s)U(s,r)= U( t , r)$.
      \item $U(t,t)= \mathbb{I}$.
      \item $U(t,s)$ is jointly strongly continuous in $t$ and $s$.
    \end{enumerate}
  \end{definition}

  The solution of the strong or weak Schrödinger equation with initial value $\Psi$ at time $s\in \mathbb{R}$ is given, in case that it exists, by the curve $\Psi(t) := U(t,s)\Psi$.
  Even if the Schrödinger equation is a linear ordinary differential equation, the existence of solutions is not guaranteed since the operators $H(t)$ may be unbounded (i.e., not continuous).
  In the most general case, the time-dependent Hamiltonian is a family of unbounded self-adjoint operators whose domains depend on $t$; these facts make the problem of the existence of solutions for the Schrödinger equation highly non-trivial.
  There are general sufficient conditions for the existence of solutions of these equations, see for instance \cite{Kato1973,Kisynski1964,Simon1971}.
  J.\ Kisyński \cite{Kisynski1964} studied the general initial value problems for linear ordinary differential equations of first order in Banach spaces, providing sufficient conditions for the existence of solutions.
  However, for the purpose of this work such an extensive treatment is not needed and we will restrict ourselves to the simpler case of Hamiltonians with constant form domain.

  \begin{definition} \label{def:hamiltonian_const_form}
    Let $I \subset \mathbb{R}$ be a compact interval, let $\mathcal{H}^+$ be a dense subspace of $\mathcal{H}$ and $H(t)$, ${t \in I}$, a family of self-adjoint operators on $\mathcal{H}$, such that for any $t\in I$ the operator $H(t)$ is densely defined on $\mathcal{D}(t)$.
    We say that $H(t)$,  ${t \in I}$, is a \emph{time-dependent Hamiltonian with constant form domain} $\mathcal{H}^+$ if:
    \begin{enumerate}[label=\textit{(\roman*)},nosep,leftmargin=*]
      \item There is $m >0$ such that, for any $t\in I$, $\langle \Phi, H(t)\Phi \rangle \geq -m \|\Phi\|^2$ for all $\Phi\in\mathcal{D}(t)$.
      \item For any $t \in I$, the domain of the Hermitian sesquilinear form $h_t$ associated with $H(t)$ (cf.\ Theorem \ref{thm:repKato}) is $\mathcal{H}^+$.
    \end{enumerate}
  \end{definition}
  These Hamiltonians with constant form domain allow to define scales of Hilbert spaces which are the key objects needed to prove the main results of this chapter.
  As it will become clear on next sections, both J.\ Kisyński and B.\ Simon use on their works the construction of scales of Hilbert spaces; let us establish the notation which will be used for the rest of this dissertation (cf.\ Section \ref{sec:hilbert_scales}).
  Let $I \subset \mathbb{R}$ be a compact interval, let $\mathcal{H}^+$ be a dense subspace of $\mathcal{H}$ and ${H(t)}$, ${t \in I}$, a time-dependent Hamiltonian with constant form domain $\mathcal{H}^+$.
  The scale of Hilbert spaces defined by $H(t)$ is the triple of Hilbert spaces
  \nomenclature[H]{$\mathcal{H}^\pm_t$}{Hilbert spaces from the scale associated to $H(t)$}
  \nomenclature[I]{$\langle \cdot, \cdot \rangle_{\pm,t}$}{Inner product on $\mathcal{H}^\pm_t$}
  \begin{equation*}
    (\mathcal{H}^+, \langle \cdot, \cdot \rangle_{+,t}) \subset
    (\mathcal{H}, \langle \cdot, \cdot \rangle) \subset
    (\mathcal{H}^-_t, \langle \cdot, \cdot \rangle_{-,t}),
  \end{equation*}
  where $\langle \Psi, \Phi \rangle_{\pm,t} := \langle (H(t) + m + 1)^{\pm\sfrac{1}{2}}\Psi, (H(t) + m + 1)^{\pm\sfrac{1}{2}} \Phi \rangle$ and $\mathcal{H}^-_t$ denotes the closure of $\mathcal{H}$ with respect to the norm defined by $\|\Phi\|_{-,t}^2 \coloneqq \langle \Phi, \Phi \rangle_{-,t}$.
  We denote $\mathcal{H}^+_t = (\mathcal{H}^+, \langle \cdot, \cdot \rangle_{+,t})$ and, with a slight abuse of notation, we use the symbol $\mathcal{H}^-_t$ to represent the Hilbert space $(\mathcal{H}^-_t, \langle \cdot, \cdot \rangle_{-,t})$.
  We denote by $(\cdot,\cdot)_t : \mathcal{H}^+_t \times \mathcal{H}^-_t \cup \mathcal{H}^-_t \times \mathcal{H}^+_t \to \mathbb{C}$ the canonical pairings.
  The unbounded operators $H(t)$ with constant form domain can be continuously extended to bounded operators $\tilde{H}(t): \mathcal{H}_t^+ \to \mathcal{H}_t^-$.
  The inverse operators, $H(t)^{-1} \in \mathcal{B}(\mathcal{H})$, can be extended to bounded operators from $\mathcal{H}_t^-$ to $\mathcal{H}_t^+$, which coincide with $\tilde{H}(t)^{-1}$.
  In order to simplify the notation we will drop the tilde, denoting the extensions and the original operators with the same symbols.
  We will also denote by $\|\cdot\|_{+,-,t}$ and $\|\cdot\|_{-,+,t}$ the norms in $\mathcal{B}(\mathcal{H}^+, \mathcal{H}^-)$ and $\mathcal{B}(\mathcal{H}^-, \mathcal{H}^+)$ respectively.

\section{Existence of dynamics} \label{sec:existence_dynamics}
\subsection{Simon's approach} \label{subsec:simon}
  Let us now introduce the main ideas used by B.\ Simon on his study \cite{Simon1971}.
  He takes the following assumptions for proving the existence of dynamics.
  \begin{assumption} \label{assump:simon}
    Let $H_0$ be a positive operator on a Hilbert space and let $\mathcal{H}^+ \subset \mathcal{H} \subset \mathcal{H}^-$ be its associated scale of Hilbert spaces.
    For $0 \leq t \leq T$, let $H(t)$ be a family of $\mathcal{H}$-symmetric operators from $\mathcal{H}^+$ to $\mathcal{H}^-$.
    We assume that there is $C>0$ independent of $t$ such that:
    \begin{enumerate}[label=\textit{(S\arabic*)},nosep,leftmargin=*]
      \item \label{assump:simon_normequiv}
        $C^{-1} (H_0 + 1) \leq H(t) \leq C(H_0 + 1)$.
      \item \label{assump:simon_regularity}
        $B(t) = \frac{\d}{\d t} H(t)^{-1}$ exists in the $\|\cdot\|$-sense and
        \begin{equation*}
          \|H(t)^{\sfrac{1}{2}} B(t) H(t)^{\sfrac{1}{2}}\| \leq C.
        \end{equation*}
    \end{enumerate}
  \end{assumption}

  The approach by Simon defines the Hamiltonian of the system through a family of operators from $\mathcal{H}^+ \to \mathcal{H}^-$.
  This simplifies some usual problems for the existence of quantum dynamics.
  First, it fixes a self-adjoint extension of $H(t)$ (see \cite[Lemma II.6]{Simon1971}) as an unbounded operator on $\mathcal{H}$.
  Also, it allows to solve the problem for some cases on which the domain of the time-dependent Hamiltonian $H(t)$ depends on $t$, sidestepping some of the technical difficulties this carries by having a common domain as an operator from $\mathcal{H}^+ \to \mathcal{H}^-$.

  Under these hypothesis, Simon proves the existence of dynamics in \cite[Appendix II.7]{Simon1971}.
  \begin{theorem}[{\cite[Thm. II.27]{Simon1971}}] \label{thm:simon}
    Let $H(t)$ satisfy Assumtion \ref{assump:simon}.
    Then, for any $\Phi_0 \in \mathcal{H}^+$, there is a unique function $\Phi(t) \in \mathcal{H}^+$ such that:
    \begin{enumerate}[label=\textit{(\roman*)},nosep,leftmargin=*]
      \item $\Phi(t)$ is continuous in the $\mathcal{H}^+$ weak topology, i.e.\ for all $\Psi \in \mathcal{H}^-$, $t \mapsto \langle \Psi, \Phi(t) \rangle$ is continuous.
      \item For any $\Psi \in \mathcal{H}^+$,
        \begin{equation*}
          \frac{\d}{\d t}\langle \Psi, \Phi(t) \rangle = -i \langle \Psi, H(t)\Phi(t) \rangle, \qquad
          \Phi(0) = \Phi_0.
        \end{equation*}
    \end{enumerate}
    Moreover:
    \begin{enumerate}[resume,label=\textit{(\roman*)},nosep,leftmargin=*]
      \item $\displaystyle \lim_{t \to t_0} \left\| \frac{\Phi(t) - \Phi(t_0)}{t - t_0} + i H(t)\Phi(t) \right\|_- = 0$.
      \item $\|\Phi(t)\| = \|\Phi_0\|$.
      \item $\Phi(t)$ is $\|\cdot\|$-continuous.
    \end{enumerate}
    Thus, the map $U(t,s): \Phi(s)\in\mathcal{H}^+ \mapsto \Phi(t)\in\mathcal{H}^+$, is unitary on $\mathcal{H}$ and can be extended to a unitary propagator.
  \end{theorem}

  Simon's proof is based on Yosida's idea \cite[pp. 425-429]{Yosida1965} whose original theorem only applies to the case on which $\dom H(t)$ does not depend on $t$.
  The proof starts defining approximate Hamiltonians $H_n(t) = H(t) (1 + n^{-1}H(t))^{-1}$, which can be seen as a bounded operator in $\mathcal{B}(\mathcal{H}^-)$.
  Therefore, the existence of unitary propagators $U_n(t,s)$ for $H_n(t)$ is provided by the Dyson expansion.
  Using these propagators, for any $\Phi_0 \in \mathcal{H}^+$ Simon constructs approximated solutions $\Phi_n(t) = U_n(t,s)\Phi_0$ for the dynamical equation for $H(t)$, and show that Assumption \ref{assump:simon} ensures that $\{\Phi_n(t)\}_{n \in \mathbb{N}} \subset \mathcal{H}^+$ has a convergent subsequence and the limit of that subsequence satisfies the properties of the Theorem.

  \subsection{Kisyński's approach} \label{subsec:kisynski}
  Kisyński's work \cite{Kisynski1964} studies the existence of solutions for the equation
  \begin{equation} \label{eq:kisynski_problem}
    \frac{\d}{\d t} \Phi(t) = A(t) \Phi(t), \quad \Phi(0) = \Phi_0,
  \end{equation}
  on a Banach space with $A(t)$ an unbounded linear operator defined on a dense domain $\dom A(t)$ which can depend on time.
  He proves first the existence of solutions for the case on which $\dom A(t)$ is constant based again on Yosida's approximation.
  Then, relying on the constant domain case, he provides sufficient conditions for the existence of solutions in the more general case on which $\dom A(t)$ does vary with $t$.
  Finally, he applies these more general results to the case on which the Banach space is a Hilbert space and the family of operators is $A(t) = -i \Lambda(t)$ with $\Lambda(t)$ a positive, self-adjoint operator such that $\dom \Lambda(t)^{\sfrac{1}{2}}$ is independent of $t$ \cite[Secs. 7, 8]{Kisynski1964}.

  In order to prove the existence of a unitary propagator for this case, J.\ Kisyński assumes the following.
  \begin{assumption}[{\cite[Hyp. 7.1]{Kisynski1964}}] \label{assump:kisynski}
    Let $\mathcal{H}$ be a Hilbert space with inner product $\langle \cdot, \cdot \rangle$, and let $\mathcal{H}^+ \subset \mathcal{H}$ be a dense subspace.
    We take the following assumptions:
    \begin{enumerate}[label=\textit{(K\arabic*)},nosep,leftmargin=*]
      \item \label{assump:kisynski_normequiv}
        For every $t \in [0,T]$, $\langle \cdot, \cdot \rangle_{+,t}$ is an inner product on $\mathcal{H}^+$, endowing it with the estructure of a Hilbert space $\mathcal{H}^+_t$ which is continuously contained on $\mathcal{H}$.
      \item \label{assump:kisynski_regularity}
        For every $t \in [0,T]$ and every $\Psi,\Phi \in \mathcal{H}^+$, the function $t \mapsto \langle \Psi, \Phi \rangle_{+,t}$ is in $C^n(0,T)$ for $n \geq 1$.
    \end{enumerate}
  \end{assumption}

  Under these hypothesis, for each $t \in [0,T]$, Kisyński constructs the scale of Hilbert spaces  $\mathcal{H}^+_t \subset \mathcal{H} \subset \mathcal{H}^-_t$ associated to $\langle \Psi, \Phi \rangle_{+,t}$ (see \cite[Lemmas 7.2-7.6]{Kisynski1964}).
  Then, the following operators are defined (cf.\ \cite[Lemmas 7.7-7.10]{Kisynski1964}).
  \begin{proposition} \label{prop:kisynski_operators}
    Under Assumption \ref{assump:kisynski}, for every $t \in [0,T]$, it holds:
    \begin{enumerate}[label=\textit{(\roman*)},nosep,leftmargin=*]
      \item Define the domain
        \begin{equation*}
          \mathcal{D}(\Lambda_0) = \left\{\Phi \in \mathcal{H}^+ \mid \sup\{|\langle \Psi, \Phi \rangle_{+,t}| \mid \Psi \in \mathcal{H}^+, \|\Psi\|=1\} <\infty \right\}.
        \end{equation*}
        Then the equality
        \begin{equation*}
          \langle \Psi, \Lambda_0(t)\Phi \rangle = \langle \Psi, \Phi \rangle_{+,t}, \qquad
          \Phi \in \mathcal{D}(\Lambda_0), \quad \Psi \in \mathcal{H}^+,
        \end{equation*}
        defines a self-adjoint, positive unbounded operator on $\mathcal{H}$.
      \item Let $\Lambda(t)$ be the closure of $\Lambda_0(t)$ on $\mathcal{H}^-_t$; then $\Lambda(t)$ is a self-adjoint, positive operator on $\mathcal{H}^-_t$ with $\dom \Lambda(t) = \mathcal{H}^+$.\
        Moreover, $\Lambda(t)^{-1}$ is the canonical isomorphism $J: \mathcal{H}^-_t \to \mathcal{H}^+_t$ (cf.\ Sec.\ \ref{sec:hilbert_scales}) for the scale of Hilbert spaces $\mathcal{H}^+_t \subset \mathcal{H} \subset \mathcal{H}^-_t$.
      \item Let $\mathcal{D}(\Lambda_1) = \{\Phi \in \mathcal{D}(\Lambda_0) \mid \Lambda_0\Phi \in \mathcal{H}^+\}$ and $\Lambda_1\Phi = \Lambda_0\Phi$ for $\Phi \in \mathcal{D}(\Lambda_1)$.
        Then $\Lambda_1$ is a self-adjoint, positive unbounded operator on $\mathcal{H}^+_t$.
    \end{enumerate}
  \end{proposition}

  Finally, in \cite[Sec. 8]{Kisynski1964} J.\ Kisyński uses his more general results to show the following theorem on the existence of unitary propagators.
  Instead of starting from a family of operators (a Hamiltonian), he starts his construction from a family of inner products (which are sesquilinear forms) and construct the associated scale of Hilbert spaces.
  Then, he focuses on the operators representing these inner products.
  This approach has the same advantages as Simon's: the self-adjointness of the operators is guaranteed by the Hermitianity of the inner products and the representation theorem (see Theorem \ref{thm:repKato}).
  Additionally, it sets the problem on a constant (form) domain $\mathcal{H}^+$.

  \begin{theorem}[{\cite[Thm. 8.1]{Kisynski1964}}] \label{thm:kisynski_original}
    Under Assumption \ref{assump:kisynski} with $n \geq 1$, there is a unique unitary propagator $U(t,s)$ for the problem \eqref{eq:kisynski_problem} such that:
    \begin{enumerate}[label=\textit{(\roman*)},nosep,leftmargin=*]
      \item The extension $U(t,s) \in \mathcal{B}(\mathcal{H}^-)$ is strongly continuous for $s,t \in [0,T]$.
      \item $U(t,s) \mathcal{H}^+ = \mathcal{H}^+$ and the restriction $U(t,s) \in \mathcal{B}(\mathcal{H}^+)$ is strongly continuous for $s \leq t$ in $[0,T]$.
      \item $t,s \mapsto U(t,s)$ is a strongly continuously differentiable function with values on $\mathcal{B}(\mathcal{H}^+, \mathcal{H}^-)$ for $t,s \in [0,T]$ and in this sense
        \begin{equation*}
          \frac{\d}{\d t} U(t,s) = -i\Lambda(t)U(t,s) \quad\text{and}\quad
          \frac{\d}{\d s} U(t,s) = -iU(t,s)\Lambda(s).
        \end{equation*}
    \end{enumerate}
    Moreover, if Assumption \ref{assump:kisynski} is satisfied with $n \geq 2$, it also holds:
    \begin{enumerate}[resume,label=\textit{(\roman*)},nosep,leftmargin=*]
      \item $U(t,s)\mathcal{D}(\Lambda_1(s)) = \mathcal{D}(\Lambda_1(t))$ for $t,s \in [0,T]$ and for $s \in [0,T]$ and $\Phi \in \mathcal{D}(\Lambda_1(s))$ fixed, the function $t \mapsto U(t,s)\Phi$ is continuously differentiable in $[0,T]$ in the sense of $\mathcal{H}^+$ and there it satisfies
        \begin{equation*}
          \frac{\d}{\d t} U(t,s)\Phi = -i \Lambda_1(t)U(t,s)\Phi.
        \end{equation*}
      \item $U(t,s)\mathcal{D}(\Lambda_0(s)) = \mathcal{D}(\Lambda_0(t))$ for $t,s \in [0,T]$ and for $s \in [0,T]$ and $\Phi \in \mathcal{D}(\Lambda_0(s))$ fixed, the function $t \mapsto U(t,s)\Phi$ is continuously differentiable in $[0,T]$ in the sense of $\mathcal{H}$ and there it satisfies
        \begin{equation*}
          \frac{\d}{\d t} U(t,s)\Phi = -i \Lambda_0(t)U(t,s)\Phi.
        \end{equation*}
    \end{enumerate}
  \end{theorem}

\subsection{Relations between Simon's and Kisyński's approaches} \label{subsec:simon_vs_kisynski}
  Let us start this section exposing the similarities of Simon's and Kisyński's approaches.
  B.\ Simon makes some comments on the relation between both approaches in his book (see footnote 21 on \cite[p. 56]{Simon1971} and his remark after Theorem II.27).
  Referring to Theorem \ref{thm:simon}, B.\ Simon says \cite[pp. 58-59]{Simon1971}:
  \begin{quote}
    ``[T]his theorem (in a slightly different form) is contained in the work of Kisyński \cite{Kisynski1964}.
    [...]
    Yosida's techniques are also behind Kisyński's approach.
    He improves upon them and thereby obtains $\|\cdot\|_{+1}$ continuity by proving $\|\cdot\|_{+1}$ convergence of the $x_n(t)$ which we construct below.
    I discovered Kisyński's paper only after completing this proof myself and have not checked carefully the differences, if any, in the details of the two proofs.''
  \end{quote}
  On this quote, $\|\cdot\|_{+1}$ is Simon's notation for the norm on $\mathcal{H}^+$ and $x_n(t)$ denotes the solutions of the approximation $H_n(t)$ (see Subsec.\ \ref{subsec:simon}).

  Even though the starting point for each of them is slightly different, the tools they use are analogous.
  The link allowing us to relate both approaches are the scales of Hilbert spaces.
  Simon starts its construction with the scale of Hilbert spaces associated to an operator $H_0$, $\mathcal{H}^+ \subset \mathcal{H} \subset \mathcal{H}^-$, and then consider the family of operators $H(t): \mathcal{H}^+ \to \mathcal{H}^-$.
  On the other hand, Kisyński starts from a family of inner products $\langle \cdot, \cdot \rangle_{+,t}$ and builds the scales of Hilbert spaces $\mathcal{H}^+_t \subset \mathcal{H} \subset \mathcal{H}^-_t$ associated to the sesquilinear form $\langle \cdot, \cdot \rangle_{+,t}$ and the family of operators $\Lambda(t): \mathcal{H}^+ \to \mathcal{H}^-$ representing it.
  Comparing Simon's and Kisyński's theorems (Thm.\ \ref{thm:simon} and Thm.  \ref{thm:kisynski_original} respectively), it is clear that the role the familly of operators $H(t)$ plays in the approach by B.\ Simon is the same that the family $\Lambda(t)$ plays in J.\ Kisyński's.
  Therefore, the problem addressed by Simon is the same addressed by Kisyński if $h_t(\cdot,\cdot) \coloneqq \langle \cdot, \cdot \rangle_{+,t}$ is the sesquilinear form associated to $H(t)$.
  For the rest of this section we are going to assume that this equality holds.

  Before further studying the relation between both approaches, let us briefly review the assumptions B.\ Simon and J.\ Kisyński made to prove the existence of a unitary propagator.
  There are two main ingredients playing a central role in their proofs: the (uniform) equivalence of the norms $\|\cdot\|_{+,t} \coloneqq \sqrt{\langle \cdot, \cdot \rangle_{+,t}}$ and the regularity of the generators $H(t)$.

  Although both of them use repeatedly this equivalence of the norms, neither Simon nor Kisyński impose it in these terms as an assumption.
  Instead, it follows from their assumptions.
  Let us first examine how this uniform equivalence of the norms appears on each of the approaches.

  \begin{proposition} \label{prop:norm_equiv_simon}
    Let $H_0, H(t)$ be as in Assumption \ref{assump:simon} and for $\Phi \in \mathcal{H}^+$ define the norms
    \begin{equation*}
      \|\Phi\|_{+,t} \coloneqq \sqrt{\left( \Phi, H(t)\Phi \right)}, \qquad
      \|\Phi\|_0 \coloneqq \sqrt{\left( \Phi, (H_0 + 1)\Phi \right)}.
    \end{equation*}
    Then \ref{assump:simon_normequiv} holds if and only if the norms $\|\cdot\|_{+,t}$ are equivalent to $\|\cdot\|_0$ uniformly on $t \in [0,T]$.
  \end{proposition}
  \begin{proof}
    It is enough to note that $\|\Phi\|_0^2 = (\Phi, (H_0 + 1)\Phi)$ and $\|\Phi\|_{+,t}^2 = (\Phi, H(t)\Phi)$.
    Therefore, taking square roots in \ref{assump:simon_normequiv}, the uniform equivalence of the norms follows.
    Conversely, taking squares on the uniform equivalence of the norms yields \ref{assump:simon_normequiv}.
  \end{proof}
  \begin{remark}
    Note that, since we have $\|\cdot\|_0 \sim \|\cdot\|_{+,t}$ with constant uniform on $t$, for any $t_0 \in [0,T]$ there is $K$ such that
    \begin{equation*}
      K^{-1} \|\cdot\|_{+,t_0} \leq \|\cdot\|_{+,t} \leq K \|\cdot\|_{+,t_0},
    \end{equation*}
    for every $t \in [0, T]$.
  \end{remark}

  \begin{proposition} \label{prop:norm_equiv_kisynski}
    Let $\langle \cdot, \cdot \rangle_{+,t}$ be inner products as in Assumption \ref{assump:kisynski}.
    Then \ref{assump:kisynski_normequiv} imples that, for every $t \in [0,T]$, the norms $\|\cdot\|_{+,t}$ are equivalent.
    If, moreover, $t \mapsto \langle \cdot, \cdot \rangle_{+,t}$ is in $C^1([0,T])$, then the equivalence of the norms is uniform on $t$.
  \end{proposition}
  \begin{proof}
    By \ref{assump:kisynski_normequiv}, $\mathcal{H}^+_t$ is continuously embedded on $\mathcal{H}$; that is, there is $K > 0$ such that
    \begin{equation*}
      \|\Phi\| \leq K \|\Phi\|_{+,t}, \qquad \forall \Phi \in \mathcal{H}^+_t.
    \end{equation*}
    Therefore, $\langle \cdot, \cdot \rangle_{+,t}$ as an Hermitian sesquilinear form densely defined on $\mathcal{H}$, is strictly positive and there exists a self-adjoint, strictly positive operator $A(t): \mathcal{H}^+ \to \mathcal{H}$ such that
    \begin{equation*}
      \langle \Psi, \Phi \rangle_{+,t} = \langle A(t)\Psi, A(t)\Phi \rangle, \qquad
      (\Psi,\Phi \in \mathcal{H}^+).
    \end{equation*}
    Since it is positive and self-adjoint, $A(t)^{-1}: \mathcal{H} \to \mathcal{H}^+$ is a bounded self-adjoint operator.
    By the closedness of $A(t')$ and the Closed Graph Theorem, $A(t')A(t)^{-1}: \mathcal{H} \to \mathcal{H}$ is a bounded operator and for $\Phi \in \mathcal{H}^+$
    \begin{equation*}
      \|\Phi\|_{+,t'} = \|A(t')A(t)^{-1}A(t)\Phi\|
      \leq C_{t,t'} \|A(t)\Phi\| = C_{t,t'}\|\Phi\|_{+,t},
    \end{equation*}
    where $C_{t,t'} \coloneqq \|A(t')A(t)^{-1}\|$.

    The rest of the proof follows \cite[Lemma 7.3]{Kisynski1964}, where it is proven using the Uniform Boundedness Principle that the constant can be chosen independently of $t$ and $t'$.
  \end{proof}

  \begin{corollary} \label{corol:kisynski_implies_S1}
    Let $\langle \cdot, \cdot \rangle_{+,t}$, $t \in [0, T]$, be inner products satisfying Assumption \ref{assump:kisynski}, and let $H(t)$ be the operator in $\mathcal{B}(\mathcal{H}^+, \mathcal{H}^-)$ such that $\langle \Psi, \Phi \rangle_{+,t} = (\Psi, H(t)\Phi)$ for every $\Psi, \Phi \in \mathcal{H}^+$.
    Then, $H(t)$ satisfies \ref{assump:simon_normequiv} of Assumption \ref{assump:simon} with $H_0 = H(t_0)$ for some $t_0$ fixed.
  \end{corollary}
  \begin{proof}
    Let $\|\cdot\|_{+,t} = \sqrt{\langle \cdot, H(t) \cdot \rangle}$; by Proposition \ref{prop:norm_equiv_kisynski} for any $t,t_0$ it holds
    \begin{equation*}
      K^{-1} \|\Phi\|_{+,t_0} \leq \|\Phi\|_{+,t} \leq K \|\Phi\|_{+,t_0}.
    \end{equation*}
    For a fixed $t_0$, define the norm $\|\Phi\|_0 \coloneqq \sqrt{\|\Phi\|_{+,t_0}^2 + \|\Phi\|^2}$; let us show that $\|\cdot\|_0$ is equivalent to $\|\cdot\|_{+,t}$.
    Since $\|\Phi\|_{+,t_0} \leq \|\Phi\|_0$, it follows $\|\Phi\|_{+,t} \leq K \|\Phi\|_0$.
    Also, since $\|\Phi\|_{+,t} \geq \|\Phi\|$, it holds
    \begin{equation*}
      K^{-2} \|\Phi\|_0^2 \leq K^{-2}\|\Phi\|_{+,t_0}^2 + \|\Phi\|^2 \leq 2 \|\Phi\|_{+,t}^2.
    \end{equation*}
    Hence, there is $\tilde{K} > 1$ such that
    \begin{equation*}
      \tilde{K}^{-1} \|\Phi\|_0 \leq \|\Phi\|_{+,t} \leq \tilde{K} \|\Phi\|_0.
    \end{equation*}
    By Proposition \ref{prop:norm_equiv_simon}, the preceding inequalities imply \ref{assump:simon_normequiv}.
  \end{proof}

  Using the structure of the scales of Hilbert spaces, from the equivalence of the norms $\|\cdot\|_{+,t}$ it follows the equivalence of the norms $\|\cdot\|_{-,t}$ (see Theorem \ref{thm:eqiv+_equiv-}).
  This uniform equivalence of the norms $\|\cdot\|_{\pm,t}$ motivates us to work with a reference norm $\|\cdot\|_\pm$ and use the uniform equivalence when a particular $\|\cdot\|_{\pm,t}$ is convenient.
  The most convenient choice is to take $\|\cdot\|_\pm = \|\cdot\|_{\pm,t_0}$ for some reference $t_0$.
  We will denote by $\langle \cdot, \cdot \rangle_{\pm}\coloneqq \langle \cdot, \cdot \rangle_{\pm,t_0}$ and by $\mathcal{H}^\pm$ the Hilbert space $\mathcal{H}^{\pm}_{t_0}$.

  Let us now review how the regularity of the generators $H(t)$ appears in both approaches.

  \begin{proposition} \label{prop:regularity_simon}
    Let $H(t)$ be as in Assumption \ref{assump:simon}.
    Then \ref{assump:simon_regularity} holds if and only if $t \mapsto H(t)$ is differentiable in the sense of $\mathcal{B}(\mathcal{H}^+, \mathcal{H}^-)$ and, for every $t \in [0, T]$,
    \begin{equation*}
      \left\| \frac{\d}{\d t} H(t) \right\|_{+,-} \leq C.
    \end{equation*}
  \end{proposition}
  \begin{proof}
    Note that, if \ref{assump:simon_regularity} holds,
    \begin{equation*}
      \|B(t)\|_{-,+} = \sup_{\Phi \in \mathcal{H}^-} \frac{\|B(t) \Phi\|_+}{\|\Phi\|_-}
      \leq K \sup_{\Psi \in \mathcal{H}} \frac{\|H(t)^{\sfrac{1}{2}} B(t) H(t)^{\sfrac{1}{2}} \Psi\|}{\|\Psi\|}
      \leq KC
    \end{equation*}
    where we have used the equivalence of the norms $\|\cdot\|_{\pm,t} \sim \|\cdot\|_\pm$ (cf.\ Prop.\ \ref{prop:norm_equiv_simon}) and the assumption $\|H(t)^{\sfrac{1}{2}} B(t) H(t)^{\sfrac{1}{2}}\| < C$.
    Therefore, $\|B(t)\|_{-,+} < KC$ and $B(t)$ can be extended to an operator in $\mathcal{B}(\mathcal{H}^-, \mathcal{H}^+)$.
    Define the operator $T_h(t) = h^{-1}[H(t+h)^{-1} - H(t)^{-1}] - B(t)$ in $\mathcal{B}(\mathcal{H}^-, \mathcal{H}^+)$; it follows
    \begin{equation*}
      \|T_h(t)\|_{-,+} = \sup_{\Phi \in \mathcal{H}^-} \frac{\|T_h(t)\Phi\|_+}{\|\Phi\|_-}
      = \|A_0^{\sfrac{1}{2}} T_h(t) A_0^{\sfrac{1}{2}}\|,
    \end{equation*}
    where $A_0$ is the strictly positive self-adjoint operator such that $\|\cdot\|_\pm = \|A_0^{\pm\sfrac{1}{2}}\cdot\|$.
    Hence $H(t)^{-1}$ is differentiable in the sense of $\|\cdot\|_{-,+}$ if $H(t)^{-1}$ is differentiable in the sense of $\|\cdot\|$.
    By the product rule this is equivalent to $H(t)$ being differentiable in the sense of $\|\cdot\|_{+,-}$.

    Conversely, assume $H(t)$ is differentiable in the sense of $\mathcal{B}(\mathcal{H}^+, \mathcal{H}^-)$.
    By the product rule, $H(t)^{-1}$ is differentiable in the sense of $\mathcal{B}(\mathcal{H}^-, \mathcal{H}^+)$.
    Denote by $B(t)$ the derivative of $H(t)^{-1}$ in the sense of $\mathcal{B}(\mathcal{H}^-, \mathcal{H}^+)$ and consider again the operator $T_h(t) = h^{-1}[H(t+h)^{-1} - H(t)^{-1}] - B(t)$.
    For $\Phi \in \mathcal{H}$, it holds
    \begin{equation*}
      \|T_h(t)\Phi\| \leq \|T_h(t)\Phi\|_+ \leq \|T_h(t)\|_{-,+} \|\Phi\|_- \leq \|\Phi\|.
    \end{equation*}
    Therefore, $\|T_h(t)|_\mathcal{H}\| \leq \|T_h(t)\|_{-,+}$ and the differentiability of $H(t)^{-1}$ in the sense of $\mathcal{B}(\mathcal{H}^-, \mathcal{H}^+)$ implies the differentiability in the sense of $\mathcal{B}(\mathcal{H})$.

    Finally,
    \begin{equation*}
      H(t)^{\sfrac{1}{2}} \frac{\d}{\d t} H(t)^{-1} H(t)^{\sfrac{1}{2}} = H(t)^{-\sfrac{1}{2}} \frac{\d}{\d t} H(t) H(t)^{-\sfrac{1}{2}},
    \end{equation*}
    and it follows $\|\frac{\d}{\d t} H(t)\|_{+,-,t} = \|H(t)^{\sfrac{1}{2}} B(t) H(t)^{\sfrac{1}{2}}\|$.
    The equivalence of the norms $\|\cdot\|_{\pm,t} \sim \|\cdot\|_{\pm}$ implies that $\|H(t)^{\sfrac{1}{2}} B(t) H(t)^{\sfrac{1}{2}}\|$ is bounded uniformly on $t$ if and only if $\|\frac{\d}{\d t} H(t)\|_{+,-}$ is bounded uniformly on $t$.
  \end{proof}
  \begin{remark} \label{remark:op_deriv_plusminus}
    Note that, in proving Proposition \ref{prop:regularity_simon}, we have also shown that:
    \begin{enumerate}[label=\textit{(\roman*)},nosep,leftmargin=*]
      \item $\frac{\d}{\d t} H(t)^{-1}$ exists in the sense of $\mathcal{B}(\mathcal{H})$ if and only if it exists in the sense of $\mathcal{B}(\mathcal{H}^-, \mathcal{H}^+)$, which holds if and only if $\frac{\d}{\d t}H(t)$ exists in the sense of $\mathcal{B(\mathcal{H}^+, \mathcal{H}^-)}$.
      \item $\|H(t)^{\sfrac{1}{2}} \frac{\d}{\d t} H(t)^{-1} H(t)^{\sfrac{1}{2}}\| = \|H(t)^{-\sfrac{1}{2}} \frac{\d}{\d t} H(t) H(t)^{-\sfrac{1}{2}}\|$.
    \end{enumerate}
  \end{remark}

  \begin{proposition} \label{prop:regularity_kisynski}
    Let $\langle \cdot, \cdot \rangle_{+,t}$ be inner products as in Assumption \ref{assump:kisynski} and let $H(t): \mathcal{H}^+ \to \mathcal{H}^-$ be the family of bounded operators defined by $(\Psi, H(t)\Phi) = \langle \Psi, \Phi \rangle_+$.
    Then, \ref{assump:kisynski_regularity} implies that $H(t)$ is differentiable in the sense of $\mathcal{B}(\mathcal{H}^+, \mathcal{H}^-)$ and that there is $C > 0$ such that
    \begin{equation*}
      \left\| \frac{\d}{\d t} H(t) \right\|_{+,-} \leq C.
    \end{equation*}
  \end{proposition}
  \begin{proof}
    This is Proposition \ref{prop:form-differentiability} applied to $v_t(\Psi, \Phi) = \langle \Psi, \Phi \rangle_{+,t}$, which is differentiable by assumption (see \ref{assump:kisynski_regularity}).
  \end{proof}

  Corollary \ref{corol:kisynski_implies_S1} and Propositions \ref{prop:regularity_simon} and \ref{prop:regularity_kisynski} yield immediately the following result showing the connection between Simon's and Kisyński's approaches.

  \begin{theorem} \label{thm:kisynski_implies_simon}
    Let $\langle \cdot, \cdot \rangle_{+,t}$ be a family of inner products satisfying Assumption \ref{assump:kisynski} and let $H(t): \mathcal{H}^+ \to \mathcal{H}^-$ be the family of bounded operators defined by $(\Psi, H(t)\Phi) = \langle \Psi, \Phi \rangle_{+,t}$.
    Then, the family $H(t)$ satisfies Assumption \ref{assump:simon} with $H_0 = H(t_0)$ for some $t_0 \in [0,T]$.
  \end{theorem}
  \begin{proof}
    By \ref{assump:kisynski_normequiv}, $H(t)$ are strictly positive.
    By Corollary \ref{corol:kisynski_implies_S1}, \ref{assump:simon_normequiv} holds.
    Proposition \ref{prop:regularity_kisynski} implies $H(t)$ is differentiable in the sense of $\mathcal{B}(\mathcal{H}^+, \mathcal{H}^-)$ with $\|\frac{\d}{\d t}H(t)\|_{+,-} <C$, and by Proposition \ref{prop:regularity_simon} this implies \ref{assump:simon_regularity}.
  \end{proof}
  \begin{remark}
    Note that the regularity obtained for $H(t)$ is higher than required by \ref{assump:simon_regularity}: not only $t \mapsto H(t)$ is differentiable, but continuously differentiable.
    This is implied by \ref{prop:op_norm_differentiability} and the fact that $t \mapsto \langle \cdot, \cdot \rangle_{+,t}$ is continuously differentiable.
    However, this point is crucial since without the derivative $\frac{\d}{\d t} \langle \cdot, \cdot \rangle_{+,t}$ being continuous, the uniform bound for $\frac{\d}{\d t}H(t)^{-1}$ would not be recovered.
  \end{remark}


  The converse implication (i.e.\ Simon's assumption implies Kisyński's) also holds if one requires the continuity of $\frac{\d}{\d t} H(t)^{-1}$.
  \begin{theorem}
    Let $H_0, H(t)$ be operators as in Assumption \ref{assump:simon}.
    If $H(t)$ satisfies Assump.\ \ref{assump:simon} and $t \in [0,T] \mapsto B(t) = \frac{\d}{\d t} H(t)^{-1} \in \mathcal{B}(\mathcal{H})$ is continuous, then $\langle \cdot, \cdot \rangle_{+,t} \coloneqq (\cdot, H(t) \cdot)$ defines a family of inner products on $\mathcal{H}^+$ satisfying Assumption \ref{assump:kisynski} with $n = 1$.
  \end{theorem}
  \begin{proof}
    From Assumption \ref{assump:simon} we have $\mathcal{H}^+$ is a dense subspace on $\mathcal{H}$.
    Moreover, $\langle \cdot, \cdot \rangle_0 \coloneqq (\cdot, (H_0 + 1)\cdot)$ endows $\mathcal{H}^+$ with the structure of a Hilbert space topologically embedded on $\mathcal{H}$.
    Since $H_0 + 1 \leq H(t)$ by assumption, $\langle \cdot, \cdot \rangle_{+,t} \coloneqq (\cdot, H(t)\cdot)$ defines a family of inner products on $\mathcal{H}^+$ and by \ref{assump:simon_normequiv} they induce on $\mathcal{H}^+$ topologies which are equivalent to that induced by $\langle \cdot, \cdot \rangle_0$.
    Therefore, $(\mathcal{H}^+, \langle \cdot, \cdot \rangle_{+,t})$ are Hilbert spaces topologically embedded on $\mathcal{H}$ and \ref{assump:kisynski_normequiv} holds.

    By Proposition \ref{prop:regularity_simon}, \ref{assump:simon_regularity} implies that $H(t)$ is differentiable in the sense of $\mathcal{B}(\mathcal{H}^+, \mathcal{H}^-)$.
    By Proposition \ref{prop:op_norm_differentiability} this implies that the limit
    \begin{equation*}
      \lim_{t \to t_0} \sup_{\substack{\Psi, \Phi \in \mathcal{H}^+ \\ \|\Psi\|_+ = 1 = \|\Phi\|_+}}
      \frac{\langle\Psi, \Phi\rangle_{+,t} - \langle\Psi, \Phi\rangle_{+,t_0}}{t - t_0}
    \end{equation*}
    exists, which clearly implies that for every $\Psi,\Phi \in \mathcal{H}^+$ fixed, $t \mapsto \langle \Psi, \Phi \rangle_{+,t}$ is differentiable.
    Moreover, if $H(t)$ is continuously differentiable in $\mathcal{B}(\mathcal{H}^+, \mathcal{H}^-)$, so is the map $t \mapsto \langle \Psi, \Phi \rangle_{+,t}$ for every $\Psi,\Phi \in \mathcal{H}^+$.
  \end{proof}

  \begin{remark}
    Kisyński's hypothesis are slightly more restrictive than Simon's:
    Kisyński needs the extra assumption of continuous differentiability to prove the strong continuity of the unitary propagator in the sense of $\mathcal{H}^+$.
    Note also that Simon does not prove the existence of solutions in the strong sense, which is show by Kisyński requiring higher regularity.
    In particular, when $t\mapsto \langle \Psi, \Phi \rangle_{+,t}$ is twice differentiable, he is able to show that the unitary propagator $U(t,s)$ solves the Schrödinger equation not only in the weak sense but also in the strong sense (and even in the sense of $\mathcal{H}^+$).
  \end{remark}

\subsection{Hamiltonians with constant form domain}
  As it has been discussed before, J.\ Kisyński's assumptions (Assump.\ \ref{assump:kisynski}) are enough to prove the existence of unitary propagators for time-dependent Hamiltonians with constant form domain (see Def.\ \ref{def:hamiltonian_const_form}).
  However, it will be more convenient to rephrase the results previously reviewed in terms more suitable to our purposes.

  As they are stated, Kisyński's assumptions apply to a family of inner products and Simon's apply to a family of positive operators from $\mathcal{H}^+$ to $\mathcal{H}^-$.
  However, given the relation between the approaches, it can be convenient to rephrase them.
  A Hamiltonian with constant form domain $H(t)$, with lower bound $-m$, will be said to satisfy Assumption \ref{assump:simon} (or \ref{assump:simon_normequiv}, or \ref{assump:simon_regularity}) if the extension of $A(t) \coloneqq H(t) + m + 1$ to a family of operators from $\mathcal{H}^+ \to \mathcal{H}^-$ does.
  Similarly, $H(t)$ will be said to satisfy Assumption \ref{assump:kisynski} (or \ref{assump:kisynski_normequiv}, or \ref{assump:kisynski_regularity}) if the inner products $\langle \cdot, \cdot \rangle_{+,t} \coloneqq \langle \cdot, (H(t) + m + 1)\cdot \rangle$ do.

  \begin{theorem} \label{thm:kisynski}
    Let $H(t)$, ${t \in I}$, be a time-dependent Hamiltonian with constant form domain $\mathcal{H}^+$. For $t\in I$ let $\mathcal{D}(t)\subset \mathcal{H}$ be the dense domain of the operator $H(t)$.
    Suppose that for any $\Phi, \Psi \in \mathcal{H}^+$ the sesquilinear form associated with $H(t)$ is such that $t \mapsto h_t(\Phi, \Psi)$ is in $C^1(I)$.
    Then, there exists a unitary propagator $U(t, s)$, $s,t\in I$, such that:
    \begin{enumerate}[label=\textit{(\roman*)},nosep,leftmargin=*]
      \item\label{thm:kisynski_i} $U(t, s) \mathcal{H}^+ = \mathcal{H}^+$.
      \item\label{thm:kisynski_ii} For every $\Phi,\Psi \in \mathcal{H}^+$ the function $t \mapsto \langle \Phi, U(t,s)\Psi \rangle$ is continuously differentiable and $U(t,s)\Psi$ solves the \emph{weak Schrödinger equation}.
    \end{enumerate}
    If, moreover, for every $\Phi,\Psi \in \mathcal{H}^+$ the function $t \mapsto h_t(\Phi, \Psi)$ is in $C^2(I)$, then the following properties also hold:
    \begin{enumerate}[resume*]
      \item\label{thm:kisynski_iii} $U(t, s) \mathcal{D}(s) = \mathcal{D}(t)$.
      \item\label{thm:kisynski_iv} For every $\Psi \in \mathcal{D}(s)$, the function $t \mapsto U(t,s)\Psi$ is strongly continuously differentiable and $U(t,s)\Psi$ solves the strong Schrödinger equation.
    \end{enumerate}
  \end{theorem}

  \begin{remark} \label{remark:existence_form_constant}
    It is worth noticing the following:
    \begin{enumerate}[label=\textit{(\roman*)},nosep,leftmargin=*,ref=\ref{remark:existence_form_constant}\textit{(\roman*)}]
      \item We drop the assumption on the positivity of the sesquilinear forms (or the operators), which is more a convenient simplification than an actual restriction.
        The results hold for Hamiltonians which are semibounded from below.
        Indeed, if $H(t)$ is bounded from below by $-m$, then $A(t) = H(t) + m + 1$ is positive, and so the original theorems by Kisyński and Simon can be applied.
        Moreover, it is easy to check that if $\tilde{\Psi}(t)$ solves the weak (strong) dynamical problem associated to $A(t)$, then $\Psi(t) = e^{(m+1)t} \tilde{\Psi}(t)$ solves the weak (strong) dynamical problem associated to $H(t)$.
      \item \label{remark:kisinski_holds}
        Assumption \ref{assump:kisynski_regularity} is explicitly required in the statement of the theorem, while \ref{assump:kisynski_normequiv} follows from the definition of Hamiltonian with constant form domain. In particular, the \emph{topological contention} of $\mathcal{H}^+_t$ on $\mathcal{H}$ follows immediately from the uniform lower semiboundedness of $H(t)$ and the definition of $\|\cdot\|_+$:
        \begin{equation*}
          \|\Psi\|_+^2 = (\Psi, H(t)\Psi) + (m + 1) \|\Psi\|^2 \geq \|\Psi\|^2.
        \end{equation*}
    \end{enumerate}
  \end{remark}

  A further generalisation of these ideas will be needed when dealing with piecewise defined evolutions.

  \begin{definition} \label{def:piecewise_differentiable}
    We will say that a function $f$ is $n$-times piecewise differentiable on $I \subset \mathbb{R}$, denoted $f \in C_p^n(I)$, if there exists a finite collection of open subintervals of $I$, $\{I_j\}_{j=1}^m$, such that they are pairwise disjoint with $I = \bigcup_{j} \overline{I_j}$, and there exists a collection of $n$-times differentiable functions $\{g_j\}_{j=1}^m \subset C^n(I)$  satisfying $f|_{I_j} = g_j|_{I_j}$ for $j = 1, \dots, m$.
  \end{definition}
  \begin{remark}
    Note that this definition guarantees that for each interval $I_j$ the function and its derivatives can be continuously extended to $\overline{I_j}$.
  \end{remark}

  \begin{definition}\label{def:piecewise_solution}
    Let $H(t)$, $t\in I$, be defined as above.
    We say that a unitary propagator $U(t,s)$, $t,s\in I$, is a \textit{piecewise solution of the Schrödinger equation} in the strong (respectively, weak) sense if, for all $\Psi_0\in \mathcal{H}^+$, there exist $t_0<t_1<\dots<t_d\in I$ and a family of solutions of the strong (respectively, weak) Schrödinger equation $\{U_i(t,s) \mid t,s \in (t_{i-1},t_i)\}_{i=1,\dots,d}$ such that  for $t\in(t_{i-1},t_i)$, $s\in(t_{j-1},t_j)$, $1\leq j<i \leq d$ the unitary propagator $U(t,s)$ can be expressed as
    \begin{equation}
      U(t,s)=U_i(t,t_{i-1})U_{i-1}(t_{i-1},t_{i-2})\cdots U_j(t_{j},s).
    \end{equation}
  \end{definition}

  We end this section with an application of Theorem~\ref{thm:kisynski}, showing the existence of piecewise solutions for form-linear Hamiltonians with $C_p^k(I)$ coefficients.

  \begin{corollary} \label{corol:dynamics_piecewise}
    Let $H(t)$, ${t \in I}$, be a time-dependent Hamiltonian with constant form domain $\mathcal{H}^+$ such that $t \mapsto h_t(\Psi,\Phi) \in C_p^1(I)$ for every $\Phi,\Psi$.
    Suppose that the collection of intervals $\{I_j\}_{j=1}^m$ such that the restrictions $t \in I_j \mapsto h_t(\Phi, \Psi) \in \mathbb{C}$ are in $C^1(I_j)$ can be chosen independently of $\Psi,\Phi$.
    Then, there exists a strongly continuous unitary propagator solving the weak Schrödinger equation (i.e., satisfying \textit{(i)} and \textit{(ii)} of Theorem~\ref{thm:kisynski}) almost everywhere.
  \end{corollary}
  \begin{proof}
    By Theorem~\ref{thm:kisynski}, for every $j$ there exists a unitary propagator $U_j(t,s)$ that solves the weak Schrödinger equation on $\overline{I_j}$.
    Let $t_1 < t_2 < \dots < t_{m+1}$ be such that $I_j = (t_j, t_{j+1})$.

    Let $s,t \in I$ with $s < t$.
    If $s,t \in I_r$ for some $r$, define $U(t,s) = U_r(t,s)$.
    If $s \in I_r$ and $t \in I_j$ for $r \neq j$ define
    \begin{equation*}
      U(t, s) \coloneqq U_{j}(t, t_{j}) U_{j-1}(t_{j}, t_{j-1}) \cdots U_{r+1}(t_{r+1}, t_r) U_{r}(t_r, s).
    \end{equation*}
    The operator $U(t,s)$ is a unitary propagator, cf.\ Definition~\ref{def:unitarypropagator}, and solves the weak Schrödinger equation for all $t \in \bigcup_{j} I_j$ by construction.
  \end{proof}

  \begin{remark}
    \phantom{v}
    \begin{enumerate}[label=\textit{(\roman*)},nosep,leftmargin=*]
      \item The \emph{solution} provided by the previous corollary satisfies the Schrödinger equation only inside the intervals $I_j$ and therefore only for almost every $t \in I$.
        To emphasise this fact, we will call these \emph{piecewise solutions} in what follows.
      \item Since $t \mapsto h_t(\Psi,\Phi)$ is not in $C^2(I)$, the propagator $U(t,s)$ defined above does not solve the Schrödinger equation with Hamiltonian $H(t)$ in the strong sense, for $U(t,s) \Psi$ might not be in $\dom H(t)$ even if $\Psi \in \dom H(s)$ unless $t$ and $s$ lie in the same $I_{j}$.
        However, since the form domain of $H(t)$ is constant and $U(t,s)$ preserves $\mathcal{H}^+$ (cf.\ Theorem~\ref{thm:kisynski}), $U(t,s)$ solves the weak Schrödinger equation for every $t \neq t_j$, $j = 1, \dots, m$.
    \end{enumerate}
  \end{remark}

\section{Stability} \label{sec:stability}
\subsection{Sloan's stability}
  This section is devoted to the development of a stability result which is the cornerstone of several proofs in the following chapters.
  The problem to be addressed on this section is the following.
  Consider a family $H_n(t)$ of time-dependent Hamiltonians, all of them sharing the same form domain $\mathcal{H}^+$.
  If $H_n(t)$ converges to $H_0(t)$ in some sense, does the corresponding unitary propagators $U_n(t,s)$ also converge?

  After a revision of the literature on this subject, the most general result we found in this direction is due to A.D. Sloan \cite{Sloan1981}.
  In his work, tools from non-standard analysis are used to prove a stability result following the ideas in Simon's existence proof.
  First, in \cite[Sec.\ II]{Sloan1981}, it is proven that for a family $A_n(t)$ of bounded self-adjoint operators with $\sup_n \sup_{t \in I} \|A_n(t)\| < \infty$ for every compact $I$, the strong convergence of $A_n(t)$ to $A(t)$ implies that the corresponding unitary propagators $U_{A_n}(t,s)$ converge strongly to the unitary propagator associated to $A(t)$.

  Then, in \cite[Sec.\ III]{Sloan1981}, Sloan studies families of operators satisfying Assumption \ref{assump:simon} (what he calls $K$-generators) and proves the following stability theorem.

  \begin{theorem}[{\cite[Cor.\ 10]{Sloan1981}}]
    Let $A_n$, $A$ satisfy Assumption \ref{assump:simon} on $I$, with the same operator $H_0$ and the same constant $C$ for $n=1, 2, \dots$.
    Suppose $A_n(t)$ converges in the strong resolvent topology to $A(t)$ for each $t$ in $I$.
    Then $U_{A_n}(t,s)$ converges strongly to $U_A(t,s)$ for every $t$ and $s$ in $I$.
  \end{theorem}
  In order to prove this theorem, Sloan uses Yosida's approximation and the stability result for bounded generators.
  Then, using techniques similar to those used by B.\ Simon, Sloan manages to prove that the convergence of the unitary propagators for the approximating generators implies the convergence of  the approximated ones.

  Although this result is quite general, it is not suitable for the purposes of this work.
  The fact that all generators $A_n(t)$ need to satisfy \ref{assump:simon_regularity} with the same constant $C$ is not compatible with the situation under study in Section~\ref{sec:controllability}.
  In the following section we generalise these results so that the constant appearing on \ref{assump:simon_regularity} is not necessary the same for all the involved generators.
  Moreover, even though the strong resolvent convergence could be kept, it is substituted by convergence in $\mathcal{B}(\mathcal{H}^+, \mathcal{H}^-)$.

\subsection{Generalised stability}
  This subsection constitutes a generalisation of Sloan's results, needed to prove the main results of the following chapters.
  The following assumptions determine the class of problems on which we will be able to obtain the stability result.
  \begin{assumption} \label{assump:stability_assumptions}
    Let $I \subset \mathbb{R}$ be a compact interval and for each $n \in \mathbf{N} \subset \mathbb{N}$ let $H_n(t)$, ${t \in I}$, be a time-dependent Hamiltonian with constant form domain $\mathcal{H}^+$.

    We assume:
    \begin{enumerate}[label=\textit{(A\arabic*)},nosep,leftmargin=*]
      \item \label{assump:uniformbound}
        There is $m > 0$ such that $H_n(t) > -m$ for every $n \in \mathbf{N}$ and every $t\in I$.
      \item \label{assump:differentiability_forms}
        For each $n \in \boldsymbol{N}$ and every $\Phi,\Psi$ in $\mathcal{H}^+$, $t \mapsto h_{n,t}(\Psi,\Phi)$ is in $C_p^2(I)$ and the collection of subintervals $\{I_j\}_{j=1}^\nu$ on which the restrictions of $t \mapsto h_{n,t}(\Psi, \Phi)$ is $C^2$ can be chosen independently of $\Psi, \Phi$.
      \item \label{assump:norm_equiv}
        Let $n_0\in \mathbf{N}$ and $t_0\in I$. There is $c > 0$ such that, for all $n\in\mathbf{N}$ and $t\in I$, the norms of the corresponding scales of Hilbert spaces satisfy
        \begin{equation*}
          c^{-1} \|\cdot\|_{\pm,n,t} \leq \|\cdot\|_{\pm,n_0,t_0} \leq c \|\cdot\|_{\pm,n,t}.
        \end{equation*}
      \item \label{assump:L1_deriv_op_bound}
        The family of real functions on $\bigcup_j I_j$ defined by
        \begin{equation*}
          \tilde{C}_n(t) \coloneqq \sup_{\substack{\Psi, \Phi \in \mathcal{H}^+ \\ \|\Psi\|_+ = \|\Phi\|_+ = 1}} \left|\frac{\d}{\d t} h_{n,t}(\Psi, \Phi)\right|
        \end{equation*}
        satisfies $\tilde{C}_n|_{I_j} \in L^1(I_j)$ for every $1 \leq j \leq \nu$ and
        \begin{equation*}
          M \coloneqq \sup_n \sum_{j=1}^\nu \|\tilde{C}_n\|_{L^1(I_j)} <\infty.
        \end{equation*}
    \end{enumerate}
  \end{assumption}

  \begin{remark} \label{remark:stability_assump}
    $\phantom{V}$
    \begin{enumerate}[label=\textit{(\roman*)},nosep,leftmargin=*]
      \item As before, the uniform equivalence of the norms motivates the use of the simplified notation $\|\Phi\|_\pm \coloneqq \|\Phi\|_{\pm,n_0,t_0}$, that we will use throughout the text. Again we consider $\mathcal{H}^\pm$ as Hilbert spaces endowed with the norms $\|\cdot\|_\pm$.
      \item For each $n \in \boldsymbol{N}$, \ref{assump:differentiability_forms} and Prop.\ \ref{prop:form_equicont_equidiff} imply $\tilde{C}_n(t)$ is a continuous function.
        Since $I$ is compact, it is therefore guaranteed that the restrictions $\tilde{C}_n|_{I_j}$ are in $L^1(I_j)$.
    \end{enumerate}
  \end{remark}

  Now we are able to present the main result of this section.
  It is worth to stress that the type $L_1$bound obtained is crucial to prove the controllability results presented in later sections.

  \begin{theorem} \label{thm:stability_bound}
    Let $I \subset \mathbb{R}$ be a compact interval, $\mathbf{N}\subset \mathbb{N}$ and $\{H_n(t)\}_{n\in \mathbf{N}}$, $t\in I$, be a family of time-dependent Hamiltonians with constant form domain $\mathcal{H}^+$ that satisfies Assumption \ref{assump:stability_assumptions}.
    Then, there is a constant $L$ depending only on $c$ and $M$ such that for any $j,k\in \mathbf{N}$ and any $t,s \in I$ it holds
    \begin{equation*}
      \|U_j(t,s) - U_k(t,s)\|_{+,-} \leq L \int_s^t \|H_j(\tau) - H_k(\tau)\|_{+,-} \, \d\tau,
    \end{equation*}
    where $U_n(t,s)$ denotes the unitary propagator solving the weak Schrödinger equation for $H_n(t)$, $n\in \mathbf{N}$, and $\|\cdot\|_{+,-}$ is the norm in $\mathcal{B}(\mathcal{H}^+,\mathcal{H}^-)$.
  \end{theorem}

  Before we prove this theorem, it is convenient to show some auxiliary results.

  \begin{lemma} \label{lemma:Ainverse_convergence}
    Let $I \subset \mathbb{R}$ be a compact interval, $\mathbf{N}\subset \mathbb{N}$ and let $\{H_n(t)\}_{n\in \mathbf{N}}$, $t\in I$, be a family of time-dependent Hamiltonians with constant form domain $\mathcal{H}^+$ that satisfy \ref{assump:uniformbound} and \ref{assump:norm_equiv}.
    Denote $A_n(t) = H_n(t) + m + 1$.
    For any $j,k\in \mathbf{N}$ the following inequality holds
    \begin{equation*}
      \|A_j(t)^{-1} - A_k(t)^{-1}\|_{-,+} \leq c^4 \|H_j(t) - H_k(t)\|_{+,-},
    \end{equation*}
    where $\|\cdot\|_{-,+}$ denotes the norm of $\mathcal{B}(\mathcal{H}^-,\mathcal{H}^+)$.
  \end{lemma}
  \begin{proof}
    By definition, for any $n\in \mathbf{N}$ we have $\|\Phi\|_{\pm,n,t} = \|A_n(t)^{\pm \sfrac{1}{2}}\Phi\|$.
    By \ref{assump:norm_equiv} it follows
    \begin{equation*}
      \|A_n(t)^{-1}\Phi\|_+ \leq c \|A_n(t)^{-1}\Phi\|_{+,n,t} = c \|\Phi\|_{-,n,t} \leq c^2 \|\Phi\|_-,
    \end{equation*}
    and therefore $\|A_n(t)^{-1}\|_{-,+} \leq c^2$.
    Hence,
  \begin{alignat*}{2}
    \|A_j(t)^{-1} - A_k(t)^{-1}\|_{-,+} &= \|A_j(t)^{-1} [A_k(t) - A_j(t)] A_k(t)^{-1}\|_{-,+} \\
    &\leq \|A_j(t)^{-1}\|_{-,+} \|A_j(t) - A_k(t)\|_{+,-} \|A_k(t)^{-1}\|_{-,+} \\
    &\leq c^4 \|H_j(t) - H_k(t)\|_{+,-}.\tag*{\qedhere}
  \end{alignat*}
  \end{proof}

  \begin{lemma} \label{lemma:bound_derivative_operators}
    Let $\{H_n(t)\}_{n\in \mathbf{N}}$, $ t\in I$, be as in Theorem~\ref{thm:stability_bound}.
    Denote $A_n(t) = H_n(t) + m + 1$.
    Then, for every $n \in \boldsymbol{N}$ the derivative $\frac{\d}{\d t} \left(A_n(t)^{-1}\right)$ exists in the sense of the norm of $\mathcal{B}(\mathcal{H})$.
    Moreover,
    \begin{equation*}
      C_n(t) \coloneqq \left\|A_n(t)^{\sfrac{1}{2}} \frac{\d}{\d t} \left(A_n(t)^{-1}\right) A_n(t)^{\sfrac{1}{2}}\right\|
    \end{equation*}
    is in $L^1(I)$ and $\|C_n(t)\|_{L^1(I)} < c^2 M$.
  \end{lemma}
  \begin{proof}
    By hypothesis, for every fixed $n$, $A_n(t)$ satisfies Kisyński's assumptions (Assump.\ \ref{assump:kisynski}).
    By Theorem \ref{thm:kisynski_implies_simon}, $\frac{\d}{\d t} \left(A_n(t)^{-1}\right)$ exists on $\mathcal{B}(\mathcal{H})$.

    By Proposition \ref{prop:op_norm_differentiability} and \ref{assump:L1_deriv_op_bound}, we have
    \begin{equation*}
      \tilde{C}_n(t) = \left\| \frac{\d}{\d t} A_n(t) \right\|_{+,-} \in L^1(I), \quad\text{and}\quad
      \|\tilde{C}_n\|_{L^1(I)} < M.
    \end{equation*}
    Finally, from Remark \ref{remark:op_deriv_plusminus}
    \begin{align*}
      \left\|A_n(t)^{\sfrac{1}{2}} \frac{\d}{\d t} A_n(t)^{-1} A_n(t)^{\sfrac{1}{2}}\right\|
      &= \left\|A_n(t)^{-\sfrac{1}{2}} \frac{\d}{\d t} A_n(t) A_n(t)^{-\sfrac{1}{2}}\right\|\\
      &\leq c \sup_{\Psi \in \mathcal{H}\smallsetminus \{0\}}  \frac{\|\frac{\d}{\d t} A_n(t) A_n(t)^{-\sfrac{1}{2}}\Psi\|_-}{\|\Psi\|}\\
      &\leq c \sup_{\Psi \in \mathcal{H}\smallsetminus \{0\}}  \left\|\frac{\d}{\d t} A_n(t)\right\|_{+,-}
      \frac{\|A_n(t)^{-\sfrac{1}{2}}\Psi\|_+}{\|\Psi\|}\\
      &\leq c^2 \left\|\frac{\d}{\d t} A_n(t)\right\|_{+,-},
    \end{align*}
    where we have used \ref{assump:norm_equiv} twice.
  \end{proof}

  \begin{lemma} \label{lemma:propagators_bound_simon}
    Let $\{H_n(t)\}_{n\in \mathbf{N}}$, $ t\in I$, be as in Theorem~\ref{thm:stability_bound}. Define $A_n(t) := H_n(t) + m + 1$ and let
    $\Phi_n(t) = U_n(t,s)\Phi$, where $\Phi \in \mathcal{H}^+$ and $U_n(t,s)$ is the unitary propagator that provides the solution of the time-dependent Schrödinger equation with generator $A_n(t)$.
    For every $n$, we have the bounds
    \begin{gather*}
      \|A_n(t)\Phi_n(t)\|_{-,n,t} =
      \|\Phi_n(t)\|_{+,n,t} \leq e^{\frac{3}{2}\int_s^t C_n(\tau)\,\d\tau}\|\Phi\|_{+,n,s} \\
      \|\Phi_n(t)\|_{-,n,t} \leq e^{\frac{1}{2}\int_s^t C_n(\tau)\,\d\tau}\|\Phi\|_{-,n,s},
    \end{gather*}
    with $C_n(t) := \|A_n(t)^{\sfrac{1}{2}} \frac{\d}{\d t} \left(A_n(t)^{-1}\right) A_n(t)^{\sfrac{1}{2}}\|$.
  \end{lemma}
  \begin{proof}
    For each $n \in \boldsymbol{N}$, let $\{I_{n,j}\}_{j=1}^{\nu_n}$ be the open intervals where $t \mapsto h_{t,n}(\Psi,\Phi)$ is differentiable for each $\Psi, \Phi \in \mathcal{H}^+$.
    For each $t \in I_{j,n}$, $1 \leq j \leq m_n$, one has $\dot{\Phi}_n(t) = -iA_n(t) \Phi_n(t)$.
    Using that $\|A_n(t) \Psi\|_{-,n,t} = \|\Psi\|_{+,n,t}$ for all $\Psi \in \mathcal{H}^+$ and $t \in I$, this result on $I_{n,j}$ is a straightforward refinement of a theorem by Simon \cite[Theorem II.27]{Simon1971}) where we are replacing $\sup_{s \leq \tau \leq t} C_n(\tau)$ by $\int_s^t C_n(\tau)d\tau$.
    The extension to the whole interval $I$ follows straightforwardly from the additivity of the integral.
  \end{proof}

  We can prove now a slightly simplified version of Theorem~\ref{thm:stability_bound}.

  \begin{theorem} \label{thm:stability_bound_c2}
    Let $I \subset \mathbb{R}$ be a compact interval, $\mathbf{N}\subset \mathbb{N}$ and $\{H_n(t)\}_{n\in \mathbf{N}}$, $t\in I$, be a family of time-dependent Hamiltonians with constant form domain $\mathcal{H}^+$ that satisfies assumptions \ref{assump:uniformbound}, \ref{assump:norm_equiv}, \ref{assump:L1_deriv_op_bound} and such that for every $\Phi, \Psi$, $t \mapsto h_{n,t}(\Psi,\Phi)$ is a function in $C^2(I)$.
    Then, there is a constant $L$ depending only on $c$ and $M$ such that for any $j,k\in \mathbf{N}$ and any $t,s \in I$ it holds
    \begin{equation*}
      \|U_j(t,s) - U_k(t,s)\|_{+,-} \leq L \int_s^t \|H_j(\tau) - H_k(\tau)\|_{+,-} \, \d\tau,
    \end{equation*}
    where $U_n(t,s)$ denotes the unitary propagator solving the weak Schrödinger equation for $H_n(t)$, $n\in \mathbf{N}$, and $\|\cdot\|_{+,-}$ is the norm in $\mathcal{B}(\mathcal{H}^+,\mathcal{H}^-)$.
  \end{theorem}

  \begin{proof}
    Let $A_n(t) := H_n(t) + m + 1$.
    Clearly, for any $\Phi\in\mathcal{H}^+$, it holds $\langle \Phi, A_n(t)\Phi \rangle \geq \|\Phi\|^2$ and therefore $\{A_n(t)\}_{n\in\mathbf{N}}$, $t\in I$, defines a family of time-dependent Hamiltonians with constant form domain $\mathcal{H}^+$.
    The family $\{A_n(t)\}_{n\in \mathbf{N}}$ satisfies trivially Assumption \ref{assump:stability_assumptions}.
    The unitary propagators that solve the weak Schrödinger equations for $A_n(t)$ and $H_n(t)$, respectively $U_n(t,s)$ and $\tilde{U}_n(t,s)$, are related through the equation
    \begin{equation*}
      \tilde{U}_{n}(t,s) = U_{n}(t,s) e^{-i(m + 1)(t-s)}
    \end{equation*}
    and therefore it suffices to show the theorem for $\{A_n(t)\}_{n\in \mathbf{N}}$.

    For $\Phi \in \mathcal{H}^+$, define $\Phi_n(t) \coloneqq U_n(t,s) \Phi$.
    By \ref{assump:norm_equiv}, we have for $j,k\in \mathbf{N}$
    \begin{equation*}
      \|\Phi_j(t) - \Phi_k(t)\|_- \leq c \|\Phi_j(t) - \Phi_k(t)\|_{-,j,t}
    \end{equation*}
    For convenience, let us denote $z(t) = \|\Phi_j(t) - \Phi_k(t)\|_{-,j,t}$ and $y(t) = z(t)^2$.
    By the definition of the $\|\cdot\|_{-,j,t}$ we have
    \begin{equation*}
      y(t) = \langle\Phi_j(t) - \Phi_k(t), A_j(t)^{-1} [\Phi_j(t) - \Phi_k(t)]\rangle,
    \end{equation*}
    and using the chain rule it follows
    \begin{equation} \label{eq:z_derivative}
      \frac{\d}{\d t} z(t) = \frac{1}{2 z(t)} \frac{\d}{\d t} y(t).
    \end{equation}
    By Lemma~\ref{lemma:bound_derivative_operators}, $A_j(t)^{-1}$ is differentiable in the sense of $\mathcal{B}(\mathcal{H})$ and
    \begin{alignat*}{2}
      \frac{\d}{\d t} y(t) &= \langle\dot{\Phi}_j(t) - \dot{\Phi}_k(t), A_j(t)^{-1} [\Phi_j(t) - \Phi_k(t)]\rangle \\
      &\phantom{=}+ \langle\Phi_j(t) - \Phi_k(t),  B_j(t)[\Phi_j(t) - \Phi_k(t)]\rangle \\
      &\phantom{=}+ \langle\Phi_j(t) - \Phi_k(t), A_j(t)^{-1}[\dot{\Phi}_j(t) - \dot{\Phi}_k(t)]\rangle,
    \end{alignat*}
    where we have denoted $\dot{\Phi}_n(t) = \frac{\d}{\d t} \Phi_n(t)$ and $B_n(t) := \frac{\d}{\d t} A_n(t)^{-1}$, $n\in \mathbf{N}$.
    Using the self-adjointness of $A_j(t)^{-1}$, we get
    \begin{alignat}{2} \label{eq:bound_y_derivative}
      \frac{\d}{\d t} y(t) &= 2 \Re\langle\Phi_j(t) - \Phi_k(t), A_j(t)^{-1}[\dot{\Phi}_j(t) - \dot{\Phi}_k(t)]\rangle \\
      &\phantom{=}+\langle\Phi_j(t) - \Phi_k(t),  B_j(t)[\Phi_j(t) - \Phi_k(t)]\rangle. \nonumber
    \end{alignat}
    Let us bound each of the terms in the equation above.
    On the one hand, we have
    \begin{alignat*}{2}
      |\langle&\Phi_j(t) - \Phi_k(t),  B_j(t)[\Phi_j(t) - \Phi_k(t)]\rangle| = \\
      &= |\langle A_j(t)^{-\sfrac{1}{2}}[\Phi_j(t) - \Phi_k(t)],  A_j(t)^{\sfrac{1}{2}}B_j(t)A_j(t)^{\sfrac{1}{2}}A_j(t)^{-\sfrac{1}{2}}[\Phi_j(t) - \Phi_k(t)]\rangle| \\
      &\leq \|A_j(t)^{\sfrac{1}{2}}B_j(t)A_j(t)^{\sfrac{1}{2}}\|
      \langle \Phi_j(t) - \Phi_k(t),  A_j(t)^{-1}[\Phi_j(t) - \Phi_k(t)]\rangle \\
      &= C_j(t) z(t)^2,
    \end{alignat*}
    where $C_j(t) \coloneqq \|A_j(t)^{\sfrac{1}{2}}B_j(t)A_j(t)^{\sfrac{1}{2}}\|$.

    On the other hand, using the weak Schrödinger Equation, it follows that
    \begin{multline*}
      \Re\langle\Phi_j(t) - \Phi_k(t), A_j(t)^{-1}[\dot{\Phi}_j(t) - \dot{\Phi}_k(t)]\rangle = \\
      \begin{alignedat}{2}
        &= -\Im\langle\Phi_j(t) - \Phi_k(t), \Phi_j(t) - A_j(t)^{-1}A_k(t)\Phi_k(t)\rangle \\
        &= -\Im \langle \Phi_j(t) - \Phi_k(t), [A_k(t)^{-1} - A_j(t)^{-1}] A_k(t) \Phi_k(t) \rangle
      \end{alignedat}
    \end{multline*}
    where we have used that $\Im\|\Phi_j(t)-\Phi_k(t)\|=0$.
    Noting that $A_j(t)^{-1}$ maps $\mathcal{H}^-$ into $\mathcal{H}^+$ one gets
    \begin{multline*}
      \Re\langle\Phi_j(t) - \Phi_k(t), A_j(t)^{-1}[\dot{\Phi}_j(t) - \dot{\Phi}_k(t)]\rangle \leq \\
      \begin{alignedat}{2}
        &\leq  \|\Phi_j(t) - \Phi_k(t)\|_- \|[A_k(t)^{-1} - A_j(t)^{-1}] A_k(t) \Phi_k(t)\|_+ \\
        &\leq  c z(t) \|A_k(t)^{-1} - A_j(t)^{-1}\|_{-,+} \|A_k(t) \Phi_k(t)\|_-.
      \end{alignedat}
    \end{multline*}
    By Assumption \ref{assump:norm_equiv}, Lemma~\ref{lemma:Ainverse_convergence} and Lemma~\ref{lemma:propagators_bound_simon}, we have
    \begin{multline*}
      \Re\langle\Phi_j(t) - \Phi_k(t), A_j(t)^{-1}[\dot{\Phi}_j(t) - \dot{\Phi}_k(t)]\rangle \leq \\
      \leq  c^7 z(t) e^{\frac{3}{2} \int_s^t C_k(\tau)\,\d\tau} \|A_j(t) - A_k(t)\|_{+,-} \|\Phi\|_+.
    \end{multline*}

    Substituting these results into Equation \eqref{eq:bound_y_derivative} and using Lemma~\ref{lemma:bound_derivative_operators}, it follows
    \begin{equation*}
      \frac{\d}{\d t} y(t) \leq C_j(t) z(t)^2 + 2c^7 e^{\frac{3}{2}c^2 M} z(t) F(t) \|\Phi\|_+,
    \end{equation*}
    where $F(t) = \|A_j(t) - A_k(t)\|_{+,-}$.
    By Equation \eqref{eq:z_derivative}, one has
    \begin{equation*}
      \frac{\d}{\d t} z(t) \leq \frac{C_j(t)}{2} z(t) + c^7 e^{\frac{3}{2}c^2 M} F(t) \|\Phi\|_+.
    \end{equation*}
  Using Grönwall's inequality with initial condition $z(s) = 0$, it follows
  \begin{equation*}
    z(t) \leq c^7e^{2c^2M} \|F\|_{L^1(s,t)} \|\Phi\|_+.
  \end{equation*}
  Hence, using again \ref{assump:norm_equiv},
    \begin{equation*}
      \|\Phi_j(t)-\Phi_k(t)\|_-\leq c^8e^{2c^2M}\|F\|_{L^1(s,t)}\|\Phi\|_+.
    \end{equation*}
    Using the definition of the operator norm in $\mathcal{B}(\mathcal{H}^+,\mathcal{H}^-)$ we get
    \begin{equation*}
      \|U_j(t,s) - U_k(t,s)\|_{+,-} \leq c^8e^{2c^2M}\|F\|_{L^1(s,t)},
    \end{equation*}
    which completes the proof.
  \end{proof}

  \begin{proof}[Proof of Theorem~\ref{thm:stability_bound}]
    Let $k,\ell \in \boldsymbol{N}$ fixed, and let $\{I_{j}\}_{j=1}^\nu$ be a collection of open intervals satisfying $\bigcup_j \overline{I_j} = I$ such that the functions $t \mapsto h_{i,t}(\Psi, \Phi)$, $\Psi,\Phi \in \mathcal{H}^+$, $i = k,\ell$, are differentiable on each $I_j$ (note that the existence of such collections follows from \ref{assump:differentiability_forms}).
    Let $t_1 < t_2 < \dots < t_{m+1}$ be such that $I_j = (t_j, t_{j+1})$.

    Note that, for $t,s \in I$, Lemma~\ref{lemma:bound_derivative_operators} implies
    \begin{equation} \label{eq:Cn_bound}
      \int_s^t C_n(\tau) \,\d\tau \leq \sum_{j=1}^\nu \|C_n(t)\|_{L^1(I_j)} \leq c^2 M \qquad (n \in \boldsymbol{N}),
    \end{equation}
    where, again, $C_n(t) = \|\frac{\d}{\d t} H_n(t)\|_{+,-,t}$, $t \in \bigcup_j I_j$.
    By Theorem~\ref{thm:stability_bound_c2}, it holds
    \begin{equation} \label{eq:stability_bound}
      \|U_k(t, s) - U_\ell(t, s)\|_{+,-} < L \int_s^t \|H_k(t) - H_\ell(t)\|_{+,-} \,\d\tau,
      \qquad (s,t \in \overline{I_j}).
    \end{equation}
    where the constant $L$ depends only on $c$ and $M$ in Assumption~\ref{assump:stability_assumptions}.

    Consider now $0 \leq s < t$ such that $t \in I_{j}$, $s \in I$, with $j \neq r$, and denote $\Psi_i(t) \coloneqq U_i(t,s)\Psi$, $i = k,\ell$.
    By Eq.~\eqref{eq:stability_bound}, we have
    \begin{alignat*}{2}
      \|\Psi_k(t) - \Psi_\ell(t)\|_{-,\ell,t}
      &= \|U_k(t,t_j) \Psi_k(t_j) - U_\ell(t,t_j) \Psi_\ell(t_j)\|_{-,\ell,t} \\
      &\leq \|[U_k(t,t_j) - U_\ell(t,t_j)]\Psi_k(t_j)\|_{-,\ell,t}\;+ \\
      &\phantom{\leq\quad} +  \|U_\ell(t,t_j)[\Psi_k(t_j) - \Psi_\ell(t_j)]\|_{-,\ell,t} \\
      &\leq cL \|F_{k,\ell}\|_{L^1(t_j,t)} \|\Psi_n(t_j)\|_+ \;+\\
      &\phantom{\leq\quad}+ \|U_\ell(t,t_j)[\Psi_k(t_j) - \Psi_\ell(t_j)]\|_{-,\ell,t},
    \end{alignat*}
    where we have defined $F_{k,\ell}(t) \coloneqq \|H_k(t) - H_\ell(t)\|_{+,-}$.

    Let us focus on each of the addends separately.
    First, using the equivalence of the norms \ref{assump:norm_equiv} and Lemma~\ref{lemma:propagators_bound_simon} one gets
    \begin{equation*}
      \|\Psi_k(t_j)\|_+
      \leq ce^{\frac{3}{2} \int_s^{t_j}C_k(\tau)\,\d\tau} \|\Psi_k(s)\|_{+,k,s}
      \leq c^2 e^{\frac{3}{2} c^2 M} \|\Psi\|_+,
    \end{equation*}
    where we have used Equation~\eqref{eq:Cn_bound}.
    On the other hand, by Lemma~\ref{lemma:propagators_bound_simon} one has
    \begin{equation*}
      \|U_\ell(t,t_j)[\Psi_k(t_j) - \Psi_\ell(t_j)]\|_{-,\ell,t}
      \leq e^{\frac{1}{2}\int_{t_j}^t C_\ell(\tau)\,\d\tau} \|\Psi_k(t_j) - \Psi_\ell(t_j)\|_{-,\ell,t_j}.
    \end{equation*}

    Therefore, it follows
    \begin{alignat*}{2}
      \|\Psi_k(t) - \Psi_\ell(t)\|_{-,\ell,t}
      &\leq Lc^3 e^{\frac{3}{2} c^2 M} \|F_{k,\ell}\|_{L^1(t_j,t)} \|\Psi\|_+ +\\
      &\phantom{\leq}+ e^{\frac{1}{2}\int_{t_j}^t C_\ell(\tau)\,\d\tau}\|\Psi_k(t_j) - \Psi_\ell(t_j)\|_{-,\ell,t_j}.
    \end{alignat*}
    Applying again the same strategy to bound $\|\Psi_k(t_j) - \Psi_\ell(t_j)\|_{-,\ell,t_j}$, we get
    \begin{multline*}
      \|\Psi_k(t) - \Psi_\ell(t)\|_{-,\ell,t} \leq Lc^3 e^{\frac{3}{2} c^2 M} \|F_{k,\ell}\|_{L^1(t_j,t)} \|\Psi\|_+ \;+\\
      + e^{\frac{1}{2}\int_{t_j}^t C_\ell(\tau)\,\d\tau}
      \left[\vphantom{e^{\frac{1}{2}\int_{t_{j-1}}^{t_j} C_\ell(\tau)\,\d\tau}}\right.
      Lc^3 e^{\frac{3}{2} c^2 M} \|F_{k,\ell}\|_{L^1(t_{j-1},t_j)} \|\Psi\|_+ \;+ \\
      \left.+ e^{\frac{1}{2}\int_{t_{j-1}}^{t_j} C_\ell(\tau)\,\d\tau}\|\Psi_k(t_{j-1}) - \Psi_\ell(t_{j-1})\|_{-,\ell,t_{j-1}} \right].
    \end{multline*}
    Using this procedure iteratively and Eq.~\eqref{eq:Cn_bound} we get, for $0 \leq s \leq t$, $s \in I_r$, $t \in I_j$,
    \begin{multline*}
      \|\Psi_k(t) - \Psi_\ell(t)\|_{-,\ell,t} \leq \tilde{L}\left(\|F_{k,\ell}\|_{L^1(t_j,t)} + \|F_{k,\ell}\|_{L^1(t_{j-1},t_j)} + \right. \\
      \left.+ \dots + \|F_{k,\ell}\|_{L^1(s,t_{n,r})}\right) \|\Psi\|_+,
    \end{multline*}
    where we have defined $\tilde{L} = L c^3e^{2c^2M}$.
    That is,
    \begin{equation*}
      \|\Psi_k(t) - \Psi_\ell(t)\|_{-,\ell,t} \leq \tilde{L}\|F_{k,\ell}\|_{L^1(s,t)}\|\Psi\|_+.
    \end{equation*}
  \end{proof}

  We end this subsection by introducing the following stability result which shows that convergence in the norm of $\mathcal{B}(\mathcal{H}^+,\mathcal{H}^-)$ implies strong convergence.

  \begin{theorem}\label{thm:stability_H}
    Let $\{H_n(t)\}_{n\in \mathbb{N}}$, $ t\in I$, be a family of time-dependent Hamiltonians with constant form domain $\mathcal{H}^+$and let $U_n(t,s)$ be the strongly continuous unitary propagators that solve their respective weak Schrödinger equations.
    Suppose that $\lim_{n\to\infty}\|U_n(t,s) - U_0(t,s)\|_{+,-}=0$. Then $U_n(t,s)$ converges strongly to $U_0(t,s)$.
  \end{theorem}

  \begin{proof}
    Suppose that $\Psi,\Phi\in\mathcal{H}^+$.
    Then
    \begin{equation*}
      \left|\langle \Psi, \left(U_n(t,s) - U_0(t,s)\right)\Phi \rangle\right|\leq \|\Psi\|_+\|\Phi\|_+\|U_n(t,s) - U_0(t,s)\|_{+,-}.
    \end{equation*}
    By the density of $\mathcal{H}^+$ in $\mathcal{H}$ and the unitarity of $U_n(t,s)$ for all $n\in \mathbb{N}$, this implies that $U_n(t,s)\Phi$ converges weakly to $U_0(t,s)\Phi$. Since $\|U_n(t,s)\Phi\|=\|\Phi\|$, the latter implies the strong convergence of $U_n(t,s)$ to $U_0(t,s)$.
  \end{proof}

\section{Form-linear Hamiltonians} \label{sec:form_linear}
  All the Hamiltonian operators that we consider in Quantum Circuits have a very particular time-dependent structure, on which we are going to focus from now on.

  \begin{definition} \label{def:formlinear_hamiltonian}
    Let $I\subset{\mathbb{R}}$ be a compact interval, $H(t)$, ${t \in I}$, a time-dependent Hamiltonian with constant form domain $\mathcal{H}^+$, cf.\ Definition \ref{def:hamiltonian_const_form}, and let $\mathcal{F}$ be a subset of the real-valued, continuous functions on $I$.
    We say that $H(t)$ is a \emph{form-linear time-dependent Hamiltonian with space of coefficients} $\mathcal{F}$ if there is a finite collection of Hermitian sesquilinear forms densely defined on $\mathcal{H}^+$, $h_0, h_1, \dots, h_N$, and functions $f_1, f_2, \dots, f_N \in \mathcal{F}$ such that the sesquilinear form $h_t$ associated with $H(t)$ is given by
    \begin{equation*}
      h_t(\Psi, \Phi) = h_0(\Psi, \Phi) + \sum_{i=1}^N f_i(t) h_i(\Psi, \Phi).
    \end{equation*}
    The collection $\{ h_i\}_{i=0}^N$ is called the \emph{structure} of $H(t)$, while $\{f_i\}_{i=1}^N$ are called the \emph{coefficients} of $H(t)$.
  \end{definition}

  As an immediate consequence of Theorem~\ref{thm:kisynski}, we have the following result on the existence of dynamics for form-linear time-dependent Hamiltonians.
  \begin{proposition} \label{prop:dynamics_formlinear}
    For every form-linear, time-dependent Hamiltonian whose space of coefficients $\mathcal{F}$ is a subset of $C_p^1(I)$ there exists a strongly continuous unitary propagator that satisfies \ref{thm:kisynski_i}, \ref{thm:kisynski_ii} of Theorem~\ref{thm:kisynski} for almost every $t \in I$.
    If, moreover, $\mathcal{F} \subset C^2(I)$, then the unitary propagator also satisfies \ref{thm:kisynski_iii} and \ref{thm:kisynski_iv} of Theorem~\ref{thm:kisynski}.
  \end{proposition}

  \begin{proposition} \label{prop:formlinear_equiv_norms}
    For every $n \in \mathbf{N}\subset \mathbb{N}$, let $H_n(t)$, $t\in I$, be a form-linear time-dependent Hamiltonian with structure $\{ h_i\}_{i=0}^N$ and coefficients $\{f_{n,i}\}_{i=1}^N$ such that $\sup_{n,t} |f_{n,i}(t)| < \infty$ for every $i$.
    Assume also that the Hamiltonians $\{H_n(t)\}_{n\in \mathbf{N}}$ have the same lower bound $m$. Then there is $c$ independent of $n$ and $t$ such that, for every $n \in \mathbf{N}$ and $t \in I$,
    \begin{equation*}
      c^{-1} \|\argdot\|_{\pm,n,t} \leq \|\argdot\|_\pm \leq c \|\argdot\|_{\pm,n,t},
    \end{equation*}
    where we have defined $\|\argdot\|_{\pm}:=\|\argdot\|_{\pm,n_0,t_0}$, for some fixed time $t_0\in I$ and $n_0 \in \boldsymbol{N}$.
  \end{proposition}
  \begin{proof}
    Let $A_n(t) := H_n(t) + m + 1$ for $n\in\mathbf{N}$.
    We have
    \begin{equation*}
      \langle \Psi, \Phi \rangle_{+,n,t} = \langle A_n(t)^{\sfrac{1}{2}} \Psi, A_n(t)^{\sfrac{1}{2}} \Phi \rangle,
    \end{equation*}
    cf.\ Section~\ref{sec:hilbert_scales}.
    By the Closed Graph Theorem, the operator defined by $T_n(t) \coloneqq A_n(t)^{\sfrac{1}{2}} A_0(t_0)^{-\sfrac{1}{2}}$ is a bounded operator on $\mathcal{H}$ and, for $\Phi\in\mathcal{H}^+$,
    \begin{equation*}
      \|\Phi\|_{+,n,t} = \|T_n(t) A_0(t_0)^{+\sfrac{1}{2}} \Phi\| \leq \|T_n(t)\| \|\Phi\|_{+}.
    \end{equation*}
    Analogously one can get  $\|\Phi\|_{+} \leq \|T_n(t)^{-1}\| \|\Phi\|_{+,n,t}$ for $\Phi \in \mathcal{H}^+$, and taking $c_{n,t} = \max \{\|T_n(t)\|^{-1}, \|T_n(t)^{-1}\|\}$ it follows
    \begin{equation} \label{eq:norm_equiv_item1}
      c_{n,t}^{-1} \|\Phi\|_{+,n,t} \leq \|\Phi\|_+ \leq c_{n,t} \|\Phi\|_{+,n,t},
      \qquad \forall \Phi \in \mathcal{H}^+.
    \end{equation}

    We will show now that the previous inequalities also hold with constants independent of $n$ and $t$.
    By Equation \eqref{eq:norm_equiv_item1} we have $c_{n,t}^{-2} \|\Phi\|_+^{2} \leq |\langle \Phi, \Phi \rangle_{+,n,t}| \leq c_{n,t}^2 \|\Phi\|_+^2$; therefore, $\langle \Phi, \Psi \rangle_{+,n,t}$ is a  closed, positive, bounded Hermitian sesquilinear form on the Hilbert space $(\mathcal{H}^+, \|\argdot\|_+)$.
    By Riesz's Theorem one can define a positive, bounded, self-adjoint operator such that
    \begin{equation*}
      \langle \Phi, \Psi \rangle_{+,n,t} = \langle Q_n(t)\Phi, \Psi \rangle_+.
    \end{equation*}
    Moreover $Q_n(t)$ is bounded away from the origin by $c_{n,t}^{-2}>0$ and therefore boundedly invertible.

    Let $M = \max_i \sup_{n,t} |f_{n,i}(t)|$, which is finite by assumption.
    Then,
    \begin{alignat*}{2}
      \|\Phi\|_{+,n,t}^2 &= h_0(\Phi,\Phi) + (m+1) \|\Phi\|^2 + \sum_{i=1}^N f_{n,i}(t) h_i(\Phi,\Phi) \\
      &\leq |h_0(\Phi,\Phi)| + (m+1) \|\Phi\|^2 + M \sum_{i=1}^N |h_i(\Phi, \Phi)|.
    \end{alignat*}
    Therefore, it is clear that we have $\sup_{n,t} \langle Q_n(t)\Phi, \Phi \rangle_+ < \infty$, and by the Uniform Boundedness Principle there is $K_1$ such that $\|Q_n(t)^{\sfrac{1}{2}}\|_{+} \leq K_1$.
    This implies
    \begin{equation*}
      \|\Phi\|_{+,n,t} \leq K_1 \|\Phi\|_+.
    \end{equation*}
    By hypothesis we have $A_n(t)>1$, and therefore
    \begin{equation*}
      \|\Phi\|_{+,n,t}^2 \geq \|\Phi\|^2 = \frac{\|\Phi\|^2}{\|\Phi\|_+^2} \|\Phi\|_+^2.
    \end{equation*}
    Therefore, for every $\Phi\in\mathcal{H}^+$ there is a constant $K_{2,\Phi}$, independent of $n$ and $t$, such that $\|Q_n(t)^{\sfrac{1}{2}}\Phi\|_+\geq K_{2,\Phi}\|\Phi\|_+$ and hence
    $$\|Q_n(t)^{-\sfrac{1}{2}}\Phi\|_+\leq K_{2,\Phi}^{-1}\|\Phi\|_+$$

    By the Uniform Boundedness Principle we have $\sup_{n,t} \|Q_n(t)^{-\sfrac{1}{2}}\|_+ = K_2 < \infty$, and
    \begin{alignat*}{2}
      \|\Phi\|_+^2 &= \langle Q_n(t)\Phi, Q_n(t)^{-1}\Phi \rangle_+ = \langle \Phi, Q_n(t)^{-1}\Phi \rangle_{+,n,t} \\
      &\leq \|\Phi\|_{+,n,t} \|Q_n(t)^{-1} \Phi\|_{+,n,t}
      = \|\Phi\|_{+,n,t} \langle \Phi, Q_n(t)^{-1}\Phi \rangle_{+}^{\sfrac{1}{2}}\\
      &\leq K_2 \|\Phi\|_{+,n,t}\|\Phi\|_+.
    \end{alignat*}
    This shows that $\|\Phi\|_+ \leq K_2 \|\Phi\|_{+,n,t},$
    and taking $c = \max\{K_1,K_2\}$ it follows
    \begin{equation*}
      c^{-1} \|\Phi\|_{+,n,t} \leq \|\Phi\|_+ \leq c \|\Phi\|_{+,n,t}.
    \end{equation*}
    Finally, by Theorem \ref{thm:eqiv+_equiv-}, the uniform equivalence of the norms $\|\cdot\|_{-,t}$ follows.
  \end{proof}

  \begin{proposition} \label{prop:formlinear_A4}
    For every $n \in \mathbf{N}\subset \mathbb{N}$, let $H_n(t)$, $t\in I$, be a form-linear time-dependent Hamiltonian with structure $\{ h_i\}_{i=0}^N$ and coefficients $\{f_{n,i}\}_{i=1}^N \subset C_p^1(I)$.
    Let $\{I_j\}_{j=1}^\nu$ be a collection of open intervals satisfying $I = \bigcup_j \overline{I_j}$ and such that $f_{n,i}$, $1 \leq i \leq N$, is differentiable on every $I_j$, $1 \leq j \leq m$.
    Assume that the derivatives of the coefficients satisfy $\sup_n \sum_{j=1}^\nu \|f_{n,i}'\|_{L^1(I_j)} < \infty$ and that there exists is a constant $K$ such that $h_i(\Psi, \Phi) \leq K \|\Psi\|_+ \|\Phi\|_+$ for every $i=1,\dots,N$.
    Then, the functions
    \begin{equation*}
      \tilde{C}_n(t) \coloneqq \sup_{\substack{\Psi, \Phi \in \mathcal{H}^+ \\ \|\Psi\|_+ = \|\Phi\|_+ = 1}} \left|\frac{\d}{\d t} h_{n,t}(\Psi, \Phi)\right|
    \end{equation*}
    are in $L^1(I)$ and $M \coloneqq \sup_n \sum_{j} \|\tilde{C}_n\|_{L^1(I_j)} <\infty$.
  \end{proposition}
  \begin{proof}
    We only need to show that there is $L > 0$ such that $\sum_j \|\tilde{C}_n\|_{L^1(I_j)} \leq L$ (see Remark \ref{remark:stability_assump}).
    For $\Psi,\Phi \in \mathcal{H}^+$, one has
    \begin{equation*}
      \left| \frac{\d}{\d t} h_{n,t}(\Psi, \Phi) \right| \leq \sum_{i=1}^N |f_{n,i}'(t)| |h_i(\Psi, \Phi)|
      \leq K \|\Psi\|_+ \|\Phi\|_+ \sum_{i=1}^N |f_{n,i}'(t)|.
    \end{equation*}

    Hence,
    \begin{equation*}
      \sum_{j=1}^\nu\|\tilde{C}_n\|_{L^1(I_j)} \leq K \sum_{i=1}^N \sum_{j=1}^\nu \|f_{n,i}'\|_{L^1(I_j)} \leq L,
    \end{equation*}
    with $L = \max_i \sup_n \sum_j \|f_{n,i}'\|_{L^1(I_j)} / NK$.
  \end{proof}

  \begin{theorem} \label{thm:stability_formlinear}
    Let $I \subset \mathbb{R}$ be a compact interval.
    For every $n \in \mathbf{N} \subset \mathbb{N}$, let $H_n(t)$, $t\in I$, be a form-linear time-dependent Hamiltonian with structure $\{h_i\}_{i=0}^N$ and coefficients $\{f_{n,i}\}_{i=1}^N \subset C_p^2(I)$.
    For a given $n \in \boldsymbol{N}$, denote by $\{I_{n,j}\}_{j=1}^{\nu_n}$ the family of open intervals such that $f_{n,i}$, $1 \leq i \leq N$, is differentiable on every $I_{n,j}$, $1 \leq j \leq m_n$.
    Assume that:
    \begin{enumerate}[label=\textit{(\roman*)},nosep,leftmargin=*]
      \item The Hamiltonians $\{H_n(t)\}_{n\in \mathbf{N}}$ have the same lower bound $m$.
      \item For every $i = 1, \dots, N$, it holds $\sup_{n,t} |f_{n,i}(t)| < \infty$.
      \item For every $i = 1, \dots, N$, it holds $\sup_n \sum_j \|f_{n,i}'\|_{L^1(I_{n,j})} < \infty$.
      \item There is $K$ such that $h_i(\Psi, \Phi) \leq K \|\Psi\|_+ \|\Phi\|_+$ for every $i=0,\dots,N$.
    \end{enumerate}
    Then, the family of form-linear time-dependent Hamiltonians $\{H_n(t)\}_{n \in \boldsymbol{N}}$ satisfies Assumption \ref{assump:stability_assumptions} and for $j,k\in\mathbf{N}$, and $s,t \in I$
    \begin{equation*}
      \|U_j(t,s) - U_k(t,s)\|_{+,-} \leq L \sum_{i=1}^N \|f_{j,i} - f_{k,i}\|_{L^1(s,t)}
    \end{equation*}
    for a constant $L$ independent of $j$, $k$, $t$ and $s$.
    If, in addition, for every $i$ we have $f_{n,i} \to f_{0,i}$ in $L^1(I)$ then, for every $s,t \in I$, $U_n(t,s)$ converges strongly to $U_0(t,s)$ uniformly on $t,s$ and also in the sense of $\mathcal{B}(\mathcal{H}^+, \mathcal{H^-})$.
  \end{theorem}
  \begin{proof}
    By hypothesis, \ref{assump:uniformbound} and \ref{assump:differentiability_forms} hold.
    The condition \ref{assump:norm_equiv} follows from Proposition~\ref{prop:formlinear_equiv_norms}, and \ref{assump:L1_deriv_op_bound} from Proposition \ref{prop:formlinear_A4}.
    Applying Theorem~\ref{thm:stability_bound} and Theorem~\ref{thm:stability_H}, the statement follows.
  \end{proof}

\end{document}


\ifSubfilesClassLoaded{
    \setcounter{chapter}{4}
  }{}
\chapter{Laplace-Beltrami operator on a Riemannian manifold} \label{ch:laplacian}
  The aim of this chapter is to introduce the families of operators that will be relevant in the subsequent chapters.
  In section \ref{sec:sobolev_laplace}, we recall the basic notions of Sobolev spaces on a Riemannian manifold and introduce the Laplace-Beltrami operator and the magnetic Laplacian.
  In Section \ref{sec:sa_extensions}, we review the method exposed in \cite{IbortLledoPerezPardo2015} which uses the relation between self-adjoint operators and sesquilinear forms to characterise (a subclass of) the self-adjoint extensions of the Laplace-Beltrami and the magnetic Laplacian.

\section{Sobolev spaces and the Laplace-Beltrami operator} \label{sec:sobolev_laplace}
  Let $\Omega$ be a Riemannian manifold with boundary $\partial \Omega$ and let $\eta$ denote the Riemannian metric.
  We will consider only orientable, compact and smooth manifolds with piecewise smooth boundary.
  This is a convenient simplification, but most of the results should hold with minor changes for more general cases, such as convex manifolds with Lipschitz boundary.
  \nomenclature[L]{$\Lambda(\Omega)$}{Space of smooth, complex-valued 1-forms on $\Omega$}
  We will denote by $\Lambda(\Omega)$ the space of smooth, complex-valued 1-forms on $\Omega$.

  For $\alpha, \beta\in \Lambda(\Omega)$ smooth, complex-valued 1-forms in $\Omega$, we define the inner product
  \begin{equation*}
    \langle \alpha, \beta \rangle_{\Lambda} = \int_\Omega \eta^{-1}(\overline{\alpha}, \beta) \, \d\mu_\eta,
  \end{equation*}
  and $\|\cdot\|_{\Lambda}$ the associated norm, which allows to define the space of differential 1-forms with square-integrable coefficients.
  For the rest of the work we will drop the subindex in the norm and inner product defined above whenever it leads to no confusion.

  We denote by $H^k(\Omega)$ the Sobolev space of class $k$ on $\Omega$ (see \cite{Taylor2011,LionsMagenes1972}), and its norm $\|\Phi\|_{H^k(\Omega)}$.
  \nomenclature[H]{$H^k(\Omega)$}{Sobolev space of order $k$ on the manifold $\Omega$}%
  \nomenclature[N]{$\Vert\cdot\Vert_k$}{Sobolev norm of order $k = 1, 2, \dots$}%
  When there is no risk of confusion about what the manifold $\Omega$ we are working with is, we will simplify the notation and write $\|\cdot\|_k \coloneqq \|\cdot\|_{H^k(\Omega)}$.

  The Laplace-Beltrami operator can be defined on the space of smooth functions $C^{\infty}(\Omega)$ in terms of the exterior differential, $\d$, and the codifferential, $\delta$, cf.\ \cite{MarsdenAbrahamRatiu1988},
  \begin{equation*}
    \Delta = \delta\d.
  \end{equation*}

  For the rest of this work all the derivatives acting on the manifold $\Omega$  are going to be considered in the weak sense. The following characterisation will be useful, cf.\ \cite[Section 7.1]{Davies1995}.
  \begin{proposition} \label{prop:sobolev_norm_equivalence}
    The Sobolev norm of order 1, $\|\cdot\|_1$, is equivalent to $$\sqrt{\|\cdot\|^2 + \|\d \cdot\|^2}.$$
  \end{proposition}

In particular we will take the former expression as our definition of the Sobolev norm of order 1, $\|\cdot\|_1$. The boundary $\partial\Omega$ of the Riemannian manifold inherits the structure of a Riemannian manifold with the metric given by the pull-back of the Riemannian metric on $\Omega$, which we will denote by $\partial\mu$.
  The well-known Trace Theorem (cf.\ \cite{LionsMagenes1972} and\ \cite{AdamsFournier2003}) relates Sobolev spaces defined over a manifold and its boundary:
  \begin{theorem}
    \label{thm:lions_trace}
    For any smooth function $\Phi \in C^\infty(\Omega)$, define the trace map $\gamma:
    C^\infty(\Omega) \to C^\infty(\partial\Omega)$ by $\gamma(\Phi) =
    \Phi|_{\partial\Omega}$. Then, there is a unique continuous extension of the trace map
    $\gamma: H^k(\Omega) \to H^{k-1/2}(\partial\Omega)$, $k > 1/2$, such that it is
    surjective.
  \end{theorem}
  Notice that, with a slight abuse of notation, we denote the trace map and its continuous extension with the same symbol.
  Finally, let us define some particular extensions of the Laplace-Beltrami operator which we will use in the following chapters.
  We refer to \cite{LionsMagenes1972} for further details.

  \begin{definition}
    \label{def:min_max_laplacian}
    Let $\Delta$ denote the Laplace-Beltrami operator on a Riemannian manifold with boundary and $\Delta_0 \coloneqq \Delta |_{C_c^\infty(\Omega^\circ)}$, its restriction to smooth functions compactly supported on the interior of $\Omega$.
    We define the following extensions:
    \begin{enumerate}[label=\textit{(\roman*)},nosep, leftmargin=*]
      \item The \emph{minimal closed extension}, $\Delta_{\text{min}}$, is the closure of $\Delta_0$.
      \item The \emph{maximal closed extension} $\Delta_{\text{max}}$ is the adjoint of the
        minimal closed extension, whose domain can be characterised by
        \begin{equation*}
          \dom\Delta_{\text{max}} = \{\Phi \in L^2(\Omega) \mid\Delta \Phi \in L^2(\Omega)\}.
        \end{equation*}
    \end{enumerate}
  \end{definition}
  For the rest of the work we will denote by Greek capital letters the functions on the spaces
  $H^k(\Omega)$ and by the corresponding lower case letter their trace.
  For $k \geq 2$ the trace of the normal derivatives is well defined as well, and in such a case we will denote it by doted lower case letters.
  That is, for $\Psi \in H^k(\Omega)$, $k \geq 2$, we denote
  \begin{equation*}
    \psi \coloneqq \gamma(\Psi), \qquad \dot{\psi} \coloneqq \gamma(\d\Psi(\nu)),
  \end{equation*}
  where $\nu \in \mathfrak{X}(\Omega)$ is a vector field pointing outwards, and the orientation of the manifold $\Omega$ and its boundary are chosen in such a way that $\mathrm{i}_\nu \d\mu_\eta = \d\mu_{\partial\eta}$.

    A \emph{magnetic potential} is a real-valued 1-form $A \in \Lambda(\Omega)$. We consider a deformed differential associated
    with a magnetic potential,
    \begin{equation*}
      \d_A: C^\infty(\Omega) \to \Lambda(\Omega), \quad \Phi \mapsto \d \Phi + i \Phi A,
    \end{equation*}
    where $\d$ is the exterior derivative and $i$ is the imaginary unit.
    Its formal adjoint,
    \begin{equation*}
      \delta_A: \Lambda(\Omega) \to C^{\infty}(\Omega),
    \end{equation*}
    can be defined by the identity
    \begin{equation*}
      \langle \d_A \Phi, \omega \rangle = \langle \Phi, \delta_A \omega \rangle, \quad
      \Phi \in C^\infty_c(\Omega^\circ),\; \omega \in \Lambda(\Omega).
    \end{equation*}

    We can now define the magnetic Laplacian $\Delta_A$ associated with the potential
    $A \in \Lambda(\Omega)$ by the formula
    \begin{equation*}
      \Delta_A = \delta_A \d_A: C^\infty(\Omega) \to C^\infty(\Omega).
    \end{equation*}
    As in the case of the Laplace-Beltrami operator, Stokes' theorem implies that $\d_A^\dagger = \delta_A$ for a
    boundaryless manifold, and in such a case the magnetic Laplacian $\Delta_A$ defined above is essentially self-adjoint.

    A straightforward calculation leads to the relation
    \begin{equation}
      \label{eq:magnetic_laplacian_expanded}
      \Delta_A \Phi = \Delta \Phi - 2i (A, \d \Phi) + \bigl(i \delta A + (A,A)\bigr) \Phi,
    \end{equation}
    where $(\alpha, \beta) = \eta^{-1}(\overline{\alpha}, \beta)$ represents the canonical scalar product on the cotangent bundle induced by the Riemannian metric.
    We shall now concentrate on the definition of the boundary conditions that make these operators self-adjoint, therefore giving rise to well defined quantum dynamics.

\section{Self-adjoint extensions and sesquilinear forms}
\label{sec:sa_extensions}

  One of the goals of the research presented in this work is to characterise the controllability of quantum systems using as space of controls the set of self-adjoint extensions of the Laplace-Beltrami operator.
  It is therefore necessary to have a convenient parametrisation of such a set.
  In the one-dimensional case, the set of self-adjoint extensions of the Laplacian is in one-to-one correspondence with
  the unitary operators on the Hilbert space of boundary data (i.e., the trace of the function and
  its normal derivative at the boundary).
  This correspondence is given through the boundary
  equation,
  \begin{equation}
    \label{eq:boundary_equation}
    \varphi - i\dot{\varphi} = U(\varphi + i \dot{\varphi}),
  \end{equation}
  which defines the domain of the self-adjoint extension associated with the unitary $U$, cf.\ \cite{AsoreyIbortMarmo2005, BruningGeylerPankrashkin2008, Kochubei1975}.

  In the more general case of a higher dimensional manifold $\Omega$, the statement above is no longer true; instead, there are
  several characterisations of the self-adjoint extensions of differential symmetric operators based on the
  boundary data.
  For instance, G.\ Grubb \cite{Grubb1968} gave a complete characterisation in terms of
  pseudo-differential operators acting on the Sobolev spaces over the boundary.
  In general it is not an easy task to connect this characterisation with the boundary data.
  However, it is also possible in the multidimensional case to characterise a large class of self-adjoint extensions of the Laplace-Beltrami operator in terms of the boundary data through Equation \eqref{eq:boundary_equation}.
  To establish this characterisation, the relation between self-adjoint operators and Hermitian sesquilinear forms can be used (see \cite{IbortLledoPerezPardo2015} for a detailed exposition of this method).
  Let us introduce some definitions before summarising the characterisation result.

  \begin{definition}
    A unitary operator $U: L^2(\partial \Omega) \to L^2(\partial \Omega)$ is said to have \emph{gap at $-1$} if
    either $U + I$ is invertible or $-1$ is in the spectrum of $U$ but it is not an accumulation
    point of $\sigma(U)$.
  \end{definition}

  Consider a unitary $U$ with gap at $-1$.
  Let $P$ denote the orthogonal projector onto the eigenspace associated with $-1$ and $P^\perp = I - P$. The \emph{partial Cayley transform} of $U$ is the linear operator on
  $L^2(\partial \Omega)$ defined by
  \nomenclature[C]{$\mathcal{C}_U$}{Partial Cayley transform of the unitary $U$}%
  \begin{equation*}
    \mathcal{C}_U \varphi \coloneqq i P^\perp \frac{U - I}{U + I} \varphi.
  \end{equation*}
  Using the spectral resolution of $U$, it can be shown that $\mathcal{C}_U$ is a bounded, self-adjoint
  operator on $L^2(\partial \Omega)$, cf.\ \cite[Prop. 3.11]{IbortLledoPerezPardo2015}.

  \begin{definition} \label{def:admissible_unitary}
    A unitary $U$ with gap at $-1$ is said to be \emph{admissible} if its partial Cayley transform,
    $\mathcal{C}_U$, leaves invariant the fractional Sobolev space $H^{1/2}(\partial \Omega)$ and it is
    continuous on it, i.e.,
    \begin{equation*}
      \|\mathcal{C}_U \varphi\|_{H^{1/2}(\partial \Omega)} \leq K \|\varphi\|_{H^{1/2}(\partial \Omega)},
      \qquad \varphi \in H^{1/2}(\partial \Omega).
    \end{equation*}
  \end{definition}

  \begin{remark} \label{remark:P_H1_cont}
    Note that the admissibility condition implies that the projection $P$ is continuous on $H^{1/2}(\partial \Omega)$ and therefore $P\circ\gamma:\mathcal{H}^1\to \mathcal{H}^{1/2}(\partial \Omega)$ is also continuous.
  \end{remark}
  We can now restrict our attention to admissible unitary operators and associate to them a sesquilinear form.
  Eventually, this sesquilinear form will allow us to associate a self-adjoint extension of the Laplace-Beltrami operator to each of such unitaries.

  \begin{definition}
    \label{def:associated_form}
    \nomenclature[Q1]{$Q_{U}$}{Sesquilinear form with unitary $U$ and zero potential}%
    \nomenclature[Q2]{$Q_{A,U}$}{Sesquilinear form with unitary $U$ and potential $A$}%
    Let $U$ be an admissible unitary operator and let $A \in \Lambda(\Omega)$ be a magnetic potential. The \emph{sesquilinear form associated with $U$ and $A$} is defined by
    \begin{alignat*}{2}
      Q_{A,U}(\Phi, \Psi) &= \langle \d_A\Phi, \d_A\Psi \rangle - \langle \varphi, \mathcal{C}_U \psi \rangle,\\
      \dom Q_{A,U} &= \{\Phi \in H^1(\Omega) \mid P \varphi = 0\},
    \end{alignat*}
    where $P$ is the projector onto the eigenspace associated with the eigenvalue $-1$ of $U$ and $\varphi = \gamma(\Phi)$, $\psi = \gamma(\Psi)$.
  \end{definition}

  Notice that for $U = - \mathbb{I}$, $Q_{A,U}$ is the sesquilinear form associated to the magnetic Laplacian with magnetic potential $A$ and Dirichlet boundary conditions, a fact that will be used later.
  Whenever $A$ is the zero magnetic potential, denoted $\mathcal{O}$, we will shorten the notation and denote by $Q_U := Q_{\mathcal{O}, U}$ the sesquilinear form
  associated with the admissible unitary $U$. For such a case, i.e., $A = \mathcal{O}$, it is proven in \cite[Thm. 4.9]{IbortLledoPerezPardo2015} that forms
  constructed this way are semibounded from below. There is $m>0$ depending only on $\Omega$ and $\|\mathcal{C}_U\|$ such that $Q_{U}(\Phi, \Phi) \geq -m \|\Phi\|^2$.

  We will show next that this is also the case for non-vanishing magnetic potentials.  Before we proceed, let us show the following technical lemma.
  \begin{lemma} \label{lemma:dphi_Q_bounded}
    For any unitary $U$ with gap at $-1$, there is a constant $C$ depending only on $\Omega$ and $\|\mathcal{C}_U\|$ such that $C$ is linear on $\|\mathcal{C}_U\|$ and the associated sesquilinear form $Q_U$ satisfies the inequality
    \begin{equation*}
      \|\d\Phi\|^2 \leq 2 Q_U(\Phi,\Phi) + (C + 1) \|\Phi\|^2.
    \end{equation*}
  \end{lemma}
  \begin{proof}
    By the Cauchy-Schwarz inequality and \cite[Theorem 1.5.1.10]{Grisvard1985}, we have
    \begin{equation*}
      |\langle \varphi, \mathcal{C}_U \varphi \rangle| \leq \|\mathcal{C}_U\| \|\varphi\|^2
      \leq \frac{1}{2} \|\Phi\|_1^2 + \frac{C}{2} \|\Phi\|^2,
    \end{equation*}
    with $C > 0$ linear on $\|\mathcal{C}_U\|$.

    Substituting into the definition of $Q_U$,
    \begin{equation*}
      Q_U(\Phi, \Phi) = \|\d\Phi\|^2 - \langle \varphi, \mathcal{C}_U \varphi \rangle
      \geq \frac{1}{2} \|\d\Phi\|^2 - \frac{1}{2}(C + 1) \|\Phi\|^2
    \end{equation*}
    from which the result follows.
  \end{proof}

  \begin{proposition}\label{prop:graph_norm_magnetic}
    For any magnetic potential $A \in \Lambda(\Omega)$ and any admissible unitary operator $U$, the following holds:
    \begin{enumerate}[label=\textit{(\roman*)},nosep, leftmargin=*]
      \item \label{prop:graph_norm_magnetic_1}
        $Q_{A,U}$ is semibounded from below.
      \item \label{prop:graph_norm_magnetic_2}
        The graph norm $\|\cdot\|_{Q_{A,U}}$ is equivalent to $\|\cdot\|_{Q_U}$.
    \end{enumerate}
  \end{proposition}

  \begin{proof}
    By definition one has
    \begin{equation*}
      \langle \d_A \Phi, \d_A \Phi \rangle = \|\d\Phi\|^2 + \|\Phi A\|^2 + 2 \Im \langle \d\Phi, \Phi A \rangle.
    \end{equation*}
    Now,
    \begin{alignat*}{2}
      Q_{A,U}(\Phi, \Phi) &= \langle \d_A \Phi, \d_A \Phi \rangle - \langle \varphi, \mathcal{C}_U \varphi \rangle \\
      &\geq Q_U(\Phi, \Phi) + \|\Phi A\|^2 - 2 |\langle \d\Phi, \Phi A \rangle|.
    \end{alignat*}
    By Cauchy-Schwarz inequality and Young inequality with $\varepsilon$ we have
    \begin{equation*}
      Q_{A,U}(\Phi, \Phi) \geq Q_U(\Phi, \Phi) + (1 - \varepsilon) \|\Phi A\|^2 - \frac{1}{\varepsilon}\|\d\Phi\|^2
    \end{equation*}
    for every $\varepsilon>0$.
    Taking $\varepsilon=4$ and applying Lemma~\ref{lemma:dphi_Q_bounded} it follows
    \begin{equation*}
      Q_{A,U}(\Phi,\Phi) \geq \frac{1}{2} Q_U(\Phi,\Phi) - \left(3 \|A\|^2 + \frac{C+1}{4}\right) \|\Phi\|^2.
    \end{equation*}
    The lower semibound of $Q_{A,U}$ follows now from that of $Q_U$.

    Let us now show the equivalence \ref{prop:graph_norm_magnetic_2}.
    On the one hand, from the previous inequality it follows immediately that for some $K>0$
    \begin{equation*}
      \|\Phi\|_{Q_U} \leq K\|\Phi\|_{Q_{A,U}}.
    \end{equation*}
    On the other hand, let $m>0$ be the lower bound of $Q_{A,U}$ and define $\alpha := \sqrt{\sup_{\Omega} \eta^{-1}(A,A)}$. Then,
    \begin{equation*}
      \begin{alignedat}{2}
        \|\Phi\|_{Q_{A,U}}^2
        &= (m + 1) \|\Phi\|^2 + \|\d\Phi\|^2 + \|\Phi A\|^2 + 2 \Im \langle \d\Phi, \Phi A \rangle - \langle \varphi, \mathcal{C}_U \varphi \rangle\\
        &\leq (m+\alpha^2+1) \|\Phi\|^2 + \|\d\Phi\|^2 + 2 \alpha \|\d\Phi\| \|\Phi\| - \langle\varphi, \mathcal{C}_U \varphi\rangle\\
        &= (m+\alpha^2+1) \|\Phi\|^2 + Q_U(\Phi, \Phi) + 2 \alpha \|\d\Phi\|\|\Phi\| \\
        &\leq K \|\Phi\|_{Q_U}^2
      \end{alignedat}
    \end{equation*}
    where $K = \max\{m + \alpha^2 + 1, 2\alpha\}$, since all the addends are proportionally smaller than $\|\Phi\|_{Q_U}^2$.
  \end{proof}

  \begin{theorem} \label{thm:quadratic_form_H1bounded}
    For any magnetic potential $A$, the sesquilinear form associated with an admissible unitary with
    gap at $-1$ is closed, i.e., $\dom Q_{A,U}$ is closed with respect to the graph norm,
    $\|\cdot\|_{Q_{A,U}}$.
  \end{theorem}
  \begin{proof}
    The admissibility condition guarantees that $\dom Q_{A,U}$ is closed in $H^1(\Omega)$ (see Remark \ref{remark:P_H1_cont}), thus it suffices to show that $\|\cdot\|_{Q_{A,U}}$ is equivalent to $\|\cdot\|_1$. By Proposition~\ref{prop:graph_norm_magnetic} it is enough to show the equivalence for $A$ equal to the zero magnetic potential.

    First, by Cauchy-Schwarz inequality we have
    \begin{equation*}
      |\langle \varphi, \mathcal{C}_U \varphi \rangle| \leq \|\varphi\| \|\mathcal{C}_U \varphi\|
      \leq \|\mathcal{C}_U\| \|\varphi\|^2
      \leq C \|\mathcal{C}_U\| \|\varphi\|_{\frac{1}{2}}^2
      \leq C' \|\mathcal{C}_U\| \|\Phi\|^2_1,
    \end{equation*}
    where we have used the Sobolev inclusions and the continuity of the trace map.

    Let $m$ be the lower bound of $Q_U$. Substituting the previous inequality into the definition of the graph norm, one gets
    \begin{equation*}
      \begin{alignedat}{2}
        \|\Phi\|_{Q_U}^2 &= \|\d \Phi\|^2 + (1 + m) \|\Phi\|^2 -
        \langle \varphi, \mathcal{C}_U \varphi \rangle \\
        &\leq \|\d \Phi\|^2 + (1 + m) \|\Phi\|^2 + |\langle \varphi, \mathcal{C}_U \varphi \rangle| \\
        &\leq K^2 \|\Phi\|_1^2,
      \end{alignedat}
    \end{equation*}
    for some $K > 0$.

    Let us prove now the reverse inequality.
    By Lemma~\ref{lemma:dphi_Q_bounded} we have
    \begin{equation*}
      Q_U(\Phi, \Phi) \geq \frac{1}{2} \|\d\Phi\|^2 - \frac{1}{2}(C + 1) \|\Phi\|^2,
    \end{equation*}
    and by Proposition~\ref{prop:sobolev_norm_equivalence} one gets
    \begin{equation*}
      Q_U(\Phi, \Phi) + \frac{1}{2}(C + 2) \|\Phi\|^2 \geq \frac{1}{2} \|\Phi\|_1^2,
    \end{equation*}
    from which the desired inequality follows.
  \end{proof}

  The previous results lead to the following corollary, which will be useful in Chapter~\ref{ch:quantum_circuits} for showing that the Hamiltonians of some families of quasi-$\delta$ boundary control systems satisfy Assumption \ref{assump:stability_assumptions}.
  \begin{corollary} \label{corol:uniform_semibound_equivalence_formnorms}
    Let $\mathcal{A}$ be a family of magnetic potentials and $\mathcal{U}$ a family of admissible unitaries such that the associated forms have common domain $\mathcal{H}^+$, satisfying
    \begin{equation*}
      \sup_{U \in \mathcal{U}} \|\mathcal{C}_U\| < \infty
      \quad\text{and}\quad
      \sup_{A \in \mathcal{A}} \|A\|_{\infty} < \infty.
    \end{equation*}
    Then:
    \begin{enumerate}[label=\textit{(\roman*)},nosep,leftmargin=*]
      \item There exists $m>0$ such that for every $A \in \mathcal{A}$ and every $U \in \mathcal{U}$ we have $Q_{A,U}(\Phi, \Phi) \geq -m \|\Phi\|^2$ for every $\Phi\in\mathcal{H}^+$.
      \item Let $A_0\in\mathcal{A}$ and $U_0\in\mathcal{U}$. There is a constant $K>0$ such that for every $A \in \mathcal{A}$ and $U \in \mathcal{U}$ we have
        \begin{equation*}
          K^{-1} \|\Phi\|_{Q_{A,U}} \leq \|\Phi\|_{Q_{A_0,U_0}} \leq K \|\Phi\|_{Q_{A,U}}, \qquad \forall\Phi \in \mathcal{H}^+,
        \end{equation*}
        for every $U \in \mathcal{U}$ and every $A \in \mathcal{A}$.
    \end{enumerate}
  \end{corollary}

  We have established that, for any admissible unitary operator and a magnetic potential satisfying the conditions above, the sesquilinear form associated with it is closed and semibounded from below. Theorem~\ref{thm:repKato} can then be applied, leading to the definition of the magnetic Laplacian.

  \begin{definition}\label{def:magnetic-laplacian}
   Let $\Omega$ be a compact, Riemannian manifold with boundary and let $U\colon L^2(\partial\Omega)\to L^2(\partial\Omega)$ be an admissible unitary operator.
   Let $A\in\Lambda(\Omega)$ be a magnetic potential.
   \nomenclature[Delta1]{$\Delta_U$}{Laplace-Beltrami operator with unitary $U$}%
   \nomenclature[Delta2]{$\Delta_{A,U}$}{Magnetic Laplacian with unitary $U$ and potential $A$}%
   The \emph{magnetic Laplacian} operator associated with the unitary $U$ and the potential $A$, denoted by $\Delta_{A,U}$, with domain $\dom \Delta_{A,U}$, is the unique self-adjoint operator associated with the closed, semibounded sesquilinear form $Q_{A,U}$, cf.\ Theorem~\ref{thm:repKato}.
   In the case in which the magnetic potential $A$ is identically zero, we will refer to this operator as the \emph{Laplace-Beltrami} operator and denote it and its domain respectively by $\Delta_U$ and $\dom \Delta_U$.
  \end{definition}

  For the Laplace-Beltrami operator associated with the unitary $U$, one can provide the following characterisation of (a part of) its domain.
  \begin{theorem}
    \label{thm:sa_extensions_laplacian}
    Let $\Delta_U$, densely defined on $\dom \Delta_U$, be the Laplace-Beltrami operator associated with the unitary $U$. Then
    \begin{equation*}
      H^2(\Omega) \cap \dom \Delta_U = \{\Phi \in H^2(\Omega) \mid \varphi - i\dot{\varphi} = U(\varphi + i \dot{\varphi})\}.
    \end{equation*}
  \end{theorem}
  \begin{proof}
    Let $\Phi \in H^2(\Omega) \cap \dom \Delta_U$ and $\Psi\in\dom{Q_U}$.
    Integration by parts in the definition of $Q_U$, cf.\ Definition~\ref{def:associated_form}, yields
    \begin{equation*}
      Q_U(\Psi, \Phi) = \langle \d\Psi, \d\Phi \rangle - \langle \psi, \mathcal{C}_U\varphi \rangle
      = \langle \Psi, \Delta\Phi \rangle + \langle \psi, \dot{\varphi} - \mathcal{C}_U\varphi \rangle.
    \end{equation*}
    From this it follows that
    \begin{equation}
      \label{eq:Delta_U_dom}
      \langle \Psi, \Delta_U \Phi - \Delta \Phi \rangle
      = \langle \psi, \dot{\varphi} - \mathcal{C}_U \varphi \rangle.
    \end{equation}
    Thus, for every $\Psi \in H^1_0(\Omega) \cap H^2(\Omega) \subset \dom Q_U$, we have
    \begin{equation*}
      \langle \Psi, \Delta_U \Phi - \Delta \Phi \rangle = 0,
    \end{equation*}
    which implies $\Delta_U \Phi = \Delta \Phi$ since $H^1_0(\Omega) \cap H^2(\Omega)$ is dense in
    $L^2(\Omega)$. Substituting this in Eq.~\eqref{eq:Delta_U_dom}, it follows
    \begin{equation*}
      \langle \psi, \dot{\varphi} - \mathcal{C}_U \varphi \rangle = 0.
    \end{equation*}
    Since $\Psi \in \dom Q_U$, $\psi = P^{\bot}\psi$ and the equation above is equivalent to
    \begin{equation*}
      P^\perp \dot{\varphi} = \mathcal{C}_U \varphi.
    \end{equation*}
    Additionally, since $\Phi \in \dom \Delta_U\subset \dom Q_U$, we have $P\varphi = 0$.

    Applying the orthogonal projector onto $\ker(U + 1)$, $P$, on both sides of the boundary equation $\varphi - i\dot{\varphi} = U(\varphi + i \dot{\varphi})$, it follows
    \begin{equation*}
      P\varphi - iP\dot{\varphi} = PU (\varphi + i \dot{\varphi}) = -P\varphi - i P\dot{\varphi},
    \end{equation*}
    which is equivalent to $P\varphi = 0$. On the other hand, projecting with $P^\perp = I - P$,
    one gets
    \begin{equation*}
      P^\perp\dot{\varphi} = iP^\perp \frac{U - I}{U + I} \varphi = \mathcal{C}_U \varphi.
      \qedhere
    \end{equation*}
  \end{proof}

  The previous result motivates the following definition.

  \begin{definition}\label{def:boundary_equation_operator}
    Let $U\colon L^2(\partial\Omega) \to L^2(\partial\Omega)$ be an admissible unitary operator and $A\in\Lambda(\Omega)$ a magnetic potential.
    The \emph{domain associated with the unitary $U$ and the magnetic potential $A$} is the set
    \nomenclature[D1]{$\mathcal{D}_{U}$}{Domain associated with unitary $U$ and zero potential}%
    \nomenclature[D2]{$\mathcal{D}_{A,U}$}{Domain associated with unitary $U$ and potential $A$}%
    $$\mathcal{D}_{U,A} = \left\{ \Phi\in\mathcal{H}^2(\Omega) \mid \varphi - i\dot{\varphi}_A = U(\varphi + i \dot{\varphi}_A) \right\},$$
    where $\dot{\varphi}_A:=\gamma\left(\mathrm{d}_A\Phi(\nu)\right)$, and $\nu$ is the normal vector field to the boundary $\partial \Omega$ pointing outwards. In the case in which the magnetic potential is identically zero, we will drop the subindex and denote it by $\mathcal{D}_U$.
  \end{definition}

  \begin{proposition} \label{prop:magnetic_form_equivalence}
    Let $A \in \Lambda(\Omega)$ be an exact, magnetic potential, i.e.\ $A = \d\Theta$ for some function $\Theta \in C^2(\Omega)$, and denote $\theta = \gamma(\Theta)$.
    Define the unitary operator $J: \Phi \in L^2(\Omega) \mapsto e^{i\Theta}\Phi \in L^2(\Omega)$.
    Then $J(\dom Q_{A,U}) = \dom Q_{e^{i\theta}Ue^{-i\theta}}$ and
    \begin{equation*}
      Q_{A,U}(\Psi, \Phi) = Q_{e^{i\theta} U e^{-i\theta}}(J\Psi, J\Phi).
    \end{equation*}
  \end{proposition}
  \begin{proof}
    Let us first show that $J(\dom Q_{A,U}) = \dom Q_{e^{i\theta}Ue^{-i\theta}}$.
    Since $\Theta \in C^2(\Omega)$, for any $\Phi \in H^1(\Omega)$ we have $J\Phi \in H^1(\Omega)$ and $\gamma(J\Phi) = e^{i\theta} \varphi$.
    Denote $\tilde{U} = e^{i\theta}Ue^{-i\theta}$ and let $\tilde{P}$ be the orthogonal projector onto $\ker(\tilde{U}+1)$.
    It is immediate to check that $\tilde{U} \gamma(J\Phi) = - \gamma(J\Phi)$ if and only if $\Phi \in \ker(U+1)$, and therefore $\tilde{P} = e^{i\theta} P e^{-i\theta}$ where $P$ is the orthogonal projector onto $\ker(U+1)$.
    A straightforward calculation shows the following relation between the partial Cayley transforms:
    \begin{equation*}
      \mathcal{C}_{\tilde{U}} = e^{i\theta} \mathcal{C}_U e^{-i\theta}.
    \end{equation*}
    From the above conditions it follows that $\tilde{U}$ is an admissible unitary operator and that $\Phi \in \dom Q_{A,U}$ if and only if $J\Phi \in \dom Q_{\tilde{U}}$.

    Finally, $\d(J\Phi) = J\d_A \Phi$, and the following identity holds:
    \begin{equation*}
      Q_{A,U}(\Psi,\Phi) = \langle J\d_A\Psi, J\d_A\Phi \rangle - \langle \psi, e^{-i\theta}\mathcal{C}_{\tilde{U}}e^{i\theta}\varphi \rangle
      = Q_{\tilde{U}}(J\Psi, J\Phi),
    \end{equation*}
    where we have used the unitarity of $J$.
  \end{proof}

   As a corollary of Theorem~\ref{thm:sa_extensions_laplacian}, one can show an equivalent result for magnetic Laplacians.

  \begin{corollary}\label{corol:magnetic_laplacian_equivalence}
    Let $A \in \Lambda(\Omega)$ be an exact, magnetic potential, i.e.\ $A = \d\Theta$ for some function $\Theta \in C^2(\Omega)$, $U: L^2(\partial\Omega)\to  L^2(\partial\Omega)$ an admissible unitary operator, $J$ the unitary operator $J: \Phi \in L^2(\Omega) \mapsto e^{i\Theta} \Phi \in L^2(\Omega)$ and $\theta = \gamma(\Theta)$. Let $\Delta_{A,U}$ be the magnetic Laplacian, densely defined on $\dom \Delta_{A,U}$, associated with $U$ and $A$.
    Then
    \begin{equation*}
      H^2(\Omega) \cap \dom \Delta_{A,U} = \mathcal{D}_{A,U}.
    \end{equation*}
    Moreover, $J(\dom \Delta_{A,U}) = \dom \Delta_{e^{i\theta}Ue^{-i\theta}}$ and $\Delta_{A, U} = J^{-1}\Delta_{e^{i\theta} U e^{-i\theta}}J$, where $\Delta_{e^{i\theta}Ue^{-i\theta}}$ is the Laplace-Beltrami operator associated with unitary $e^{i\theta} U e^{-i\theta}$.
  \end{corollary}
  \begin{proof}
    Denote $\tilde{U} = e^{i\theta} U e^{-i\theta}$.
    From Proposition~\ref{prop:magnetic_form_equivalence} we have
    \begin{equation*}
      J(\dom Q_{A,U}) = \dom Q_{e^{i\theta}Ue^{-i\theta}}
      \quad\text{and}\quad
      Q_{A,U}(\Psi, \Phi) = Q_{\tilde{U}}(J\Psi, J\Phi).
    \end{equation*}
    Therefore, cf.\ Theorem~\ref{thm:repKato}, $\Phi \in \dom \Delta_{A,U}$ if and only if $J\Phi \in \dom \Delta_{\tilde{U}}$ and it follows that $\Delta_{A, U} = J^{-1}\Delta_{\tilde{U}}J.$

    Assume that $\Phi\in \mathcal{H}^2(\Omega) \cap \dom \Delta_{A, U}$.
    Then $J\Phi \in \mathcal{H}^2(\Omega) \cap \dom \Delta_{\tilde{U}} = \mathcal{D}_{\tilde{U}}$, which implies
    $$\gamma(J\Phi) - i\gamma (\mathrm{d}(J\Phi)(\nu)) = \tilde{U} (\gamma(J\Phi) + i\gamma (\mathrm{d}(J\Phi)(\nu))),$$
    with $\nu$ the normal vector field to the boundary.
    Now noticing that {$\mathrm{d}(J\Phi)(\beta) = J\mathrm{d}_A\Phi(\beta)$}, $\beta\in\mathfrak{X}(\Omega)$,  we get
    $e^{i\theta}\varphi-i e^{i\theta}\dot{\varphi}_A = \tilde{U}(e^{i\theta}\varphi+i e^{i\theta}\dot{\varphi}_A).$
    The converse inclusion is proven analogously.
  \end{proof}
\end{document}

\ifSubfilesClassLoaded{
    \setcounter{chapter}{5}
  }{}
\chapter{Quantum circuits} \label{ch:quantum_circuits}
  The aim of this chapter is to introduce Quantum Circuits, a generalisation of the notion of Quantum Graphs \cite{KostrykinSchrader2003,Kuchment2004}.
  A Quantum Graph is a (metric) graph equipped with a differential operator defined on its edges together with appropriate boundary conditions determining a self-adjoint extension.
  Quantum Graphs are useful for describing one-dimensional quantum systems, while Quantum Circuits generalise the construction in such a way that allows to describe higher dimensional systems.
  Section \ref{sec:quantum_circ_def} we introduce the concept of a Quantum Circuit and describe which are the boundary conditions (and therefore the self-adjoint extensions of the Laplacians) which are compatible with the topology of the circuit.
  Finally, a particular subclass of boundary conditions, quasi-$\delta$ boundary conditions, are studied in Section \ref{sec:quasi_delta_BC}.

\section{Definition and boundary conditions} \label{sec:quantum_circ_def}
  A large subclass of the family of self-adjoint extensions of the Laplace-Beltrami operator on a Riemannian manifold with smooth boundary can be described with the formalism depicted in the previous section. As already stated in the introduction, our main motivation is to study the Quantum Control at the Boundary scheme in circuit-like settings. We are interested in presenting results in higher-dimensional analogues of Quantum Graphs.

  Before we proceed further, we shall give a precise definition of this generalisation of circuits.
  We will consider circuits as graph-like structures where the edges are going to be identified with $n$-dimensional Riemannian manifolds; this generalisation contains the theory of one-dimensional Quantum Circuits, i.e.\ Quantum Graphs, as a special case to which the results can also be applied.

  \begin{definition}\label{def:quantum_circuit}
    An \emph{$n$-dimensional Quantum Circuit} is a triple $(G, \Omega, \Gamma)$ where:
    \begin{enumerate}[label=\textit{(\roman*)},nosep, leftmargin=*]
      \nomenclature[G]{$G=(V, E)$}{Finite graph with vertex set $V$ and edge set $E$}%
      \item $G = (V, E)$ is a finite, connected, directed graph with vertex set $V$ and edge set $E$.
      \nomenclature[O]{$\Omega, \Omega_e$}{Non-connected manifold with connected components $\Omega_e$}%
      \item $\Omega$ is a non-connected $n$-dimensional Riemannian manifold with as many connected components as the edges of the graph, i.e., $\Omega=\cup_{e\in E}\Omega_e$ where each $\Omega_e$ is a connected Riemannian manifold with non-trivial boundary.
      \nomenclature[G]{$\Gamma_{\text{ext},e}$}{External part of the boundary of $\Omega_e$}
      \nomenclature[G]{$\Gamma_{v,e}$}{Part of $\partial\Omega_e$ participating on the junction $v$}
      \item $\Gamma$ is a partition of the boundary $\partial\Omega$ defined as follows.
        Let $V_e\subset V$ denote the set of vertices joined by the edge $e\in E$.
        Let $\Gamma_{\mathrm{ext},e}$ and $\Gamma_{v,e}$, with $e \in E$ and $v\in V_e$, be simply connected submanifolds of $\partial \Omega_e$, the boundary of the oriented manifold $\Omega_e$.
        We will require that for any $\Gamma_{v,e}$, $e\in E$, $v\in V_e$, there exists a simply connected open neighbourhood $\mathcal{V}_{v,e}$, with $\overline{\Gamma_{v,e}}\subset\mathcal{V}_{v,e}$, such that the latter are pairwise disjoint, i.e.\ $\mathcal{V}_{v,e}\cap \mathcal{V}_{v',e} = \emptyset$ for $e\in E$ and $v,v'\in V_e$ with $v\neq v'$.
        Then
        \begin{equation*}
          \partial \Omega= \bigsqcup_{e\in E} \Gamma_{\mathrm{ext},e} \sqcup \left(\bigsqcup_{v\in V_e} \Gamma_{v,e}\right).
        \end{equation*}
    \end{enumerate}
  \end{definition}
  \begin{remark}
    The standard definition of Quantum Graphs includes the choice of a self-adjoint operator.
    However, since our aim is to study the controllability of a quantum system by modifying the self-adjoint extension of its Hamiltonian, it is more convenient to define Quantum Circuits without including one.
  \end{remark}
  At the end of this subsection there are some meaningful examples of this construction; see also Fig.~\ref{fig:examples}.

  This definition fully determines the Quantum Circuit. On the one hand, the associated graph
  schematically describes its topology: edges represent wires while vertices represent
  connections between different wires.
  On the other hand, the manifolds $\Omega_e$ describe each wire and the partition $\Gamma$ allows to characterise completely the connections between different wires. This definition can be generalised to treat more complex topologies.
  However, Def.\ \ref{def:quantum_circuit} will keep the simplicity of the mathematical treatment of the control problems under study.
  When it leads to no confusion, we will abuse the notation and denote by $\Omega$ the Quantum Circuit itself. Also, for the
  sake of simplicity we will assume from now on that every $\Omega_e$ of a Quantum Circuit is a convex, compact differentiable manifold with piecewise smooth and compact boundary.

  We will only consider \emph{finite} Quantum Circuits, i.e.\ having finitely many wires; in this case, the relevant Hilbert spaces defined for the complete circuit are just the direct
  sum of the corresponding Hilbert spaces for its wires:
  \begin{equation*}
    L^2(\Omega) = \bigoplus_{e \in E} L^2(\Omega_e);\qquad
    \langle \Phi, \Psi \rangle_{L^2(\Omega)} = \sum_{e \in E} \langle \Phi_e, \Psi_e \rangle_{
    L^2(\Omega_e)},
  \end{equation*}
  and similarly for the Sobolev spaces.

  It is also worth to make some comments about the Hilbert space at the boundary.
  It is clear that one has $L^2(\partial\Omega) = \bigoplus_{e \in E} L^2(\partial\Omega_e)$; however, for the boundary space, a further decomposition will be needed.
  \nomenclature[E]{$E_v$}{Subset of edges that share the vertex $v$}
  Let $v\in V$ and let $E_v\in E$ be the subset of edges that share the vertex $v$.
  \nomenclature[G]{$\Gamma_v$}{Part of $\partial\Omega$ participating on the junction $v$}
  \nomenclature[G]{$\Gamma_{\text{ext}}$}{Exterior part of the boundary of $\Omega$}
  Denote by $\Gamma_v = \bigcup_{e \in E_v} \Gamma_{v,e}$  the part of the boundary of $\Omega$ that is involved in the connection represented by the vertex $v \in V$, and by $\Gamma_{\text{ext}} = \bigcup_{e \in E} \Gamma_{\text{ext},e}$ the subset of the boundary of $\Omega$ that is not part of a junction; i.e., the external boundary of the Quantum Circuit.
  In the one-dimensional case, the latter is formed by the exterior vertices of the circuit.
  We can write the following decomposition:
  \begin{equation*}
    L^2(\partial\Omega) = L^2(\Gamma_{\text{ext}}) \oplus \bigoplus_{v \in V} L^2(\Gamma_v).
  \end{equation*}
  Particles inside the circuit need to be able to move \emph{freely} throughout the \emph{wires} and to \emph{jump} from a piece of wire to an adjacent one: therefore, we are going to consider the quantum evolution given by the free Hamiltonian, that is, the Laplacian $\Delta$ which acts wire by wire, i.e.\, for any $\Phi = \bigoplus_{e \in E} \Phi_e$ in $H^2(\Omega)$, $$\Delta\Phi = \bigoplus_{e \in E} \Delta_e \Phi_e,$$ with $\Delta_e$ the Laplace-Beltrami operator in $\Omega_e$.
  Since we are considering the circuit $\Omega$ as the collection of the different wires $\Omega_e$, we will model the transitions from one $\Omega_e$ to another by imposing some particular boundary conditions involving the parts of the boundary which are (physically) connected.
  It is natural to assume that allowing the particles to move through this boundary regions should result in a continuous probability of finding them at the junctions $v \in V$.
  Hence, the boundary conditions imposed should implement this requirement.
  Furthermore, particles are not supposed to escape the circuit through $\Gamma_{\text{ext}}$ which makes natural to impose Dirichlet boundary conditions in $\Gamma_{\text{ext}}$.
  Other types of local boundary conditions on $\Gamma_{\text{ext}}$, such as Neumann or Robin, could be considered as well; the choice of these other types of boundary conditions would not alter significantly the results presented in what follows.
  This discussion motivates considering boundary conditions which do not relate the boundaries of the manifolds $\Omega_e$ and $\Omega_{e'}$ for $e \neq e'$, with only one exception: they can relate the values of functions on different edges inside each $\Gamma_v$.
  Hence, the unitaries implementing the physically relevant boundary conditions (see Section \ref{sec:sa_extensions}) can be written as
  \begin{equation} \label{eq:total_unitary_structure}
    U = U_D \oplus \bigoplus_{v \in V} U_v,
  \end{equation}
  where $U_D: L^2(\Gamma_{\text{ext}}) \to L^2(\Gamma_{\text{ext}})$ is given by $U_D = -I_{L^2(\Gamma_{\text{ext}})}$ and $U_v: L^2(\Gamma_v) \to L^2(\Gamma_v)$ are the unitaries specifying the boundary conditions on each $\Gamma_v$.
  \begin{remark}
    Note that there can be some vertices which do not take part on any \emph{junction} (i.e.\ that do not connect different edges).
    This subset of vertices is called the external vertices of $G$, and for modelling certain situations it might be useful consider them as a special part of the external boundary of the circuit, $\Gamma_{\text{ext}}$.
    However, for simplicity we treat all the vertices in the same way, specifying for each of them appropriate unitaries in order to implement the boundary conditions.
  \end{remark}

  In order to find the unitaries imposing some concrete boundary conditions, some practical information can be extracted from the proof of Theorem~\ref{thm:sa_extensions_laplacian}.
  It has been established that the equation defining $\mathcal{D}_U$, which we call \emph{boundary equation} since it fixes conditions on the boundary data of $H^2(\Omega)$ functions in the domain, is equivalent to
  \begin{equation*}
    P \varphi = 0, \qquad P^\perp \dot{\varphi} = \mathcal{C}_U \varphi.
  \end{equation*}
  Therefore, a consequence of Theorem~\ref{thm:sa_extensions_laplacian} is that, for functions in $H^2(\Omega)$, boundary conditions involving only the trace of the functions in the domain must be implemented through the equation $P\varphi=0$.
  That is, these conditions should fix the eigenspace associated with the eigenvalue $-1$ of the unitary $U$.

  A particular family of self-adjoint extensions preserving the topology of a Quantum Circuit,
  in the sense described previously, is what we call quasi-$\delta$ boundary conditions, which are a generalisation of the standard periodic boundary conditions (Kirchhoff vertex conditions in the case of Quantum Graphs, see \cite{BerkolaikoKuchment2012}).
  In the simplest case, i.e.\ a vertex connecting two edges, Kirchhoff boundary conditions consist on identifying the two joined boundaries, by requiring the functions to be continuous at that vertex; the term quasi-Kirchhoff is used when allowing for a relative phase change. Quasi-$\delta$ is a further generalisation allowing the derivatives of the function to be discontinuous at the vertex. We will be more specific in the following.

  In order to keep the mathematical description as simple as possible, from now on we are going to focus on a particular way of connecting the wires.
  Let $v\in V$ and denote by $E_v\subset E$ the subset of edges that share the vertex $v$.
  We will assume that, for every $e\in E_v$, $\Gamma_{v,e}$ is diffeomorphic to a reference polyhedron, $\Gamma_0$, and thus there is a diffeomorphism $g_{v,e}: \Gamma_0 \to \Gamma_{v,e}$.
  Let $\d\mu_0$ be a fixed volume element in $\Gamma_0$; using the diffeomorphism we can define an isometry $T_{v,e}: L^2(\Gamma_{v,e}) \to L^2(\Gamma_0, \d\mu_0)$ by
  \begin{equation} \label{eq:definition_Te}
    T_{v,e} \varphi \coloneqq \sqrt{|J_{v,e}|} (\varphi \circ g_{v,e}),
  \end{equation}
  where
  $|J_{v,e}|$ is the Jacobian determinant of the transformation $g_{v,e}$; i.e., the proportionality factor between the pull-back of the induced Riemannian metric at the boundary and the reference volume at $\Gamma_0$, $(g_{v,e})^\dagger \d\mu_{\Gamma_{v,e}} = |J_{v,e}|\d\mu_0$.
  Using these isometries we can define the unitaries $\mathbb{T}_v: L^2(\Gamma_v) \to \bigoplus_{e \in E_v} L^2(\Gamma_0)$ as the direct sum $\mathbb{T}_v = \bigoplus_{e \in E_v} T_{v,e}$.
  In the case of quasi-Kirchhoff boundary conditions (and also quasi-$\delta$), the structure of the unitaries $U_v$ in \eqref{eq:total_unitary_structure} can be easily written using this notation.

  \begin{definition} \label{def:quasi_delta_BC}
    \nomenclature[C]{$\chi_{v,e}, \delta_v$}{Parameters defining quasi-$\delta$ boundary conditions}
    For every $v \in V$, and every $e \in E_v$, let $\chi_{v,e}\in [0,2\pi)$, and $\delta_{v} \in (-\pi,\pi)$.
    We call \emph{quasi-$\delta$ boundary conditions with parameters} $\chi_{v,e},\, \delta_{v}$ to the ones given by the unitary (see Equation \eqref{eq:total_unitary_structure}),
    \begin{equation*}
      U = U_D \oplus \bigoplus_{v \in V} U_v, \quad\text{with}\quad
      U_v = \mathbb{T}_v^\dagger\left( (e^{i \delta_{v}} + 1) P_{v}^\perp - I \right) \mathbb{T}_v,
    \end{equation*}
    where $P_v^\perp: \bigoplus_{e \in E_v} L^2(\Gamma_0) \to \bigoplus_{e \in E_v} L^2(\Gamma_0)$ is given by the blocks
    \begin{equation*}
      (P_{v}^\perp)_{ee'} = \frac{1}{|E_v|} e^{i (\chi_{v,e} - \chi_{v,e'})},\quad e,e'\in E_v.
    \end{equation*}
  \end{definition}

  It is a straightforward calculation to check that  $P_v^\bot$ is an orthogonal projection on $L^2(\Gamma_v)$ and that $P_v:=I-P_v^\bot$ satisfies $\mathbb{T}^\dagger_vP_v\mathbb{T}_vU_v = U_v\mathbb{T}_v^\dagger P_v \mathbb{T}_v=-\mathbb{T}^\dagger_vP_v\mathbb{T}_v$ and $\mathbb{T}^\dagger_vP^\bot_v\mathbb{T}_vU_v = U_v\mathbb{T}_v^\dagger P_v^\bot \mathbb{T}_v = e^{i\delta_v}\mathbb{T}_v^\dagger P_v^\bot \mathbb{T}_v$ which shows that $U_v$ has eigenvalues $-1$ and $e^{i\delta_v}$. Consequently,
  \begin{equation*}
    P = I_{L^2(\Gamma_{\text{ext}})} \oplus \bigoplus_{v \in V} \mathbb{T}_v^\dagger (1 - P_v^\perp) \mathbb{T}_v.
  \end{equation*}
  is the orthogonal projector onto the eigenspace of $U$ associated with $-1$, and
  \begin{equation*}
    P^\perp = 1-P = 0_{L^2(\Gamma_{\text{ext}})} \oplus \bigoplus_{v \in V} \mathbb{T}_v^\dagger P_{v}^\perp \mathbb{T}_v.
  \end{equation*}

  The unitaries above define closable sesquilinear forms associated to self-adjoint extensions of the Laplace-Beltrami operator; we refer to \cite[Section~5]{IbortLledoPerezPardo2015} for further details. Note that our use of \emph{boundary conditions} is different from the usual one: instead of using the boundary conditions to define the domain of a differential operator, by using the unitary operator $U$ we define a closed Hermitian sesquilinear form and consider the unique self-adjoint extension of the differential operator associated with it. However, as discussed at the beginning of this section, for functions in $\mathcal{H}^2 \cap \dom \Delta_U$ the usual sense of boundary conditions can be recovered and the boundary equation of Definition~\ref{def:boundary_equation_operator} is equivalent to the equations,
  \begin{equation}\label{eq:boundary_eqs}
    P \varphi = 0, \qquad P^\perp \dot{\varphi} = \mathcal{C}_U \varphi.
  \end{equation}
  The partial Cayley transform is
  \begin{equation} \label{eq:circuits_cayley}
    \mathcal{C}_U = 0_{L^2(\Gamma_{\text{ext}})} \oplus \bigoplus_{v \in V} \mathbb{T}_{v}^\dagger \left[-\tan\left( \frac{\delta_{v}}{2} \right) P_{v}^\perp \right]
    \mathbb{T}_{v}.
  \end{equation}
  Because of this block structure, the equations \eqref{eq:boundary_eqs} hold \emph{block by block}, yielding the following conditions for $\Phi \in \mathcal{H}^2 \cap \dom \Delta_U$.
  For any vertex $v\in V$, choose one of the adjacent edges $e$ and define $\varphi|_v := e^{-i\chi_{v,e}}T_{v,e} \varphi_{v,e} \in H^{\sfrac{1}{2}}(\Gamma_0)$ and then
  \renewcommand\arraystretch{1.33}
  \begin{equation*}
    \begin{array}{c c}
      \displaystyle
      \varphi|_{\Gamma_{\text{ext}}} = 0, \\
      \displaystyle
      e^{-i\chi_{v,e'}}T_{v,e'} \varphi_{v,e'} = \varphi|_{v}, \quad e'\in E_v \smallsetminus \{e\},\\
      \displaystyle
      \sum_{e' \in E_v} \frac{1}{|E_v|}e^{-i\chi_{v,e'}} T_{v,e'} \dot{\varphi}_{v,e'} = -\tan \left( \frac{\delta_v}{2} \right) \varphi|_v.
    \end{array}
  \end{equation*}
  \renewcommand\arraystretch{1}
  Analogously, for $\Phi \in \mathcal{H}^2 \cup \dom \Delta_{A,U}$ one gets the boundary conditions
  \renewcommand\arraystretch{1.33}
  \begin{equation*}
    \begin{array}{c c}
      \displaystyle
            \varphi|_{\Gamma_{\text{ext}}} = 0, \\
      \displaystyle
      e^{-i\chi_{v,e'}}T_{v,e'} \varphi_{v,e'} = \varphi|_{v}, \quad e'\in E_v \smallsetminus \{e\},\\
      \displaystyle
      \sum_{e' \in E_v} \frac{1}{|E_v|}e^{-i\chi_{v,e'}} T_{v,e'} (\dot{\varphi}_A)_{v,e'} = -\tan \left( \frac{\delta_v}{2} \right) \varphi|_v
    \end{array}
  \end{equation*}
  \renewcommand\arraystretch{1}
  where, again, $\dot{\varphi}_A = \gamma(\d_A\Phi(\nu))$.

  Once the relations to be satisfied by the functions on the junctions between different edges have been written explicitly, let us review the role of the parameters $\delta_v$ and $\chi_{v,e}$.
  It is clear that whenever $\delta_v = 0$, the boundary conditions are quasi-Kirchhoff boundary conditions (see \cite{BerkolaikoKuchment2012}).
  When, in addition to $\delta_v = 0$, we have $\chi_{v,e} = 0$ the boundary conditions are just
  Kirchhoff boundary conditions imposing continuity of the function $\Phi$ and the conservation of the flux of the normal derivatives at that point.
  Regarding $\delta_v$, note that, when $\chi_{v,e} = 0$, the
  boundary conditions impose continuity for the function $\Phi$ on the connections but a net flux of the normal derivatives proportional to the trace of $\Phi$ at that vertex exists; in other words,
  $\delta_v$ represents a delta-like interaction supported on the connections.

  To illustrate these situations, let us now introduce some concrete systems to exemplify the construction above.
  In all the examples we will consider the canonically flat metric.
  \begin{figure}[t]
    \begin{subfigure}[b]{0.45\textwidth}
      \centering
      \includegraphics[width=\textwidth]{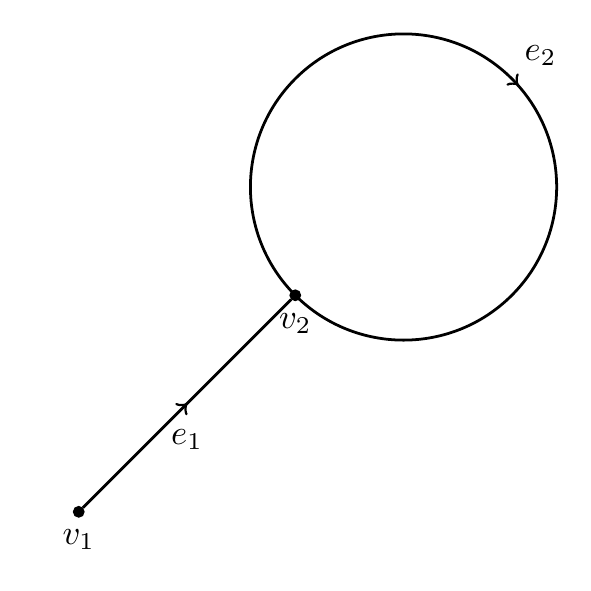}
      \caption{}\label{fig:example_graph}
    \end{subfigure}
    \begin{subfigure}[b]{0.45\textwidth}
      \centering
      \includegraphics[width=\textwidth]{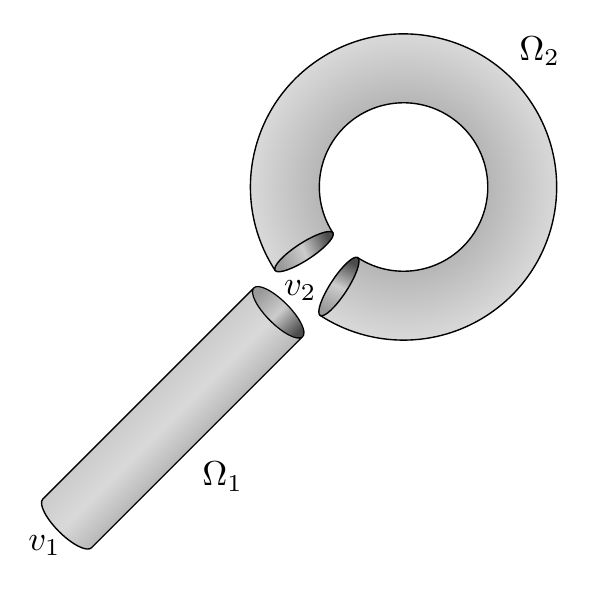}
      \caption{}\label{fig:example_cylinders}
    \end{subfigure}
    \caption{
      Examples of Quantum Circuits defined on the same graph $G$.
      The Quantum Circuit (A) is 1-dimensional (Quantum Graph).
      The Quantum Circuit (B) is 2-dimensional.
    }
    \label{fig:examples}
  \end{figure}
  \begin{example}
    Consider the graph $G$ depicted in Fig~\ref{fig:example_graph}.
    For this first example we are going to consider the simplest class of Quantum Circuits: those of dimension 1.
    For the manifold $\Omega$, consider two non-connected copies of the unit interval $\Omega_1 = [0,1]$, $\Omega_2 = [0,1]$.
    It is clear that $\partial\Omega_e = \{0_e, 1_e\}$, and the elements in the partition are the following:
    \begin{gather*}
      \Gamma_{v_1,e_1} = \{0_{e_1}\}, \quad
      \Gamma_{v_2,e_1} = \{1_{e_1}\}, \quad
      \Gamma_{v_2,e_2} = \{0_{e_2}, 1_{e_2}\}, \\
      \Gamma_{\text{ext},e} = \emptyset, \qquad (e = e_1, e_2).
    \end{gather*}
    In such a case, our Quantum Circuit reduces to a Quantum Graph, and our definition of quasi-$\delta$, quasi-Kirchhoff and Kirchhoff boundary conditions correspond with the vertex conditions equally named (cf.\ \cite{BerkolaikoKuchment2012}).
    In this simple case $L^2(\partial\Omega) = \mathbb{C}^4$, and there is no need for the unitaries $\mathbb{T}_v$, as all the pieces of the partition of the boundary are canonically diffeomorphic.
    At the vertex $v_1$, which is connected only to one edge, the boundary conditions become Robin boundary conditions.
  \end{example}
  \begin{example}
    Consider the situation depicted in Fig.~\ref{fig:example_cylinders}.
    The graph associated with this Quantum Circuit is the same $G$ (depicted in Fig. \ref{fig:example_graph}) from the previous example; however, the manifold is now given by two cylindrical surfaces, $\Omega_1$ and $\Omega_2$.
    Now $\partial\Omega_e = c_{e,-}\cup c_{e,+}$, where $c_{e,\pm}$ denote the circumferences at the edges of the cylindrical surfaces.
    As before, $\Gamma_{\text{ext},e} = \emptyset$ for both edges, and now
    \begin{equation*}
      \Gamma_{v_1,e_1} = c_{e_1,-}, \quad
      \Gamma_{v_2,e_1} = c_{e_1,+}, \quad
      \Gamma_{v_2,e_2} = c_{e_2,-} \cup c_{e_2,+}.
    \end{equation*} If both cylinders have the same radius, then the isometries $T_{v,e}$ defining the unitaries $\mathbb{T}_v$ are just the natural ones transforming $c_{e,\pm}$ into the unit circle. If the radii are different, then the isometries include a scaling factor.
  \end{example}
  \begin{example}
    Consider a circuit with the same graph $G$ as in the previous examples. $\Omega$ is the disjoint union of two rectangles. The boundary of each rectangle is the disjoint union of four intervals. Two opposite sides of each rectangle will provide the elements $\Gamma_{v_1,e}$, $e = e_1,e_2$, of the partition while the other four sides, two for each rectangle, will be the elements in $\Gamma_{\text{ext},e}$, $e = e_1,e_2$.
  \end{example}

\section{Quasi-$\delta$ boundary conditions} \label{sec:quasi_delta_BC}
  In order to show the viability of the Quantum Control at the Boundary scheme, we are going to restrict ourselves to a subfamily of the set of self-adjoint extensions of the Laplace-Beltrami operator on Quantum Graphs: the ones given by quasi-$\delta$ boundary conditions (cf.\ \ref{def:quasi_delta_BC}).

  The Hamiltonians corresponding with quasi-$\delta$ boundary conditions do not have time-independent form domain, and therefore the results exposed in Chapter \ref{ch:H_form_domain} cannot be applied to them directly; however, the following theorem shows that those results can be applied to a related system.

  \begin{theorem} \label{thm:quasi_delta_BCS_equivalence}
    Let $(G,\Omega,\Gamma)$ be a Quantum Circuit.
    Let $V$ be the vertex set of the graph $G$ and for $v\in V$ let $E_v$ be the set of edges in $G$ that share the vertex $v$.
    Let $U(t)$, $t\in\mathbb{R}$, be a time-dependent family of unitary operators defining quasi-$\delta$ boundary conditions with parameters $\chi_{v,e}(t)\in C^1(\mathbb{R})$ and $\delta_v \in (-\pi,\pi)$, $v\in V$ and $e\in E_v$.
    Let $H(t)$ be the time-dependent Hamiltonian defined by the family of Laplace-Beltrami operators $\Delta_{U(t)}$, and let $\Psi(t)$ be a solution of the strong Schrödinger equation determined by $H(t)$.
    Then there exists a family of unitary operators $\{J(t)\}_{t\in\mathbb{R}}$ such that the curve $J(t)\Psi(t)$ is a solution of the strong Schrödinger equation determined by a time-dependent Hamiltonian of the form
    $$\tilde{H}(t) = \Delta_{A(t),\tilde{U}} + \Theta'(t),$$
    where $\tilde{U}$ is the unitary operator associated with quasi-$\delta$ type boundary conditions with parameters $\tilde{\chi}_{v,e} = 0$, $\tilde{\delta}_v = \delta_v$, $v\in V$, $e\in E_v$, $A(t) \in \Lambda(\Omega)$ is a time-dependent magnetic potential, and $\Theta(t)\in C^{\infty}(\Omega)$ such that $A(t) = \d\Theta(t)$ and $\Theta'(t) = \frac{\d}{\d t} \Theta(t)$.
  \end{theorem}

  \begin{proof}
    For a given family of parameters $\chi_{v,e}(t)$, $\delta_v$, the  evolution associated to $H(t)$ is given by the Schrödinger equation
    \begin{equation} \label{eq:schrodinger_quasi_delta}
      \frac{\d}{\d t} \Phi(t) = -i \Delta_{U(t)} \Phi(t),
    \end{equation}
    where $U(t)$ is the unitary defining quasi-$\delta$ boundary conditions with parameters $\chi_{v,e}(t), \delta_v$, cf.\ Def. \ref{def:quasi_delta_BC}.

    By definition, the closure of each $\Gamma_{v,e}\in \Gamma$, $e\in E$, $v\in V_e$ is contained in a simply connected open neighbourhood $\mathcal{V}_{v,e}$ and these are pairwise disjoint; therefore, for each $e\in E$ there exists $\Theta_e\in C^\infty(\Omega_e)$ such that  for any vertex $v \in V_e$ one has $T_{v,e}(\gamma(\Theta_e)|_{\Gamma_{v,e}}) = \chi_{v,e}$ where $T_{v,e}$ is the isometry defined in Eq.~\eqref{eq:definition_Te}.
    Moreover, if $\chi_{v,e}(\cdot)\in C^{a}(\mathbb{R})$ for $a\in\mathbb{N}$, then $\Theta_e$ can be chosen such that $\Theta_e: \mathbb{R} \to C^\infty(\Omega_e)$ is $a$ times continuously differentiable.
    Let $\Theta(t) := \bigoplus_{e \in E} \Theta_e(t) \in C^\infty(\Omega)$, $t\in\mathbb{R}$.
    By construction $T_{v,e}(\gamma(\Theta(t))|_{\Gamma_{v,e}}) = \chi_{v,e}(t)$.
    Define the magnetic potential $A(t) = \d \Theta(t) = \bigoplus_{e \in E} \d \Theta_e(t)$ and denote by $J(t)$ the family of unitary transformations on $L^2(\Omega)$ given by $J(t) = \bigoplus_{e \in E} J_e(t)$, where $J_e(t) : \Phi_e \in L^2(\Omega_e) \mapsto e^{-i\Theta_e(t)} \Phi_e \in L^2(\Omega_e)$.
    Some consequences follow straightforwardly:
    \begin{enumerate}[label=\textit{(\roman*)}]
      \item For every $\Phi \in H^1(\Omega)$, the product rule yields
        \begin{equation*}
          \d_{A(t)} (J(t) \Phi) = \d(J(t)\Phi) + i A(t) J(t)\Phi = J(t) \d \Phi.
        \end{equation*}
      \item For $\Phi \in \dom \Delta_{A(t),U(t)}$, $\Psi = J(t) \Phi$ satisfies $\delta$-type boundary conditions with the parameters $\delta_v$. This is a direct application of Corollary~\ref{corol:magnetic_laplacian_equivalence}.
      \item For a curve $\Phi(t) \in \dom \Delta_{U(t)}$ satisfying the strong Schrödinger equation
        \eqref{eq:schrodinger_quasi_delta}, $\Psi(t) = J(t) \Phi(t)$ satisfies
        \begin{equation*}
          \begin{alignedat}{2}
            \frac{\d}{\d t} \Psi(t)
            &= \left( \frac{\d}{\d t}J(t) \right) \Phi(t) + J(t) \frac{\d}{\d t} \Phi(t) \\
            &= -i \left[ J(t) \Delta_{U(t)} J(t)^{-1} + \Theta'(t) \right] \Psi(t).
          \end{alignedat}
        \end{equation*}
        By Corollary~\ref{corol:magnetic_laplacian_equivalence}, this is equivalent to
        \begin{equation*}
          \frac{\d}{\d t} \Psi(t) = -i [\Delta_{A(t), \tilde{U}(t)} + \Theta'(t)] \Psi(t),
        \end{equation*}
        where $\tilde{U}(t)$ is the unitary associated with $\delta$-type boundary conditions with
        parameters $\delta_v$. \qedhere
    \end{enumerate}
  \end{proof}

  This theorem allows us to relate quantum systems whose Hamiltonians are the Laplace-Beltrami operator with quasi-$\delta$ type boundary conditions and systems whose Hamiltonians are given by magnetic Laplacians.
  Part of the time-dependence in the initial problem is moved to the analytical form of the Hamiltonian, but we still have a system with time-dependent operator domain: even if $\tilde{U}(t)$ in the previous theorem does not depend on $t$, since the magnetic potential $A(t)$ depends on $t$, $\mathcal{D}_{U(t),A(t)}$ (cf.\ Def.\ \ref{def:boundary_equation_operator} and Corollary \ref{corol:magnetic_laplacian_equivalence}).
  However, the domains of the family of associated sesquilinear forms do not depend on the parameter $t$ as shown in the next result.

  \begin{corollary} \label{prop:magnetic_system_form_domain}
    Let $\tilde{H}(t)$, $t\in I$, be as in Theorem \ref{thm:quasi_delta_BCS_equivalence}.
    Then $\tilde{H}(t)$ is a time-dependent Hamiltonian with constant form domain, cf.\ Def. \ref{def:hamiltonian_const_form},
    \begin{equation*}
      \mathcal{H}^+ = \{\Phi \in H^1(\Omega) \mid P\varphi = 0\}.
    \end{equation*}
    Moreover, if $\chi_{v,e}(t)$ is three times continuously differentiable, there exists a unitary propagator solving the associated strong Schrödinger equation.
  \end{corollary}
  \begin{proof}
    Since the graph has a finite number of vertices, $\inf_{v\in V} |\delta_v \pm \pi| > 0$. From \cite[Prop. 3.11]{IbortLledoPerezPardo2015} it follows that the corresponding Cayley transforms satisfy the inequality $\sup_t \|\mathcal{C}_{U(t)}\| < \infty$, where $U(t)$ denotes the unitary associated with quasi-$\delta$ boundary conditions with parameters $\chi_{v,e}(t)$ and $\delta_v$.
    Since $\Omega$ is compact, $\Theta'(t)$ is a bounded operator on $L^2(\Omega)$. For each $t\in I$, $A(t)$ is a differentiable form on $\Omega$ and, since $I$ is a compact interval, $\sup_{t\in I} \|A\|_{\infty} < \infty$. Corollary~\ref{corol:uniform_semibound_equivalence_formnorms} implies the uniform lower bound for $\tilde{H}(t)$.
    The form domain of $\tilde{H}(t)$ is
    \begin{equation*}
      \mathcal{H}^+ = \dom Q_{A(t),U(t)} = \{\Phi \in H^1(\Omega) \mid P\varphi = 0\},
    \end{equation*}
    where $P$ is the orthogonal projector onto the eigenspace of $U(t)$ associated with the eigenvalue -1 of the unitary operator (see Definition \ref{def:quasi_delta_BC}), which only depends on the topology of the graph $G$ associated with the Quantum Circuit.

    Theorem~\ref{thm:kisynski} ensures the existence of solutions for the Schrödinger equation, since the associated form is given by
    \begin{equation*}
      \tilde{h}_t(\Phi,\Phi) = \|\d\Phi\|^2 + \|\Phi A(t)\|^2 + 2\Im\langle \d\Phi, \Phi A(t) \rangle - \langle \varphi, \mathcal{C}_{U(t)}\varphi \rangle + \langle \Phi, \Theta'(t)\Phi \rangle
    \end{equation*}
    with $\Theta(t)$ and $A(t)$ defined in Thm.~\ref{thm:quasi_delta_BCS_equivalence} and $\mathcal{C}_{U(t)}$ defined in Eq.~\eqref{eq:circuits_cayley}.
  \end{proof}
\end{document}

\ifSubfilesClassLoaded{
    \setcounter{chapter}{6}
  }{}
\chapter{Quantum boundary control systems} \label{ch:control_systems}

The previous sections have developed the tools and notions needed to finally address the central problem of this work: the viability of que Quantum Control at the Boundary scheme.
In in Section \ref{sec:control_systems_def}, two control systems on Quantum Circuits are introduced: the quasi-$\delta$ boundary control system and the quantum induction control system.
The relation between them is exploited in order to prove the existence of dynamics for the boundary control system.
After proving they have well defined dynamics, Section \ref{sec:controllability} addresses the problem of (approximate) controllability for both control systems.
Applying the stability results proven in Chapter \ref{ch:quantum_circuits} and the results in \cite{ChambrionMasonSigalottiEtAl2009} we are able to prove first approximate controllability for the quantum induction control system.
This controllability result can be transferred to quasi-$\delta$ boundary control systems by means of the relation in Thm.\ \ref{thm:quasi_delta_BCS_equivalence} and the stability result.

\section{Control systems on quantum circuits} \label{sec:control_systems_def}
  The boundary control systems studied in this section are a subset of the quantum systems described in Section \ref{sec:quasi_delta_BC}: Quantum Circuits whose Hamiltonian have time-dependent quasi-$\delta$ boundary conditions with a particular time dependence on the parameters.

  \begin{definition}\label{def:boundary_control_system}
    Let $(G,\Omega,\Gamma)$ be a Quantum Circuit.
    Let $V$ be the vertex set of the graph $G$ and for $v\in V$ let $E_v$ be the set of edges in $G$ that share the vertex $v$.
    Let $\bar{\chi}_{e,v} \in [0,2\pi]$, $v\in V$, $e \in E_v$.
    A \emph{quasi-$\delta$ boundary control system} is the quantum control system with controls $\mathcal{C} = \mathbb{R}$ and family of Hamiltonians given by the Laplace-Beltrami operator on $\Omega$ with quasi-$\delta$ boundary conditions with parameters $\chi_{e,v} = c \bar{\chi}_{e,v} $, $c\in \mathcal{C}$, and $\delta_v \in (-\pi, \pi)$, $v\in V$, $e\in E_v$.
  \end{definition}

  Theorem \ref{thm:quasi_delta_BCS_equivalence} relates de dynamics of a quasi-$\delta$ boundary control system with the following one.

  \begin{definition}\label{def:induction_control_system}
    Let $(G,\Omega,\Gamma)$ be a Quantum Circuit.
    Let $V$ be the vertex set of the graph $G$ and for $v\in V$ let $E_v$ be the set of edges in $G$ that share the vertex $v$.
    Let $r$ be a positive real number, $A_0 \in \Lambda(\Omega)$ a smooth differential one-form and $\Theta_0 \in C^\infty (\Omega)$ a smooth function on $\Omega$ such that $\d\Theta_0  =A_0$.
    Let  $U$ be the unitary operator defining quasi-$\delta$ boundary conditions with parameters $\chi_{e,v} =0$ and $\delta_v \in (-\pi, \pi)$, $v\in V$, $e\in E_v$.

    A \emph{quantum induction control system} is a quantum control system with space of controls $\mathcal{C} = \{ (a,b)\in\mathbb{R}^2\mid b <r\}$ and family of Hamiltonians $H(a,b) = \Delta_{aA_0,U} + b\Theta_0$, where $\Delta_{aA_0,U}$ is the self-adjoint extension of the magnetic Laplacian associated with the magnetic potential $aA_0\in\Lambda(\Omega)$ and the unitary $U$, and such that the control function is of the type
    $$\begin{array}{rccc}u&:\mathbb{R} &\to& \mathcal{C}\\ & t &\mapsto & (a(t), b(t)),\end{array}$$
    with $b(t) = \frac{\d a}{\d t}(t)$ almost everywhere.
  \end{definition}

  The reason for letting $b(t)$ be the derivative of $a(t)$ only almost everywhere is because we shall consider functions that are piecewise differentiable and thus the derivative might be undefined at certain points.
  The term quantum induction refers to the fact that the Schrödinger equation associated with quantum induction control systems is the one corresponding to a particle moving in the Quantum Circuit, subject to the action of a time-dependent magnetic field that is concentrated on the loops of the graph and an electric field whose strength is proportional to the variation of the magnetic field in the same way that is described by Faraday's Induction Law.

  Theorem \ref{thm:quasi_delta_BCS_equivalence} associates to a quasi-$\delta$ boundary control systems with parameters $\chi_{v,e}(t) = f(t)\chi_{v,e}$, $\delta_{v}$, where $\chi_{v,e}$ is fixed, a quantum induction control system with $\Theta(t) = f(t)\Theta_0 \in C^2(\Omega)$ and $A(t) = f(t) A_0$, where $A_0 = \d \Theta_0$.
  Let us end this section applying some results from Chapter \ref{ch:H_form_domain} to this particular case.

  \begin{proposition} \label{prop:magnetic_family_formlinear}
    Let $I \subset \mathbb{R}$ be a compact interval, let $A_0$ be a magnetic potential and let $\Theta_0$ be a function such that $\d\Theta_0 = A_0$.
    For $n \in \boldsymbol{N} \subset \mathbb{N}$, let $u_n(t), v_n(t) \in C_p^2(I)$ such that $\sup_{n,t} |u_n(t)| < \infty$ and $\sup_{n,t} |v_n(t)| < \infty$.
    For each $t \in I$, denote by $\Delta_{u_n(t) A_0}$ the magnetic Laplacian with potential $u_n(t)A_0$ and $\delta$-type boundary conditions with $t$-independent parameters.
    Define the Hamiltonians $H_n(t) \coloneqq \Delta_{u_n(t) A_0} + v_n(t)\Theta_0$.
    The following statements hold:
    \begin{enumerate}[label=\textit{(\alph*)},nosep,leftmargin=*,ref=\ref{prop:magnetic_family_formlinear}\alph*]
      \item \label{prop:magnetic_formlinear_dynamics}
        $\{H_n(t)\}_{n \in \boldsymbol{N}}$ is a family of form-linear, time-dependent Hamiltonians, and for each $n \in \boldsymbol{N}$ there exists a unitary propagator $U_n(t,s)$ solving the weak Schrödinger equation for $H_n(t)$.
      \item \label{prop:magnetic_formlinear_normequiv}
        There is $c$ independent of $n$ and $t$ such that
        \begin{equation*}
          c^{-1} \|\cdot\|_{\pm,n,t} \leq \|\cdot\|_{\pm} \leq c\|\cdot\|_{\pm,n,t}.
        \end{equation*}
      \item \label{prop:magnetic_formlinear_stability}
        If, in addition, $\sup_{n \in \boldsymbol{N}} \sum_j \|u_n'\|_{L^1(I_j)} < \infty$ and $\sup_{n \in \boldsymbol{N}} \sum_j \|v_n'\|_{L^1(I_j)} < \infty$, where $\{I_j\}_{j=1}^\nu$ denotes the family of open intervals on which $u_n, v_n$ are differentiable, then there is a constant $L$ such that for every $n \in \boldsymbol{N}$ and $t, s \in I$ it holds
        \begin{alignat*}{2}
          \|U_n(t,s) - U_{n'}(t,s)\|_{+,-}
          \leq L &\left(\|u_n - u_{n'}\|_{L^1(s,t)}
          + \|u_n^2 - u_{n'}^2\|_{L^1(s,t)} \right. \\
          &\left.+ \|v_n - v_{n'}\|_{L^1(s,t)}\right).
        \end{alignat*}
    \end{enumerate}
  \end{proposition}
  \begin{proof}
    The sesquilinear form defined by $H_n$ can be written as
    \begin{equation*}
      h_{n,t}(\Phi,\Phi) = h_0(\Phi,\Phi) + u_n(t)^2 \|\Phi A_0\|^2 + 2u_n(t) \Im\langle \d\Phi, \Phi A_0 \rangle + v_n(t) \langle \Phi, \Theta_0 \Phi \rangle,
    \end{equation*}
    where $h_0(\Phi,\Phi) = \|\d\Phi\|^2 - \langle \varphi, \mathcal{C}_{U(t)} \varphi \rangle$ does not depend on $t$ since $\delta_v$ is constant (see Eq.~\eqref{eq:circuits_cayley}).
    Since $\Omega$ is compact, $\Theta_0$ and $A_0$ are bounded and the boundedness of $u_n(t), v_n(t)$ and Corollary~\ref{corol:uniform_semibound_equivalence_formnorms} shows that the Hamiltonians $H_n(t)$ are semibounded from below uniformly.
    By Proposition~\ref{prop:dynamics_formlinear}, \emph{(a)} follows.

    Property \emph{(b)} follows from applying Proposition~\ref{prop:formlinear_equiv_norms}.

    Now, since $\Omega$ is bounded, $h_1(\Phi,\Phi) \coloneqq \|\Phi A_0\|^2$ and $h_3(\Phi,\Phi)\coloneqq \langle \Phi, \Theta_0 \Phi \rangle$ are bounded with respect to the norm $\|\cdot\|\leq \|\cdot\|_+$.
    Finally, since $\|\cdot\|_{+} \sim \|\cdot\|_1$ (cf.\ Theorem~\ref{thm:quadratic_form_H1bounded}), the form $h_2(\Phi,\Phi) \coloneqq 2 \Im \,\langle \d\Phi, \Phi A_0 \rangle$ is bounded with respect to $\|\cdot\|_+$.
    Therefore, there is $K$ such that $h_i(\Phi,\Phi) \leq K \|\Phi\|_+$ for $i = 1,2,3$.

    Since $\sup_{n,t} |u_n(t)| <\infty$, $I$ being compact, and $\sup_{n} \sum_j \|u_n'(t)\|_{L^1(I_{j})} <\infty$, it follows $\sup_{n} \sum_j \|\frac{\d}{\d t}[u_n]^2\|_{L^1(I_j)} <\infty$.
    Therefore, Theorem~\ref{thm:stability_formlinear} applies, which concludes the proof.
  \end{proof}
  
\section{The controllability problem} \label{sec:controllability}

  The approximate controllability problem for a quantum boundary control system consists in answering whether it is possible to drive the system from any initial state to a small neighbourhood of any target state by only modifying its boundary conditions.
  In our particular setting, i.e.\ quasi-$\delta$ boundary control systems, this is done by choosing a family of curves $\chi_{v,e}(t)$.

  By Theorem~\ref{thm:quasi_delta_BCS_equivalence}, the control problem for a quasi-$\delta$ boundary control system is closely related to the control problem for the associated induction control problem, given by the magnetic Hamiltonian
  \begin{equation} \label{eq:magnetic_controlled_H}
    H(t) = \Delta_{A(t)} + \Theta'(t),
  \end{equation}
  with controls $A(t), \Theta'(t)$ such that $A(t) = \d\Theta(t)$ and with $\delta$ boundary conditions.

  We will prove now Theorem~\ref{thm:controllability_piecewise_smooth}, i.e., weak approximate controllability of quantum induction control systems; this is a stronger version of a theorem proven in \cite{BalmasedaPerezPardo2019} in the case of Quantum Graphs.
  In order to do that, we rely on Theorem~\ref{thm:chambrion_controllability} and the stability proven in Theorem~\ref{thm:stability_bound} and Proposition~\ref{prop:magnetic_family_formlinear}.
  Notice that, even if the Hamiltonian of a quantum induction control system is similar to that of a normal bilinear control system (cf.\ \ref{def:normal_bilinear_CS}), the fact that the control function appears with its derivative does not allow for a direct application of Theorem~\ref{thm:chambrion_controllability}.
  To circumvent this issue we proceed in two steps.
  First, we will define an auxiliary system to which Chambrion et al.'s theorem applies; then, we will use the controls provided by Theorem~\ref{thm:chambrion_controllability} to construct a sequence of Hamiltonians converging, in the sense of Proposition~\ref{prop:magnetic_formlinear_stability}, to the quantum induction one.

  \begin{theorem}\label{thm:controllability_piecewise_smooth}
    Let $r \in \mathbb{R}$ be a positive real number and $\mathcal{C} = \{(a,b) \in \mathbb{R}^2 \mid b < r\}$.
    A quantum induction control system is weakly approximately controllable with control function
    \begin{equation*}
      \begin{array}{rccc}
        u&:[0,T] &\to& \mathcal{C}\\
         & t &\mapsto & (a(t), b(t)),
      \end{array}
    \end{equation*}
    such that $a(t)$ is piecewise linear and $b(t) = \frac{d a}{d t}(t)$ almost everywhere.
  \end{theorem}
  \begin{proof}
    For any $u_0 > 0$, define the auxiliary system with Hamiltonian
    \begin{equation*}
      H_0(t) = \Delta_{u_0 A_0} + v(t)\Theta_0.
    \end{equation*}
    for some magnetic potential $A_0$ and $\Theta_0$ such that $d\Theta_0=A_0$.
    We have omitted the subindex $U$ denoting the boundary conditions of the magnetic Laplacian as it will remain fixed.
    Since the Thick Quantum Graph is defined on a compact manifold, $\Theta_0$ defines a bounded potential.
    Moreover, since $u_0A_0$ is fixed, the operator domain of $H_0(t)$ does not depend on $t$, and $\Delta_{u_0 A_0}$ has compact resolvent since the Thick Quantum Graph is a compact manifold.
    We will assume that the conditions on the eigenvalues and eigenfunctions of Theorem~\ref{thm:chambrion_controllability} are met.
    These conditions are met generically in the systems under study.

    Hence, for every initial and target states $\Psi$, $\Psi_T$ with $\|\Psi_T\|=\|\Psi\|$, every $\varepsilon > 0$ and every $r > 0$, there exists $T>0$ and $v(t): [0, T] \to (0, r)$ piecewise constant such that the evolution induced by $H_0(t)$, $\Psi_0(t) = U_0(t,0)\Psi$, satisfies $\Psi_0(0) = \Psi$ and
    \begin{equation*}
      \left\| \Psi_0(T) - \Psi_T \right\| < \frac{\varepsilon}{2}.
    \end{equation*}
    Note that, since the operator domain is fixed, the dynamics induced by $H_0(t)$, with a piecewise constant $v(t)$, is defined by products of unitary operators determined by Stone's Theorem.

    Now we will construct a sequence of Hamiltonians whose dynamics will converge to the auxiliary one.
    For each $n \in \mathbb{N}$, divide the time interval $I=[0,T]$ into $n$ pieces of length $\tau = T/n$.
    Let $\{I_{n,j}\}_{j=1}^{m_n}$ be the coarsest refinement of the partition $\{((k-1)\tau, k\tau)\}_{k=1}^{n}$ such that $v(t)$ is constant with value $v_{n,j}$ on $I_{n,j}$ for $j = 1, 2, \dots, n$.
    Define $t_{n,j}$ such that $I_{n,j} = (t_{n,j}, t_{n,j+1})$.
    By construction it follows that $t_{n, j+1} - t_{n,j} \leq \tau$ for $1\leq j \leq m_n$.

    For $1 \leq j \leq m_n$, define the functions $u_{n,j}: [0,T] \to \mathbb{R}$ by
    \begin{equation*}
      u_{n,j}(t) =
      \begin{cases}
        0 & \text{if }t \notin I_{n,j}, \\
        u_0 + \int_{t_{n,j}}^t v(s) \,\d s &\text{if } t \in I_{n,j}.
      \end{cases}
    \end{equation*}
    Take $u_n(t) = \sum_{j=1}^{m_n} u_{n,j}(t)$.
    By Lemma~\ref{prop:magnetic_formlinear_dynamics}, there is a unitary propagator
    $U_n(t,s)$, $t,s \in I$, solving the weak Schrödinger equation with Hamiltonian
    \begin{equation*}
      H_n(t) = \Delta_{u_n(t)A_0} + v(t)\Theta_0.
    \end{equation*}
    Note that, by definition, $u_n'(t) = v(t)$ for almost every $t \in I$ and for every $n$ and every $t \in I$ we have $|u_n'(t)| = |v(t)| < r$, $|u_n(t)| \leq u_0 + r T$ and $v'(t) = 0$.

    For every $t \in \bigcup_{j=1}^{m_n} I_{n,j}$ we have 
    \begin{equation*}
      |u_n(t) - u_0| = \left|\int_{t_{n,j}}^{t} v(s)\,\d s \right|
      \leq \frac{rT}{n},
    \end{equation*}
    and
    \begin{equation*}
      |u_n^2(t) - u_0^2| = 2u_0\int_{t_{n,j}}^t v(s)\,\d s + \left( \int_{t_{n,j}}^t v(s)\,\d s \right)^2
      \leq u_0 \frac{rT}{n} + \frac{r^2T^2}{n^2}.
    \end{equation*}
    It follows
    \begin{equation} \label{eq:convergence_un}
      \|u_n - u_0\|_{L^1(I)} \leq \frac{rT^2}{n} \quad\text{and}\quad
      \|u_n^2 - u_0^2\|_{L^1(I)} \leq  u_0 \frac{rT^2}{n} + \frac{r^2T^3}{n^2}.
    \end{equation}
    Finally, Lemma~\ref{prop:magnetic_formlinear_stability} applies yielding
    \begin{equation*}
      \|U_0(t, s) - U_n(t, s)\|_{+,-} < L (\|u_n(t) - u_0\|_{L^1(s,t)} + \|u_n(t)^2 - u_0^2\|_{L^1(s,t)}),
    \end{equation*}
    where the constant $L$ is independent of $t, s, n$.
    By Equation~\eqref{eq:convergence_un} it follows
    \begin{equation*}
      \lim_{n \to \infty} \|U_0(t, s) - U_n(t, s)\|_{+,-} = 0
    \end{equation*}
    uniformly on $s,t \in I$.

    Applying now an $\varepsilon/2$ argument we conclude that, for $n$ large enough,
    \begin{equation*}
      \|\Psi_n(T) - \Psi_T\|_- \leq \varepsilon. \qedhere
    \end{equation*}
  \end{proof}

  \begin{proposition}\label{prop:induction_app_controllability_H-}
    Let $r \in \mathbb{R}$ be a positive real number, $u_1,u_0\in\mathbb{R}$ and $\mathcal{C} = \{(a,b) \in \mathbb{R}^2 \mid b<r\}$.
    A quantum induction control system is weakly approximately controllable with control function
    \begin{equation*}
      \begin{array}{rccc}
        u&:[0,T] &\to& \mathcal{C}\\
         & t &\mapsto & (a(t), b(t)),
      \end{array}
    \end{equation*}
    piecewise linear such that $b(t) = \frac{\d a}{\d t}(t)$ almost everywhere, $a(T)=u_1$ and $a(0)=u_0$.
  \end{proposition}
  \begin{proof}
    Let $H(a,b)$ be the family of Hamiltonians of the quantum induction control system, and define $\tilde{H}(a) \coloneqq H(a,0)$ for $a\in \mathbb{R}$.
    A direct application of Theorem~\ref{thm:stability_bound} with $H_1(t) = 0$ and $H_2(t) = \tilde{H}(a)$, for $t\in\mathbb{R}$, shows that for any $\Phi\in\mathcal{H}^+$,
    \begin{equation*}
      \lim_{p\to0} \left\|(e^{-i\tilde{H}(a)p} - \mathbb{I})\Phi\right\|_- = 0.
    \end{equation*}
    Now take $\Psi_0,\Psi_T\in\mathcal{H}^+$ with $\|\Psi_T\| = \|\Psi_0\|$ and $q>0$.
    By Theorem~\ref{thm:controllability_piecewise_smooth}, for any $\varepsilon>0$ there exist $\tilde{T}>0$ and a piecewise linear control function $u\colon[q,\tilde{T}]\to\mathbb{R}$, with $|\frac{du}{dt}|\leq r$ almost everywhere and such that the solution $\tilde{U}(\tilde{T},q)$ of the quantum induction control problem satisfies
    \begin{equation*}
      \left\|\Psi_T - \tilde{U}(\tilde{T},q)e^{-i\tilde{H}(u_0)q}\Psi_0\right\|_- < \frac{\varepsilon}{3}.
    \end{equation*}
    Define $\xi\in\mathcal{H}^+$ by $\xi:= \Psi_T - \tilde{U}(\tilde{T},q)e^{-i\tilde{H}(u_1)q}\Psi_0$.
    For any $p>0$ we have
    \begin{multline*}
      \left\|\Psi_T - e^{-i\tilde{H}(u_1)p}\tilde{U}(\tilde{T},q)e^{-i\tilde{H}(u_0)q}\Psi_0\right\|_-
      \leq \left\|(e^{-i\tilde{H}(u_1)p} - \mathbb{I})\Psi_T\right\|_- + \\
      + \left\|(e^{-i\tilde{H}(u_1)p} - \mathbb{I})\xi\right\|_- + \|\xi\|_-.
    \end{multline*}
    Taking $p$ small enough the right hand side can be made smaller than $\varepsilon$. Defining $T:=\tilde{T}+p$ the statement follows.
  \end{proof}

  \begin{theorem}\label{thm:controllability_boundary}
    Let $r\in\mathbb{R}$ be a positive real number and $u_1,u_0\in\mathbb{R}$.
    A quasi-$\delta$ boundary control system is weakly approximately controllable with piecewise linear control function $u:[0,T]\to\mathbb{R}$ satisfying $|\frac{du}{dt}|<r$, $u(T)=u_1$ and $u(0)=u_0$.
  \end{theorem}
  \begin{proof}
    Let $V$ be the vertex set of the Thick Quantum Graph and, for $v\in V$, let $E_v$ be the set of edges of the Thick Quantum Graph that share the vertex $v$.
    Let $\delta_v\in (-\pi,\pi)$ for $v\in V$.
    Let $U_0$ be the admissible unitary operator defining quasi-$\delta$ boundary conditions with parameters $\chi_{v,e} = u_0\bar{\chi}_{v,e}$ and $\delta_v$, $v\in V$ and $e\in E_v$; $U_1$ is defined analogously with $u_0$ replaced by $u_1$.
    Let $\Psi_i\in\dom(\Delta_{U_i})$, $\|\Psi_i\|=1$, $i=0,1$.

    By Proposition~\ref{prop:magnetic_form_equivalence}, the sesquilinear forms associated to the Laplace-Beltrami operators $\Delta_{U_i}$, $i=0,1$, are unitarily equivalent to the ones associated to magnetic Laplacians with magnetic potential $u_iA_0$ and quasi-$\delta$ boundary conditions determined by $\chi_{v,e}=0$ and $\delta_v$, $v\in V$ and $e\in E_v$.
    The unitary operators that implement the equivalence are completely determined by the values of the parameters $u_0$ and $u_1$, and will be respectively denoted by $J_{u_0}$ and $J_{u_1}$. 

    Take $\Phi_i=J_{u_i}\Psi_i\in \dom (\Delta_{u_i A_0})$, $i=0,1$.
    By Theorem~\ref{thm:quasi_delta_BCS_equivalence} the solutions of the Schrödinger equation of the quasi-$\delta$ boundary control system are isomorphic to the solutions of a quantum induction control system, the isomorphism being a time-dependent unitary operator $J(t)$ (note that it is also unitary in $\mathcal{H}^-$ by construction, see Proposition \ref{prop:magnetic_form_equivalence}).
    Applying Theorem~\ref{thm:controllability_piecewise_smooth} to the initial and target states $\Phi_0$ and $\Phi_1$ it follows that, for any $\varepsilon>0$, there exists $T>0$ and a piecewise linear control function $u:[0,T]\to\mathbb{R}$ with $|\frac{du}{dt}| < r$, $u(T)=u_1$ and $u(0)=u_0$ such that 
    \begin{equation*}
      \|\Phi_1 - U(T,0)\Phi_0\|_- <\varepsilon,
    \end{equation*}
    where $U$ is the unitary propagator that solves the Schrödinger equation of the quantum induction system.
    The curve $J^\dagger(t)U(t,0)\Phi_0$ is a solution of the Schrödinger equation of the quasi-$\delta$ boundary control system.
    For each $t\in[0,T]$, the unitary operator $J(t)$ depends only on the value of the magnetic potential at time $t$ and in fact $J(T)=J_{u_1}$ and $J(0) = J_{u_0}$.
    Therefore, we have
    \begin{equation*}
      J^\dagger(0)U(0,0)\Phi_0 = J^\dagger_{u_0}\Phi_0= \Psi_0
    \end{equation*}
    and using the fact that $J_{u_1}$ is a bounded operator in $\mathcal{H}_{-}$
    \begin{alignat*}{2}
      \|\Psi_1 - J^\dagger(T)U(T,0)\Phi_0\|_- 
      &\leq K\|J_{u_1}\Psi_1 - U(T,0)\Phi_0\|_- \\
      &= K\|\Phi_1 - U(T,0)\Phi_0\|_-
      < K\varepsilon,
    \end{alignat*}
    as we wanted to show.
  \end{proof}

  \begin{remark}
    The controllability results provided above hold only on the weak sense, since the time-dependence of the Hamiltonians' domain hinders the existence of solutions in the strong sense.
    The main obstruction is the fact that the controls obtained by applying Theorem~\ref{thm:chambrion_controllability} are piecewise constant.
    If they were smooth, then Theorem~\ref{thm:kisynski} would ensure that the weak solutions are also strong solutions.
    A first attempt to get strong approximate controllability would be therefore to use the stability results developed in Section~\ref{sec:stability} to extend Theorem~\ref{thm:controllability_piecewise_smooth} to the case of smooth controls.
    A straightforward application of this procedure is not possible for these systems, since piecewise constant functions do not have $L^1$ integrable weak derivatives and the fact that both the control and its derivative appear on the Hamiltonian of quantum induction systems forbids the application of Theorem~\ref{thm:stability_formlinear}.
  \end{remark}

\end{document}

\backmatter
\renewcommand{\theequation}{\arabic{chapter}.\arabic{equation}}

\ifSubfilesClassLoaded{
    \setcounter{chapter}{7}
    \backmatter
  }{}
\chapter{Conclusions and further work}
\stepcounter{chapter}
  This dissertation provides a first proof of the viability for the Quantum Control at the Boundary scheme, showing that a certain family of quantum control systems can be indeed controlled (in the weak sense) using as space of controls the set of self-adjoint extensions of the symmetric Hamiltonian defining its free dynamics.

  In the process of proving weak approximate controllability for these systems, we have developed a number of concepts and techniques with a wider application potential.
  In Chapter \ref{ch:H_form_domain}, the details on the relation between B.\ Simon's and J.\ Kisyński's approaches to the existence of solutions for the Schrödinger equation with form-constant Hamiltonians were explored.
  Also, relying on Simon's ideas \cite{Simon1971}, in Sec.\ \ref{sec:stability} a stability result is obtained ensuring the strong convergence of the unitary propagators generated by Hamiltonians sharing a constant form domain $\mathcal{H}^+$, converging in $\mathcal{B}(\mathcal{H}^+, \mathcal{H}^-)$.
  This result generalising that of A.D.\ Sloan \cite{Sloan1981} is, to the best of our knowledge, the most general stability result for Hamiltonians with constant form domain.
  Moreover, it can be applied to a class of Hamiltonians general enough to fit the description not only of our control systems but also other interesting physical systems such us point-like interactions on quantum graphs or the Friedrichs-Lee model (cf.\ \cite{FacchiLigaboLonigro2021,AlbeverioGesztesyHoeghKrohnEtAl1988}).

  As mentioned in the introduction, Quantum Circuits defined in Chapter \ref{ch:quantum_circuits} can be used to provide genuine infinite-dimensional models for some physical situations.
  For such systems, the discussion on the self-adjoint extensions of the Laplace-Beltrami operator applies, providing a natural family of boundary conditions compatible with the topology of the underlying graph.

  Showing the viability of Quantum Control at the Boundary is the main aim of this dissertation, and therefore the importance of the controllability result for quasi-$\delta$ boundary control systems is clear.
  However, this is not the only important controllability result obtained in Chapter \ref{ch:control_systems}: as an intermediate step, approximate controllability for induction control systems has been shown.
  This constitutes an important result on itself, due to its applications and possible physical implementations.

  The most restrictive part of our approach is the controllability result we base on, by Chambrion et al.\ (cf. Thm.\ \ref{thm:chambrion_controllability}), since it imposes a very strict time-dependence structure of the Hamiltonian.
  Chambrion et al.'s result holds only for Hamiltonians of the form
  \begin{equation*}
    H(t) = H_0 + u(t) H_1,
  \end{equation*}
  with $H_0$, $H_1$ fixed self-adjoint operators and $u(t)$ a bounded piecewise constant control function.
  This piecewise structure has an important consequence for our boundary control systems, since with this kind of controls the existence of strong solutions of the Schrödinger equation is not guaranteed.
  This is the reason why we cannot obtain approximate controllability in the strong sense.

  In addition, this rigid structure is already present on the definition of the quantum boundary control systems used to show the feasibility of the Quantum Control at the Boundary scheme.
  In particular, the definition of quasi-$\delta$ boundary control systems restricts the space of controls to a one-parameter subfamily of the space of self-adjoint extensions of the Laplace-Beltrami operator compatible with the Quantum Circuit.
  More concretely, the time-dependent structure of the control functions $\chi_{v,e}(t)$ must be of the form $\chi_{v,e}(t) = f(t) \bar{\chi}_{v,e}$, with $\bar{\chi}_{v,e}$ fixed.
  With this time-dependence, the associated induction control system (see Thm.\ \ref{thm:quasi_delta_BCS_equivalence}) is of the form
  \begin{equation*}
    \tilde{H}(t) = H_0(u(t)) + u'(t) H_1.
  \end{equation*}
  This structure allows us to define convenient auxiliary systems for which Chambrion et al.'s result hold, while a more complicated time-dependence of $\chi_{v,e}(t)$ would lead to Hamiltonians of the form $\tilde{H}(t) = H_0(u(t)) + H_1(u(t))$, for which the same strategy cannot be applied.

  Therefore, it would be desirable to generalise Chambrion et al.'s controllability result in such a way that it allows us to use the ideas presented on this dissertation to more general cases.
  In particular, this generalisation could allow for the use of a general control curve on the set of $\delta$-type boundary conditions, with $\chi_{v,e}(t)$ depending on time in a generic way.
  Moreover, such a generalisation could enable the study of the controllability of systems whose dynamics is described by a family of Hermitian sesquilinear forms
  \begin{equation*}
    h_t(\Psi,\Phi) = h_0(\Psi, \Phi) + v_t(\Psi,\Phi), \qquad \Psi,\Phi \in \mathcal{H}^+,
  \end{equation*}
  where $h_0$ and $v_t$ are Hermitian sesquilinear forms densely defined on $\mathcal{H}$ such that $h_t$ is semibounded from below and closed.
  It defines a semibounded time-dependent Hamiltonian with constant form domain, and a similar strategy to the one used in Chapter~\ref{ch:control_systems} can be applied.
  Note that this structure captures not only a more general class of boundary control systems but also some other interesting quantum systems like those presented in \cite{AlbeverioGesztesyHoeghKrohnEtAl1988,FacchiLigaboLonigro2021}.

  Another step going further on the research on Quantum Control at the Boundary would be to find optimal control results and develop constructive methods providing the explicit control curves on the space of self-adjoint extensions for driving the system from a given initial state to a desired target state.
\end{document}

\printnomenclature[2cm]
\cleardoublepage
\bibliographystyle{plain}
\bibliography{tesis}

\end{document}